\documentclass[12pt]{article}

\usepackage{graphicx}
\usepackage[]{natbib}
\usepackage{tikz}
\usepackage{multirow}
\usepackage{dcolumn,booktabs}
\usepackage{algorithm,algorithmicx,algpseudocode,algcompatible}
\usepackage{mathtools} 
\usepackage{authblk} 
\usepackage{url} 
\usepackage{enumitem}
\usepackage{setspace} 
\usepackage[font=small]{caption} 
\usepackage{amsmath,amssymb,bm,amsthm} 
\allowdisplaybreaks
\usepackage{comment} 
\usepackage{longtable} 
\usepackage{subcaption} 
\usepackage{multicol} 
\usepackage{makecell}

\newtheorem{proposition}{Proposition}

\algrenewcommand{\algorithmiccomment}[1]{\hfill$\blacktriangleright$ #1}

\graphicspath{{Figures}}

\DeclareMathOperator{\diag}{diag}

\newcommand{\rt}{t} 

\newcommand{\hsigma}{\widehat{\sigma}}

\newcommand{\tr}{\tilde{r}}
\newcommand{\hx}{\widehat{x}}

\newcommand{\bzero}{\boldsymbol 0}
\newcommand{\bone}{\boldsymbol 1}

\newcommand{\be}{\boldsymbol e}

\newcommand{\bu}{\boldsymbol u}

\newcommand{\btu}{\boldsymbol{\tilde{u}}}
\newcommand{\bv}{\boldsymbol v}
\newcommand{\btv}{\boldsymbol{\tilde{v}}}
\newcommand{\bhv}{\boldsymbol{\hat{v}}}

\newcommand{\bx}{\boldsymbol x}

\newcommand{\bE}{\boldsymbol E}
\newcommand{\bI}{\boldsymbol I}

\newcommand{\bO}{\boldsymbol O}

\newcommand{\bR}{\boldsymbol R}
\newcommand{\btr}{\boldsymbol{\tilde{r}}}
\newcommand{\btR}{\boldsymbol{\widetilde{R}}}

\newcommand{\bU}{\boldsymbol U}
\newcommand{\btU}{\boldsymbol{\widetilde{U}}}
\newcommand{\bhU}{\boldsymbol{\widehat{U}}}

\newcommand{\bV}{\boldsymbol V}
\newcommand{\btV}{\boldsymbol{\widetilde{V}}}

\newcommand{\bhV}{\boldsymbol{\widehat{V}}}
\newcommand{\bW}{\boldsymbol W}

\newcommand{\bX}{\boldsymbol X}
\newcommand{\bhX}{\boldsymbol{\widehat{X}}}

\newcommand{\hmu}{\hat{\mu}}

\newcommand{\bmu}{\boldsymbol \mu}

\newcommand{\bhmu}{\boldsymbol{\hat{\mu}}}

\newcommand{\bSigma}{\boldsymbol \Sigma}
\newcommand{\bhSigma}{\boldsymbol{\widehat{\Sigma}}}

\newcommand{\hlambda}{\hat{\lambda}}

\newcommand{\tnu}{\tilde{\nu}}

\newcolumntype{M}[1]{>{\centering\arraybackslash}m{#1}}

\definecolor{orange1}{RGB}{255,128,0}
\definecolor{purple2}{RGB}{102,0,204}
\definecolor{blue}{RGB}{0,0,255}
\definecolor{red}{RGB}{255,0,0}

\usepackage{xr}
\def\spacingset#1{\renewcommand{\baselinestretch}{#1}\small\normalsize}
\begin{document}
\spacingset{1}

\newcommand{\blind}{0}


\if0\blind
{

\title{\bf Cellwise, Blockwise, and Casewise Robust Multiblock PCA 
for Sustainable and Inclusive Wellbeing in the EU
                     }
\author[1,2]{Arthur Daisomont}
\author[1]{Fabio Centofanti}
\author[1]{Mia Hubert}

\affil[1]{Section of Statistics and Data Science, Department 
           of Mathematics, KU Leuven, Belgium}
\affil[2]{Joint Research Center, European Commission, Ispra, Italy}

\setcounter{Maxaffil}{0}
\renewcommand\Affilfont{\itshape\small}
\date{October 2, 2026} 
  \maketitle
} \fi

\if1\blind
{
  \phantom{abc}\\
  \vspace{20mm}
  
  \begin{center}
    {\LARGE \bf Cellwise, Blockwise, and Casewise Robust \\Multiblock PCA 
for Sustainable and\\ 
\vskip4mm Inclusive Wellbeing in the EU}\\

    \vskip22mm
    \large{October 1, 2026} 
\end{center}
  \medskip
} \fi

\begin{abstract}
 
Gross Domestic Product (GDP) is widely used to guide economic and social decision-making, but it provides only a partial view of wellbeing. For this reason, the European Union (EU) has launched initiatives to monitor sustainable and inclusive wellbeing beyond GDP, developing indicator frameworks that cover dimensions such as health, education, environment, and social inclusion. These indicators are naturally grouped into thematic areas, and policymakers are interested in understanding how these areas contribute to global wellbeing and which indicators explain differences across countries. However, such data are high-dimensional, contain missing values, and may include anomalies affecting entire observations, specific thematic areas, or individual indicators.
We introduce blockwise outliers and propose bloccPCA, a robust multiblock PCA method that simultaneously handles casewise, blockwise, and cellwise outliers, as well as missing values. The method provides robust global components to summarize the overall structure of wellbeing, while preserving thematic-area contributions through robust blockcomponents. It also yields diagnostic tools to identify whether anomalies arise at the case, block, or cell level. Monte Carlo simulations and an application to the EU wellbeing dataset show that bloccPCA provides valuable insights into sustainable and inclusive wellbeing.

\end{abstract}


\newpage
\spacingset{1.75}
\setlength{\abovedisplayskip}{6pt}
\setlength{\belowdisplayskip}{6pt}
\setlength{\abovedisplayshortskip}{4pt}
\setlength{\belowdisplayshortskip}{4pt}
\section{Introduction}
Gross Domestic Product (GDP) is widely used by governments, policymakers, and international institutions as the main indicator for guiding social and economic decision-making \citep{costanza-2009,stiglitz-2010,costanza-2014a}.
While GDP describes the economic performance of a country, it provides limited information on overall wellbeing, understood as the broader quality of people’s lives, including their health, education, social relationships, environmental conditions, security, and ability to live a meaningful and fulfilling life \citep{hoekstra-2019, costanza-2014b}.
In response, several initiatives have emerged to promote a more comprehensive understanding of wellbeing beyond GDP \citep{jansen-2024,UN-2024,OECD-2025}.
In the European Union (EU), the European Commission has contributed to this
effort through the 2023 Strategic Foresight Report \citep{EC-2023}, which
initiated the development of metrics for sustainable and inclusive wellbeing
\citep{benczur-2024,benczur-2025}. As part of this initiative, a broad set of
indicators, hereafter referred to as the \textit{EU wellbeing dataset}, has been
compiled to capture different dimensions of wellbeing across EU countries over
time. These indicators provide the basis for the construction of composite
indicators and other summary measures aimed at supporting communication and
policy use.

However, although composite indicators are easy to interpret,
they may provide limited information because they summarize a large number of
variables into a single measure. For this reason, complementary statistical approaches, such as Principal Component Analysis (PCA), are useful for exploring the underlying structure of the data in greater detail.  Rather than forcing all information into a single composite
score, PCA identifies the main directions of variation in the data, allowing
different underlying dimensions of wellbeing to emerge from the correlations
among the indicators.

The indicators in the EU wellbeing dataset are organized into seven blocks, corresponding to thematic areas that capture distinct dimensions of wellbeing: GDP, wellbeing today, social and economic aspects of sustainability, natural and planetary boundaries, wellbeing tomorrow, inclusiveness, and institutional capacity and quality. Further details on the nature of these areas and the EU wellbeing dataset are provided in Section~\ref{sec:data}.
The block information is meaningful because each group of indicators represents a specific dimension of wellbeing, and ignoring this structure may obscure how these dimensions contribute to the overall patterns observed.
To this end, we consider multiblock PCA methods, which are specifically designed for data in which variables are organized into blocks. These methods perform a PCA-type decomposition simultaneously at the global level and at the block level, allowing us to identify common patterns of variation across all indicators while also preserving and interpreting the contribution of each  block.
Early multiblock PCA methods were developed as extensions of classical PCA \citep{smilde-2003}, and were later followed by more advanced methodologies \citep{abdi-2013} and unifying frameworks \citep{tenenhaus-2017,bougeard-2024}.

Real datasets often contain outliers and missing values, which can affect the results of statistical analyses. Outliers or anomalies are data entries that behave differently from what is expected under the main structure of the data. They may correspond either to errors that should be corrected or removed, or to informative deviations that reveal relevant phenomena.
Robust methods produce estimates that are only mildly affected by the presence of outliers, which can then be detected through their large deviations from the estimated model.

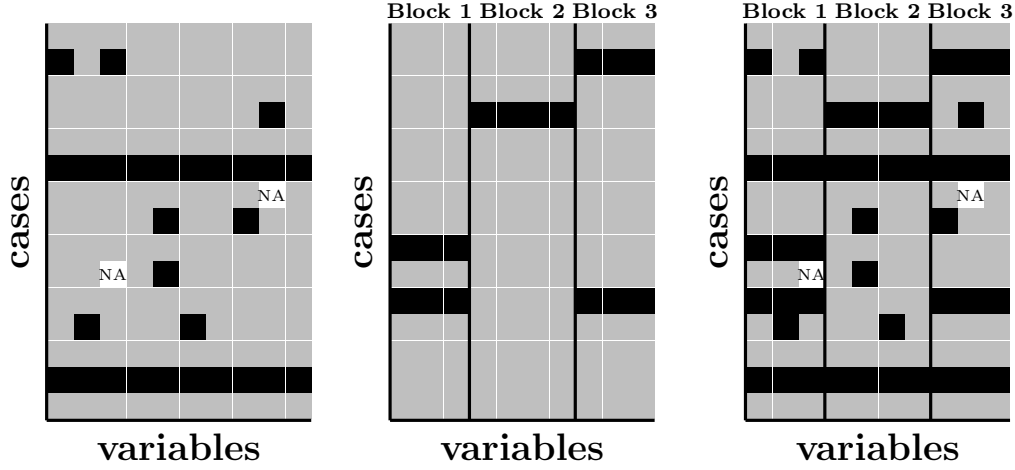
\begin{figure}[ht]
    \centering
    
    \begin{tabular}{ccc}
        \begin{tikzpicture}[scale=0.35]
            \def\ncols{10}
            \def\nrows{15}

            \foreach \i in {1,...,\ncols} {
                \foreach \j in {1,...,\nrows} {
                    \fill[gray!50] (\i + 0.05, \j + 0.05) rectangle ++(0.95, 0.95);
                }
            }

            \foreach \i in {1,...,\ncols} {
                \fill[black] (\i + 0.05, 2 + 0.05) rectangle ++(0.95,0.95);
                \fill[black] (\i + 0.05, 10 + 0.05) rectangle ++(0.95,0.95);
            }

            \fill[black] (2 + 0.05, 4 + 0.05) rectangle ++(0.95,0.95);
            \fill[black] (6 + 0.05, 4 + 0.05) rectangle ++(0.95,0.95);
            \fill[black] (5 + 0.05, 6 + 0.05) rectangle ++(0.95,0.95);
            \fill[black] (5 + 0.05, 8 + 0.05) rectangle ++(0.95,0.95);
            \fill[black] (8 + 0.05, 8 + 0.05) rectangle ++(0.95,0.95);
            \fill[black] (9 + 0.05, 12 + 0.05) rectangle ++(0.95,0.95);
            \fill[black] (1 + 0.05, 14 + 0.05) rectangle ++(0.95,0.95);
            \fill[black] (3 + 0.05, 14 + 0.05) rectangle ++(0.95,0.95);

            \fill[white] (3 + 0.05, 6 + 0.05) rectangle ++(0.95,0.95);
            \node at (3.5, 6.5) {\tiny NA};
            \fill[white] (9 + 0.05, 9 + 0.05) rectangle ++(0.95,0.95);
            \node at (9.5, 9.5) {\tiny NA};
            
            \node[rotate=90] at (0, \nrows/2 + 1) {\large \textbf{cases}};
            \node at (\ncols/2 + 1, 0) {\large \textbf{variables}};
            
            \draw[very thick] (1,1) -- (1, \nrows+1);
            \draw[very thick] (1,1) -- (\ncols+1, 1);
        \end{tikzpicture}
        
        &  
        
        \begin{tikzpicture}[scale=0.35]
            \def\ncols{10}
            \def\nrows{15}

            \foreach \i in {1,...,\ncols} {
                \foreach \j in {1,...,\nrows} {
                    \fill[gray!50] (\i + 0.05, \j + 0.05) rectangle ++(0.95, 0.95);
                }
            }

            \foreach \i in {1,...,3} {
                \fill[black] (\i + 0.05, 5 + 0.05) rectangle ++(0.95,0.95);
                \fill[black] (\i + 0.05, 7 + 0.05) rectangle ++(0.95,0.95);
            }
            \foreach \i in {4,...,7} {
                \fill[black] (\i + 0.05, 12 + 0.05) rectangle ++(0.95,0.95);
            }
            \foreach \i in {8,...,10} {
                \fill[black] (\i + 0.05, 5 + 0.05) rectangle ++(0.95,0.95);
                \fill[black] (\i + 0.05, 14 + 0.05) rectangle ++(0.95,0.95);
            }

            \node[rotate=90] at (0, \nrows/2 + 1) {\large \textbf{cases}};
            \node at (\ncols/2 + 1, 0) {\large \textbf{variables}};
            \node at (2.5, \nrows + 1.5) {\scriptsize \textbf{Block 1}};
            \node at (5 + 1, \nrows + 1.5) {\scriptsize	\textbf{Block 2}};
            \node at (8.5 + 1, \nrows + 1.5) {\scriptsize \textbf{Block 3}};
            
            \draw[very thick] (1,1) -- (1, \nrows+1);
            \draw[very thick] (4,1) -- (4, \nrows+1);
            \draw[very thick] (8,1) -- (8, \nrows+1);
            \draw[very thick] (1,1) -- (\ncols+1, 1);
        \end{tikzpicture}
        
        &  
        
        \begin{tikzpicture}[scale=0.35]
            \def\ncols{10}
            \def\nrows{15}

            \foreach \i in {1,...,\ncols} {
                \foreach \j in {1,...,\nrows} {
                    \fill[gray!50] (\i + 0.05, \j + 0.05) rectangle ++(0.95, 0.95);
                }
            }

            \foreach \i in {1,...,\ncols} {
                \fill[black] (\i + 0.05, 2 + 0.05) rectangle ++(0.95,0.95);
                \fill[black] (\i + 0.05, 10 + 0.05) rectangle ++(0.95,0.95);
            }

            \fill[black] (2 + 0.05, 4 + 0.05) rectangle ++(0.95,0.95);
            \fill[black] (6 + 0.05, 4 + 0.05) rectangle ++(0.95,0.95);
            \fill[black] (5 + 0.05, 6 + 0.05) rectangle ++(0.95,0.95);
            \fill[black] (5 + 0.05, 8 + 0.05) rectangle ++(0.95,0.95);
            \fill[black] (8 + 0.05, 8 + 0.05) rectangle ++(0.95,0.95);
            \fill[black] (9 + 0.05, 12 + 0.05) rectangle ++(0.95,0.95);
            \fill[black] (1 + 0.05, 14 + 0.05) rectangle ++(0.95,0.95);
            \fill[black] (3 + 0.05, 14 + 0.05) rectangle ++(0.95,0.95);

            \foreach \i in {1,...,3} {
                \fill[black] (\i + 0.05, 5 + 0.05) rectangle ++(0.95,0.95);
                \fill[black] (\i + 0.05, 7 + 0.05) rectangle ++(0.95,0.95);
            }
            \foreach \i in {4,...,7} {
                \fill[black] (\i + 0.05, 12 + 0.05) rectangle ++(0.95,0.95);
            }
            \foreach \i in {8,...,10} {
                \fill[black] (\i + 0.05, 5 + 0.05) rectangle ++(0.95,0.95);
                \fill[black] (\i + 0.05, 14 + 0.05) rectangle ++(0.95,0.95);
            }

            \fill[white] (3 + 0.05, 6 + 0.05) rectangle ++(0.95,0.95);
            \node at (3.5, 6.5) {\tiny NA};
            \fill[white] (9 + 0.05, 9 + 0.05) rectangle ++(0.95,0.95);
            \node at (9.5, 9.5) {\tiny NA};

            \node[rotate=90] at (0, \nrows/2 + 1) {\large \textbf{cases}};
            \node at (\ncols/2 + 1, 0) {\large \textbf{variables}};
            \node at (2.5, \nrows + 1.5) {\scriptsize \textbf{Block 1}};
            \node at (5 + 1, \nrows + 1.5) {\scriptsize	\textbf{Block 2}};
            \node at (8.5 + 1, \nrows + 1.5) {\scriptsize \textbf{Block 3}};
            
            \draw[very thick] (1,1) -- (1, \nrows+1);
            \draw[very thick] (4,1) -- (4, \nrows+1);
            \draw[very thick] (8,1) -- (8, \nrows+1);
            \draw[very thick] (1,1) -- (\ncols+1, 1);
        \end{tikzpicture}\\[6pt]
    \end{tabular}
     \vspace{-0.5cm}
    \caption{Contaminated data with cellwise and casewise outliers, together with missing values (left). 
Contaminated data with blockwise outliers only (middle).
Contaminated data with cellwise, blockwise, and casewise outliers, together with missing values (right).}
    \label{tab:con_models}
\end{figure}

Until now, two types of outliers are usually considered. 
\textit{Casewise outliers} correspond to entire observational units that deviate from the majority of the cases. 
This issue is particularly relevant in the analysis of the EU wellbeing dataset.
For example, the COVID-19 years may appear as casewise outliers for some countries, because the crisis substantially disrupted the relationships among several indicators. 
\textit{Cellwise outliers} contaminate only specific entries of the data matrix. Thus, only a few variables may be anomalous
for a given case, while the remaining variables still follow the main structure
of the data. For example, if we consider the variable ``Climate-related economic
losses'' in the EU wellbeing dataset, Belgium in 2021 can be regarded as an outlying cell because of the
major flooding that affected the country in that year. However, this observation
is not necessarily a casewise outlier, since only that specific variable was
severely affected, whereas most of the other indicators may have remained
consistent with the general pattern. The concepts of casewise and cellwise outliers, together with missing values, are illustrated in the left panel of Figure~\ref{tab:con_models}, where each row and each column represent a case and a variable, respectively.

Casewise robust methods are designed to cope with casewise outliers. They search for the model that best fits the majority of the observations, hence they require that at least half of the cases are clean \citep{maronna2019robust}. The outlying cases are downweighted
or entirely discarded. Examples include the robust PCA methods of \cite{locantore-1999}, \cite{Croux:Proj} and \cite{hubert-2005}. 

When data contain cellwise outliers, these casewise robust methods may become ineffective as they also downweight all the uncontaminated cells of cases with some anomalous cells. 
Moreover, even a small proportion of contaminated cells can lead to a large proportion of contaminated cases such that casewise robust methods may fail completely. 
Cellwise robust methods downweight or discard only the anomalous cells and do not need an outlier-free subset of observations. 
Recent contributions include methods for covariance estimation
\citep{Agostinelli2015, cellMCD, centofanti2025cellwise}, regression
\citep{CRlasso, Raymaekers:cellLTS,centofanti2026cellwise}, clustering \citep{zaccaria2025}, classification \citep{aerts2017cellwise, Centofanti:cellDA} and time series analysis
\citep{centofanti2025multivariate}. See \cite{hubert2026cellwise} for a recent review. 

Being able to cope with cellwise outliers is especially important for PCA as it is typically applied to high-dimensional data \citep{Pfeiffer:cellsparsePCA}. Moreover, as shown by \cite{alqallaf-2009},
cellwise contamination can quickly propagate through linear combinations of the
original variables. Hence, a small number of contaminated cells may significantly distort the principal components and the resulting low-dimensional representation.
Only a limited number of methods can simultaneously account for
cellwise outliers, casewise outliers, and missing values. A first contribution
in this direction is MacroPCA \citep{macropca}. More recently,
\cite{Centofanti2026cellPCA} proposed cellPCA.

When variables are organized into blocks, the block structure may also induce
specific patterns of outlying behavior. In particular, variables belonging to the
same block may behave anomalously together within a given observation, while the
variables in the other blocks remain consistent with the main structure of the
data. This type of contamination is not fully captured by either casewise or
cellwise outliers, since the anomaly affects neither the entire case nor only
isolated cells, but rather a specific group of variables.
In this article, we introduce this form of contamination as \textit{blockwise
outliers}. In addition to improving the detection of
anomalous observations, this perspective enhances the interpretation of the
anomalous behavior, since it identifies the specific dimension of the data in
which the deviation occurs. In the EU wellbeing dataset, for instance, Sweden
may appear as outlying in the block ``Natural and planetary boundaries'' for
several years, because
its approach to environment-related policy making differs
from that of most other EU member states. At the same time, Sweden may not be
outlying with respect to the remaining wellbeing dimensions.
Blockwise outliers are illustrated in the second panel of
Figure~\ref{tab:con_models}. In this example, the variables are partitioned
into three blocks of sizes 3, 4, and 3, respectively. The fifth case from the bottom contains two blockwise outliers, one in the first block and one in the third block.
Similarly, the second case contains a blockwise outlier in the third block.
We consider the more general contamination setting, visualized in the right panel, where the dataset
may contain cellwise, blockwise, and casewise outliers, together with missing
values.

The aim of this paper is to provide new insights into the broad notion of
wellbeing in the European Union, moving beyond an analysis based solely on GDP.
To this end, we consider a broad set of indicators and account for both the
information carried by the individual variables and their organization into
blocks defined by thematic areas. 
More specifically, this paper addresses the following questions:
\begin{enumerate}[label=\textbf{Q\arabic*.}]
    \item How do the thematic areas of wellbeing contribute to the global
    structure of EU wellbeing, and which indicators explain similarities and
    differences across countries within the most relevant thematic areas?

    \item Which countries and years exhibit anomalous behavior, and can these
    anomalies be interpreted as casewise, blockwise, or cellwise outliers?
\end{enumerate}

To answer these questions, we propose bloccPCA 
(BLOckwise Cellwise and Casewise PCA), a robust multiblock PCA method
designed for data in which variables are organized into meaningful blocks and may
contain cellwise, blockwise, and casewise outliers, as well as missing values.
Unlike a standard PCA applied to all indicators jointly, bloccPCA provides
low-dimensional representations at both the global and block levels through
robust supercomponents and blockcomponents. The method extends weighted
consensus PCA (CPCA-W), one of the most widely used multiblock PCA approaches,
which is particularly attractive because of its interpretability and the direct
connection it establishes between global and blockwise components \citep{smilde-2003}. 
This structure makes it possible to identify
patterns that are shared across the whole dataset, while also detecting
area-specific structures that may be hidden in a purely global analysis.
Moreover, by explicitly accounting for blockwise outliers, the method is able to
detect and limit the influence of observations that are anomalous only within a
specific thematic area, without necessarily treating the entire 
observation as outlying.

The article is organized as follows. Section~\ref{sec:data} describes the EU
wellbeing initiative and dataset. Section~\ref{sec:method} introduces bloccPCA, presenting the
objective function, the algorithm used for its minimization, and the robust
estimation of supercomponents and blockcomponents. 
Section~\ref{sec:simulation} evaluates the
performance of the proposed approach on synthetic contaminated data through a
simulation study. Section~\ref{sec:realdata} applies bloccPCA to the EU
wellbeing dataset. Finally, Section~\ref{sec:conc} concludes the article.

\section{The EU wellbeing initiative and dataset}
\label{sec:data}

Since the Treaty on European Union identifies the wellbeing of its people as one
of the Union's main objectives, recent crises have reinforced the need for a
broader understanding of progress and have made wellbeing an increasingly
relevant objective of European policymaking. This has motivated the European
Union to engage with the \textit{Beyond GDP} agenda, which recognizes the value
of GDP and its growth as economic indicators, while emphasizing that they are
not sufficient to capture the full range of societal progress.
In this context, the European Commission published the 2023 Strategic Foresight
Report, entitled ``\textit{Sustainability and wellbeing at the heart of
Europe's Open Strategic Autonomy}''. Based on the identification of potential challenges that Europe may face, the
report proposes several actions to strengthen Europe's sustainability and
wellbeing. Among these, it announced the launch of the European Commission's
work on Sustainable and Inclusive Wellbeing (SIWB) metrics
\citep{EC-2023,benczur-2024}. The SIWB initiative provides an integrated framework for analysing the
wellbeing of current and future generations, as well as that of the planet. It
goes beyond the construction of additional indicators by offering a conceptual
structure that brings together current wellbeing, future wellbeing,
sustainability, inclusiveness, resilience, planetary boundaries, and
institutional capacity. The Commission's work follows two main directions:
the development of augmented-GDP metrics which adjust the GDP by converting indicators 
of wellbeing into monetary values \citep{benczur-2023}, and the construction of a
SIWB dashboard based on a medium-sized and comprehensive set of
indicators \citep{benczur-2024, benczur-2025}.
The dashboard is intended to capture the main aspects of a
country's progress and to support the development of headline indicators,
composite measures, and other statistical summaries suitable for communication
and policy use.
Our work is positioned within this second direction.

The EU wellbeing dataset underlying the SIWB dashboard was developed from an
initial collection of a large number of indicators drawn from sources within
the European Commission, such as Eurostat and the Eurobarometer, as well as
external sources, such as the World Health Organization and the OECD. This
initial collection was intended to provide broad coverage of the different
dimensions of wellbeing. It was subsequently condensed by removing overlapping,
statistically redundant, and unavailable indicators, resulting in a final set of
140 indicators grouped into seven thematic areas.
In this paper, we analyse a subset of the EU wellbeing dataset. In particular,
we further select indicators according to two criteria: they must be available
for at least half of the observations and approximately normally distributed
after robust transformation \citep{Raymaekers:Transfo}. The resulting analysis
dataset contains 104 indicators, partitioned into the seven thematic areas. These indicators
are observed for the 27 EU member states over the period 2010--2022. Each
observation therefore corresponds to a unique country--year pair,
leading to a total of 351 cases.
The first block is GDP. The second
block is called  ``Wellbeing today'' and contains
29 variables related to quality of life and how 
different populations perceive it, such as ``Inability to make ends meet'' and
``Inactive population due to caring responsibilities by sex''.
The third block ``Social and economic resources for future wellbeing'' 
contains 13 variables related to economic, health, education
and social capital such as ``Gross domestic expenditure on R\&D''
and ``Healthy life years at age 65''. The fourth block 
``Resilience: societal challenges and sustainability transitions''
contains 26 variables related to the social and economic,
digital, geopolitical, and demographic aspects such as
``Government debt'' and ``High-speed internet coverage''.
The fifth block ``Nature and planetary boundaries''
has 25 variables related to ecosystems and biodiversity,
climate change, adaptation and sustainable use of resources,
and green economy and policies such as ``Air pollution pm10''
and ``Climate-related economic losses''. The sixth block  ``Inclusiveness'' contains 3 variables related to 
inequalities such as ``Gender employment gap''.
Finally, the seventh and last block 
``Institutional quality'' contains 7 variables related to
government and institutions such as ``Control of corruption''
and ``Political stability''. In the following,
some block names are shortened for practicality.
More information about the indicators
and the thematic areas can be found in Section~\ref{app:block_and_ind} of the Supplementary Material and in \cite{benczur-2025}.

\section{The bloccPCA method} \label{sec:method}

\subsection{Multiblock PCA via weighted
consensus PCA}

The observed data are represented by a $n \times p$ supermatrix
$\bX$, where columns correspond to variables and rows to observations. It is assumed that there is a 
meaningful partition of the variables into $B$ blocks of
size $p_b$ for $b=1,\dots,B$. The data matrix 
$\bX$ can be equivalently written as the concatenation of the block 
 matrices
$\bX^{(b)}=[x_{ij}^{(b)}]$ for $i=1,\dots,n$ and $j=1,\dots,p_b$, namely $\bX=[\bX^{(1)}, \dots,\bX^{(B)}]$,
where $\bX^{(b)}$ denotes the $n \times p_b$ data matrix
restricted to the $b$-th block of variables.
For a fixed rank $k<p$, multiblock PCA aims to represent each block
$\bX^{(b)}$ through a low-dimensional
approximation of the form
$\bX^{(b)} = \bone_n (\bmu^{(b)})^T+\bU^{(b)}(\bV^{(b)})^T
+ \bE^{(b)}$, for $b=1,\dots,B$,
where $\bU^{(b)}=\big[\bu_1^{(b)},\dots,\bu_n^{(b)}\big]^T
\in \mathbb{R}^{n\times k}$ contains the blockscores as rows, \sloppy
$\bV^{(b)}=\big[\bv_1^{(b)},\dots,\bv_{p_b}^{(b)}\big]^T =\big[\bv^{(b)1},\dots,\bv^{(b)k}\big]\in \mathbb{R}^{p_b \times k}$ contains the blockloadings or blockcomponents as columns,
$\bmu^{(b)}=\left(\mu^{(b)}_1,\dots,\mu^{(b)}_{p_b}\right)^T \in \mathbb{R}^{p_b}$ is the block center, and $\bE^{(b)}$
is the corresponding residual matrix. Here, $\bone_n$ denotes a vector of
ones of length $n$.
This directly translates to a global low-dimensional
representation of the supermatrix $\bX$
given by
\begin{equation}\label{eq:model}
\bX = \bone_n \bmu^T +\bU \bV^T + \bE,
\end{equation}
where $\bU=\lbrace u_{i\ell} \rbrace=\big[\bu^1,\dots,\bu^k\big]=
\big[\bu_1,\dots,\bu_n\big]^T \in \mathbb{R}^{n \times k}$ contains the superscores as rows,
$\bV=\left[
\big(\bV^{(1)}\big)^T,\dots,
\big(\bV^{(B)}\big)^T
\right]^T \in \mathbb{R}^{p \times k}$ contains the superloadings as columns,
$\bmu=\left(
\big(\bmu^{(1)}\big)^T,\dots,
\big(\bmu^{(B)}\big)^T
\right)^T \in \mathbb{R}^{p}$ is the  center, and
$\bE = [\bE^{(1)},\dots,\bE^{(B)}]$ is the residual matrix.

The weighted consensus PCA (CPCA-W)  method of \cite{smilde-2003} performs multiblock PCA through the following steps.
\begin{enumerate}
\item Pre-process the block matrices $\bX^{(b)}$ by first standardizing the variables to unit variance. This avoids that the first components are dominated by the variables with the largest
variability \citep{jolliffe-2002}. 
The second preprocessing step balances the contribution of the different data
blocks. If one block contains substantially more
variables than the others, it may dominate the first estimated components.
Following the default block scaling proposed by \cite{smilde-2003},
each block matrix is rescaled as $\bX^{(b)}/\sqrt{p_b}$ for $b=1,\dots, B$.
  \item Estimate the superloadings $\bV^T= \left[
\big(\bV^{(1)}\big)^T,\dots,
\big(\bV^{(B)}\big)^T
\right]$, the superscores $\bU$ and the center $\bmu^T=\left(
\big(\bmu^{(1)}\big)^T,\dots,
\big(\bmu^{(B)}\big)
\right)^T$ by solving the optimization problem
    \begin{equation}\label{eq:cpca_w}
        \min_{\bU,\bV,\bmu} \; 
        \sum_{b=1}^{B}
        \left\| \bX^{(b)} - \bU \big(\bV^{(b)}\big)^T -  \bone_n \big(\bmu^{(b)}\big)^T
        \right\|_F^2,
    \end{equation}
    under the constraint that $\bV^T\bV=\bI_k$, where $\bI_k$ is the $k\times k$ identity  matrix, and
    $\|\cdot\|_F$ denotes the Frobenius norm. 
    This step can be performed by applying classical PCA to the supermatrix $\bX$, yielding $\bV$ and the center $\bmu = \overline{\bX}$. Then, the superscore matrix is $\bU  =
    \sum_{b=1}^{B}\left(\bX^{(b)}-\bone_n\big(\bmu^{(b)}\big)^T\right)
    \bV^{(b)}$.

    \item Obtain the
    blockloadings  by normalizing the columns of $\bV^{(b)}$ as
    $\btV^{(b)} = \bV^{(b)}
    \diag\left(1/\|\bv^{(b)1}\|,\dots, 1/\|\bv^{(b)k}\| \right)$,  for $b=1,\dots,B$.
    Compute the blockscores  $\btU^{(b)} =
    \left(\bX^{(b)}-\bone_n\big(\bmu^{(b)}\big)^T\right)
    \btV^{(b)}$, for $b=1,\dots,B$.
\end{enumerate}
The superscores thus describe the main patterns of
variation across all blocks of variables, whereas the blockscores describe how these global
patterns are expressed within each individual block.

\subsection{Preprocessing}\label{subsec:prepro}

Since the data may be contaminated by outliers, bloccPCA first robustly scales each variable by computing
a M-estimator of scale defined as the solution of
\begin{equation}\label{eq:mscale}
    \dfrac{1}{n}\sum_{i=1}^n \rho\left( \dfrac{x_{i}-m}{a \sigma}\right) = \delta,
\end{equation}
where $x_i$ are observations for 
a given variable centered through their median $m$ 
and $a$ is a consistency factor 
such that $E\left[\rho(x/a) \right]
= \delta$. The function $\rho$ has $\rho(0) = 0$, is even, not
decreasing for $x \geqslant 0$, and bounded. This limits the impact of outliers and makes the scale estimates robust. We make the choice of the hyperbolic tangent function $\rho_{b,c}$, which is defined as 
\begin{equation}
\label{eq:hyperbolic}
	\rho_{b,c}(z) =
	\begin{cases} 
		\frac{z^2}{2} & \text{if } 0 \leqslant |z| \leqslant b, \\
		d - \frac{q_1}{q_2} \ln(\cosh(q_2(c - |z|))) & \text{if } 
		b \leqslant |z| \leqslant c, \\
		d & \text{if } c \leqslant |z|.
	\end{cases}
\end{equation}
We use the default parameter values  $b=1.5$,
$c=4$, $q_1= 1.540793$, and $q_2=0.8622731$. 
This function was first introduced in \cite{hampel1981} 
as the solution to a constrained statistical optimization
problem, and was later used
in \cite{raymaekers-2021} and
in \cite{Centofanti2026cellPCA}.

Next, the block-size scaling is applied as in CPCA-W. 
From now on, $\bX$ stands for the preprocessed supermatrix.

\subsection{The optimization problem} \label{subsec:objfun}

The second step of CPCA-W consists of estimating $\bV$, $\bU$, and $\bmu$
by minimizing the least squares objective in \eqref{eq:cpca_w}, equivalently by
applying classical PCA to the preprocessed supermatrix.  
Because of the least squares type objective function, this step is highly sensitive to anomalies, and the resulting components can be strongly affected by outlying
cells, blocks, or cases, and may break down in the presence of extreme
contamination. This sensitivity will be illustrated through simulations in
Section~\ref{sec:simulation}.

Instead, bloccPCA obtains estimates  $\bhV_{o}$, $\bhU_{o}$, and $\bhmu_{o}$ of  $\bV$, $\bU$, and $\bmu$ by minimizing
	\begin{multline}\label{eq:objP}
		L_{\rho_1,\rho_2,\rho_3}(\bX,\bV,\bU,\bmu) := \\
		\frac{\hsigma_3^2}{o}\sum_{i=1}^{n}
		o_i\rho_3\!\left(\frac{1}{\hsigma_{3}}\sqrt{\frac{1}{o_i}
			\sum_{b=1}^{B} o^{(b)}_{i} 
			\left(\hsigma_{2}^{(b)}\right)^2\rho_2\! 
			\left(\frac{1}{\hsigma_{2}^{(b)}}\sqrt{\frac{1}
				{o^{(b)}_{i}}
				\sum_{j=1}^ {p_b} o^{(b)}_{ij}\, 
				\left(\hsigma_{1,j}^{(b)}\right)^2\,
				\rho_1\!\left(\frac{r_{ij}^{(b)}}
				{\hsigma_{1,j}^{(b)}}\right)}\, \right)}\right).
	\end{multline}

The observed-data indicator $o_{ij}^{(b)}$  is equal to $1$ if the cell $x_{ij}^{(b)}$ is observed
and $0$ if missing. Moreover, for each case, $o_{i}^{(b)}=\sum_{j=1}^{p_b}o_{ij}^{(b)}$ and $o_i=\sum_{b=1}^{B}o_{i}^{(b)}$
count the available cells within block $b$ and across all blocks, respectively, and $o=\sum_{i=1}^{n} o_i$ is the total number of observed cells\,. The objective \eqref{eq:objP} is a function of the \textit{cellwise residuals} defined
as
$r_{ij}^{(b)} := x_{ij}^{(b)} - \hx_{ij}^{(b)}$,
where $\hx_{ij}^{(b)} =  \mu_j^{(b)}+\bu_i^T \bv_j^{(b)}$. 
The scale estimate $\hsigma_{1,j}^{(b)}$  standardizes the corresponding cellwise residuals.
Then, for each block within a case, the \textit{blockwise deviation}
is defined as
	$d_{i}^{(b)} := \sqrt{\frac{1}{o_{i}^{(b)}} \sum_{j=1}^{p_b}
		o_{ij}^{(b)}\, \left(\hsigma_{1,j}^{(b)}\right)^2\, \rho_1\!\left(
		\frac{r_{ij}^{(b)}}{\hsigma_{1,j}^{(b)}} \right)}\;$.
For each block, $\hsigma_{2}^{(b)}$ is estimated to scale the 
corresponding blockwise deviations. 
Finally, the \textit{casewise total deviations}
	$t_i := \sqrt{\frac{1}{o_i} \sum_{b=1}^{B}
		o_{i}^{(b)}\, \left(\hsigma_{2}^{(b)}
		\right)^2\,\rho_2\!\left(
		\frac{d_{i}^{(b)}}{\hsigma_{2}^{(b)}} \right)}\;$
are standardized by the scale estimate $\hsigma_3$.
The robustness of the method highly depends on the choice of the 
three functions $\rho_1$, $\rho_2$, and $\rho_3$. Indeed, if
$\rho_1(z) = \rho_2(z) = \rho_3(z) = z^2$, the objective function \eqref{eq:objP}
simplifies to classical PCA when all cells are observed. 
We use the hyperbolic tangent function $\rho_{b,c}$ as defined
in \eqref{eq:hyperbolic}.

The combination of $\rho_1$, $\rho_2$, and $\rho_3$ makes bloccPCA robust
against cellwise, blockwise, and casewise outliers. Indeed, a cellwise outlier
in the cell $(i,j)$ of block $b$ yields a cellwise residual
$r_{ij}^{(b)}$ with a large absolute value, but the boundedness of $\rho_1$
bounds its effect on the estimation. Similarly, a blockwise outlier in block
$b$ for observation~$i$ yields a large blockwise deviation $d_i^{(b)}$, but the
boundedness of $\rho_2$ limits its contribution. Finally, a casewise outlier
yields a large casewise total deviation $\rt_i$, whose effect on the objective
function is controlled by the boundedness of $\rho_3$.
This nested construction also controls the interaction between the different
types of outliers. Since blockwise deviations are computed from cellwise
residuals, the boundedness of $\rho_1$ prevents a single extreme cell from
dominating the blockwise deviation. Similarly, since the casewise total
deviation is computed from the blockwise deviations, the boundedness of
$\rho_2$ prevents a single extreme block from dominating the casewise
contribution.
Formal mathematical models for cellwise, blockwise, and casewise contamination
are provided in Section~\ref{subsec:blockoutmodel} of the  Supplementary Material, whereas
methods for imputation and out-of-sample prediction are described in Section~\ref{sec:pracext} of the
Supplementary Material.

\subsection{The algorithm} \label{subsec:algo}
The solution of the minimization in \eqref{eq:objP} must verify the first-order necessary conditions
of optimality that are obtained by setting to zero the gradients of the objective function $L_{\rho_1,\rho_2,\rho_3}$ with respect to
$\bv_j^{(b)}$, $\bu_i$, and $\bmu^{(b)}$ for $b=1,\dots,B$, $j=1,\dots,p_b$ and $i=1,\dots,n$. Their derivation can be found in Section \ref{app:firstorder}
of the Supplementary Material.
This yields
\begin{align} 
	\label{eq:condi1}
	(\bU^T\bW^{(b)j}\bU)\bv^{(b)}_{j}& =\bU^T\bW^{(b)j}
	(\bx^{(b)j}-\mu^{(b)}_j\bone_n),  
	\quad b=1,\dots,B, \;\; j=1,\dots,p_b,\\
	\label{eq:condi2}  
\sum_{b=1}^{B}\big((\bV^{(b)})^T\bW^{(b)}_i\bV^{(b)}\big)\bu_i & =
	\sum_{b=1}^{B}(\bV^{(b)})^T\bW^{(b)}_i(\bx^{(b)}_i-
	\bmu^{(b)}),  
	\quad i=1,\dots,n,\\
	\label{eq:condi3}
	\sum_{i=1}^n \bW^{(b)}_i\, \bV^{(b)} \bu_i & =  \sum_{i=1}^n \bW^{(b)}_i(\bx^{(b)}_i - \bmu^{(b)}),\quad 
	b=1,\dots,B,
\end{align}
where $\bx_i^{(b)}$ and $\bx^{(b)j}$ are the rows and columns of 
$\bX^{(b)}$, respectively.
Furthermore, the $n\times p_b$ weight matrices $\bW^{(b)}$  are given by
\begin{equation} \label{eq:updateW}
	\bW^{(b)}= [ w_{i j}^{(b)} ]=
	\bW_{\text{cell}}^{(b)}\odot \bW_{\text{block}}^{(b)} \odot \bW_{\text{case}}^{(b)} \odot \bO^{(b)}, 
    \quad b=1,\dots,B,
\end{equation}
where $\odot$ is the Hadamard product.
The diagonal entries of the $p_b \times p_b$  diagonal matrix $\bW_i^{(b)}$ are equal to the 
$i$th row of $\bW^{(b)}$. Similarly,
the $n\times n$ matrix $\bW^{(b)j}$ is a diagonal matrix
whose diagonal is the $j$th column of $\bW^{(b)}$.
Moreover, $\bO^{(b)} = [ o_{ij}^{(b)} ]$, and $\bW_{\text{cell}}^{(b)}=[ w_{ij}^{(b)\text{cell}}]$ is an $n\times p_b$ matrix that contains
the \textit{cellwise weights}, which are defined as
\begin{equation}\label{eq:cellweight}
	w_{ij}^{(b)\text{cell}}=\psi_1\!\left(
	\frac{r_{ij}^{(b)}}{\hsigma_{1,j}^{(b)}}\right)
	\Big/ \frac{r_{ij}^{(b)}}{\hsigma_{1,j}^{(b)}}\; , 
	\quad i=1,\dots,n,\quad b=1\dots,B,\quad j=1\dots,p_b,
\end{equation}
where $\psi_1=\rho_1'$ with the
convention that $w_{ij}^{(b)\text{cell}} = 1$ if $r_{ij}^{(b)}=0$. The 
matrix $\bW_{\text{block}}^{(b)}$ has constant 
rows that contains the \textit{blockwise weights}
\begin{equation}\label{eq:blockweight}
	w_{i}^{(b)\text{block}}=\psi_2\!\left(
	\frac{d_{i}^{(b)}}{\hsigma_{2}^{(b)}}\right)
	\Big/ \frac{d_{i}^{(b)}}{\hsigma_{2}^{(b)}}\; , 
	\quad i=1,\dots,n\,\quad b=1,\dots,B,\
\end{equation}
where $\psi_2=\rho_2'$ and $w_{i}^{(b)\text{block}} = 1$ if $d_{i}^{(b)}=0$. Finally, $\bW_{\text{case}}^{(b)}$
has constant rows that contain the 
\textit{casewise weight}
\begin{equation}\label{eq:caseweight}
	w_{i}^{\text{case}}=\psi_3\!\left(
	\frac{t_i}{\hsigma_3}\right)
	\Big/ \frac{t_i}{\hsigma_3}\; , 
	\quad i=1,\dots,n,\,
\end{equation}
where $\psi_3=\rho_3'$ and $w_{i}^{\text{case}} = 1$ if $t_i=0$.

To solve the nonlinear system of equations \eqref{eq:condi1}-\eqref{eq:condi3}, we consider an Iteratively Reweighted Least Squares Algorithm (IRLS).
The algorithm starts from initial estimates $\bV_{(0)}$, $\bU_{(0)}$, and $\bmu_{(0)}$, obtained by applying the cellPCA method to $\bX$. This initial fit is also used to estimate all the $\hsigma_{1,j}^{(b)}$, $\hsigma_{2}^{(b)}$, and $\hsigma_{3}$ through the M-scale estimator defined in \eqref{eq:mscale}.
The algorithm is provided in
Section~\ref{app:algo} of the Supplementary Material, and yields new estimates $\bV_{(\ell + 1)}$, $\bU_{(\ell + 1)}$,  and $\bmu_{(\ell + 1)}$  at step $\ell + 1$ based on the previous ones $\bV_{(\ell)}$, $\bU_{(\ell)}$, and $\bmu_{(\ell)}$ obtained at step $\ell$.
The theoretical convergence of the method
relies on the monotonicity of the objective function as stated by Proposition~\ref{prop:mono}.
\begin{proposition}\label{prop:mono}
Each iteration step of the algorithm decreases the objective function \eqref{eq:objP},
that is,
$L_{\rho_1,\rho_2,\rho_3}(\bX,\bV_{(\ell+1)},\bU_{(\ell+1)},\bmu_{(\ell+1)})
\leqslant L_{\rho_1,\rho_2,\rho_3}(\bX,\bV_{(\ell)},\bU_{(\ell)},\bmu_{(\ell)})$.
\end{proposition}
\noindent The proof can be found in Section \ref{app:proofmono} of the Supplementary Material,
and is based on the property that the three functions $z \rightarrow \rho_1(\sqrt{z})$,
$z \rightarrow \rho_2(\sqrt{z})$ and $z \rightarrow \rho_3(\sqrt{z})$ are differentiable
and concave. The convergence property follows directly, as the algorithm decreases
the objective function at each step and it has a lower bound of zero.

\subsection{Estimation of the superloadings, superscores and center}\label{subsec:estcenter}

The iterative algorithm of Section~\ref{subsec:algo} provides estimates $\bhV_{o}$, $\bhU_{o}$ and $\bhmu_o$. Note that $\bhV_{o}$ 
does not necessarily have orthogonal columns, as
required by the CPCA-W framework.
This is solved by computing the SVD
of $\bhV_{o}$ and putting in the matrix $\btV_o$ the
right singular vectors of this decomposition. From this, the superscores matrix is
 updated as $\btU_o= \bhU_o \bhV_o^T \btV_o$. 

The columns of $\btV_o$ are an orthonormal basis of the estimated
$k$-dimensional subspace, but they do not necessarily correspond to directions
ordered according to the variability of the projected observations within that
subspace. Thus, the resulting scores may be
difficult to interpret, since the first score directions would not necessarily
represent the dominant direction of variation in the projected data. Therefore, we consider the projected observations represented by the rows of $\btU_o$ and estimate their
center and covariance matrix. The eigenvectors of this covariance matrix are
then used to rotate the basis within the estimated subspace. The center and
covariance matrix are estimated by means of the MCD estimator
\citep{hubert2012}, a robust estimator of location and scatter. This robust
step is important to limit the influence of good leverage points, that is,
observations that lie close to the estimated subspace but behave as outliers
within it. More details can be found in Appendix A.1 of \cite{macropca}.
Let $\bhmu_{\btU_o}$ and $\bhSigma_{\btU_o}$ denote the MCD estimates of
location and scatter computed from $\btU_o$.
The spectral decomposition of 
$\bhSigma_{\btU_o}$ yields a $k \times k$ 
loading matrix $\bhV_{\btU_o}$ and eigenvalues 
$\hat{\lambda}_1 \geqslant \hat{\lambda}_2 
\geqslant\ldots \geqslant \hat{\lambda}_k$\,. 
The final bloccPCA estimates
are then defined as $\bhV = \btV_o\bhV_{\btU_o}$, $\bhU = (\btU_o-\bone_n\bhmu_{\btU_o}^T)\bhV_{\btU_o}$ and
$\bhmu= \bhmu_o+\btV_o\bhmu_{\btU_o}$.

{Note that the resulting fitted values $\bhX = \bone_n \bhmu^T + \bhU\bhV^T$ do not correspond with the orthogonal projection of the observations $\bX$ on the subspace defined by $\bhmu$ and $\bV$. As shown in Section~\ref{sec:pracext} of the Supplementary Material, the fitted values are the orthogonal projections of cell-imputed cases $\bX^{\text{imp}}$ whose outlyingness is reduced following \eqref{eq:imp}.  }

\subsection{Selecting the rank $k$} \label{subsec:selectrank}
The rank $k$ of the model in \eqref{eq:model} has to be selected. 
Similarly to \cite{Centofanti2026cellPCA}, we define the explained variance
in the supermatrix $\bX$ as $1-\nu_s/\nu_0$, where $\nu_s$ denotes the final value
attained at convergence by the objective function \eqref{eq:objP} when using
rank $s$. The quantity $\nu_0$ is the value of \eqref{eq:objP} obtained from
the constant fit $\hx_{ij}^{(b)}=m_j^{(b)}$, for
$i=1,\dots,n$, $b=1,\dots,B$, and $j=1,\dots,p_b$, where
$m_j^{(b)}$ is the sample median of the $j$th variable in $b$th block. 
As commonly done in PCA \citep{jolliffe-2002}, the rank $k$ can then be chosen either as the smallest rank for which the
explained variance in the supermatrix exceeds a predefined threshold, typically
80\% or 90\%, or as the elbow point in the curve of $\nu_s$ as a function of
the rank $s$. In our implementation, the elbow point is selected automatically using the Kneedle
algorithm of \cite{satopaa-2011}.

\subsection{Estimation of the blockloadings and blockscores} \label{subsec:block}
Similarly to CPCA-W, bloccPCA defines the blockloadings as the corresponding normalized
block of the superloadings. This operation does not require
additional robustification, since it is performed on the already robustly
estimated superloadings. This yields
\begin{equation}\label{eq:updateVblock}
    \btV^{(b)} =
\bhV^{(b)} \diag\left(\dfrac{1}{||\bhv^{(b)1}||}, \dots, 
\dfrac{1}{||\bhv^{(b)k}||}\right),\quad b=1,\dots,B.
\end{equation}

In CPCA-W, the blockscores are computed by multiplying each centered block matrix by the corresponding blockloading matrix. However,
when $\bX^{(b)}$ contains outliers or missing values, this operation may either
be infeasible or lead to non-robust blockscores.
For this reason, bloccPCA estimates
the rows
 $\btu_{i}^{(b)}$ of $\btU^{(b)}$ for $i=1,\dots,n$ and $b=1,\dots,B$ by minimizing, with respect to $\bu$, the following
function
\[
 L_{i}^{(b)}(\bu) :=o^{(b)}_{i} 
\left(\hsigma_{2}^{(b)}\right)^2
\rho_2\! 
\left(
\frac{1}{\hsigma_{2}^{(b)}}
\sqrt{
\frac{1}{o^{(b)}_{i}}
\sum_{j=1}^{p_b} 
o^{(b)}_{ij}
\left(\hsigma_{1,j}^{(b)}\right)^2
\rho_1\!
\left(
\frac{x_{ij}^{(b)}-\hmu_j^{(b)}-\bu^T \btv_j^{(b)}}
{\hsigma_{1,j}^{(b)}}
\right)
}
\right),
\]
where $\bhmu^{(b)}=\left(\hmu^{(b)}_1,\dots,\hmu^{(b)}_{p_b}\right)^T$ and $\btv_j^{(b)}$ are the rows of $\btV^{(b)}$.
That is
\begin{equation}\label{eq:updateUblock}
\btu_{i}^{(b)}
=
\left(
(\btV^{(b)})^T \bW^{*(b)}_{i} \btV^{(b)}
\right)^{\dagger}
(\btV^{(b)})^T \bW^{*(b)}_{i}
\left(\bx_i^{(b)}-\bhmu^{(b)}\right),
\end{equation}
where $\bW^{*(b)}_{i}$ is a $p_b \times p_b$ diagonal matrix whose diagonal
entries are given by the $i$th row of
$\bW_{\mathrm{cell}}^{(b)} \odot \bO^{(b)}$.

To assess the relevance of the estimated blockcomponents, it could be useful to
evaluate how much variability they explain within each block. 
This could be done by using the explained variance within block $b$, which is defined as
$1-\tnu_s^{(b)}/\tnu_0^{(b)}$ where
$\tnu_s^{(b)} = \sum_{i=1}^{n} L_i^{(b)}(\btu_{i,s}^{(b)})/o^{(b)}$,
$o^{(b)}=\sum_{i=1}^{n}\sum_{j=1}^{p_b} o_{ij}^{(b)}$ is the number of
observed cells in block $b$, and $\btu_{i,s}^{(b)}$ denotes the value of
$\btu_i^{(b)}$ obtained using rank $s$. Moreover,
$\tnu_0^{(b)}=\sum_{i=1}^{n} L_{i,0}^{(b)}/o^{(b)}$, where
$L_{i,0}^{(b)}=L_i^{(b)}(\bzero_k)$, with the columnwise medians
$m_j^{(b)}$ used in place of $\hmu_j^{(b)}$.


\section{Simulation study} \label{sec:simulation}
In this section, we assess the performance of the proposed method through a Monte Carlo simulation study. We focus on the estimation accuracy of the superloadings, as blockloadings are directly derived from them through \eqref{eq:updateVblock}.

Clean data are generated from a zero-mean $p$-dimensional multivariate Gaussian distribution. 
The covariance matrix $\bSigma$ is obtained by modifying the eigenvalues of the $p\times p$ matrix whose $(j,k)$-th entry is $(-0.9)^{|j-k|}$, while keeping its eigenvectors unchanged, so that the sum of the two largest eigenvalues accounts for 90\% of the sum of all eigenvalues.
{We further convert it to a correlation matrix.}
Two scenarios are considered. The first has $n=100$, {$p=60$}, and {$B=4$}, while the second has $n=100$, $p=200$, and {$B=5$}. We use blocks of equal size, hence {$p_b=15$} in the first setting and {$p_b=40$} in the second one. 

We consider six contamination scenarios: cellwise contamination, blockwise
contamination, casewise contamination, combined cellwise-blockwise
contamination, combined blockwise-casewise contamination, and combined
cellwise-blockwise-casewise contamination.
Cellwise contamination is obtained by randomly selecting
$20\%$ of the cells and {replacing them with}
$\gamma_{\text{cell}}$,
where $\gamma_{\text{cell}}$ takes values from $0$ to {$9$}. When 
$\gamma_{\text{cell}}=0$, the data are uncontaminated. 
In the blockwise contamination scenario, $20\%$ of the blocks are randomly selected and replaced by observations drawn from
$N\left(\gamma_{\mathrm{block}}(\be_1^{(b)}+\be_3^{(b)}),\bSigma^{(b)}/2\right)$.
 The matrix $\bSigma^{(b)}$ is 
the restriction of $\bSigma$ on
the block $b$, and $\be_1^{(b)}$ and $\be_3^{(b)}$ are
the first and third normalized restricted eigenvectors of $\bSigma$.
The parameter $\gamma_{\text{block}}$ varies from $0$ to $60$ 
in the case where $p=60$ and from $0$ to $100$ when $p=200$.
In the scenario of casewise contamination, 
$20\%$ of the cases are generated from the 
contaminated distribution $N\left(\gamma_\text{case} (\be_1+\be_3), 
\bSigma/1.5\right)$ where $\be_1$ and $\be_3$ are the first
and third eigenvectors of $\bSigma$, and $\gamma_{\text{case}}$ 
takes the same values as $\gamma_{\text{block}}$. 
For the combined cellwise-blockwise and blockwise-casewise contamination
scenarios, the outliers are generated as described above, but with a proportion
of {$15\%$} for each type of contamination. In the cellwise-blockwise scenario,
{we set $\gamma_{\mathrm{block}}=15\gamma_{\mathrm{cell}}$ 
and, in the blockwise-casewise scenario, we set
$\gamma_{\mathrm{case}}=\gamma_{\mathrm{block}}/1.5$}. For the combined cellwise-blockwise-casewise contamination scenario, each type of contamination affects {$10\%$} of the corresponding units, namely cells, blocks, or cases, and the $\gamma$ parameters follow the same relationships described above. To assess the performance in the presence of missing data, each contamination scenario is also considered after randomly replacing $20\%$ of the cells by missing values.

The angle between the subspace spanned by the true superloadings and the one spanned by the estimated superloadings is used as the performance measure. This angle is computed using the function \verb|subspace()| from the R package \verb|pracma| \citep{pracma}.
bloccPCA is compared with 
cellPCA \citep{Centofanti2026cellPCA}, MacroPCA \citep{macropca} and {CPCA-W.} 
The first two methods are robust against cellwise and casewise outliers and can handle missing values. {CPCA-W} is not robust and is included as a benchmark. In the presence of missing values, we use the ICPCA method of \citet{kiers1997} {for the computation of the supercomponents. All the methods perform a preprocessing step by scaling the variables and} are implemented by retaining $k=2$ principal components, which account for 90\% of the 
total variation.

\begin{figure} [h]
	\centering
	\begin{tabular}{ccc}
		\textbf{Cellwise}  & \textbf{Blockwise} & \textbf{Casewise}\\[-12pt]
        
		\includegraphics[width=0.28\textwidth]{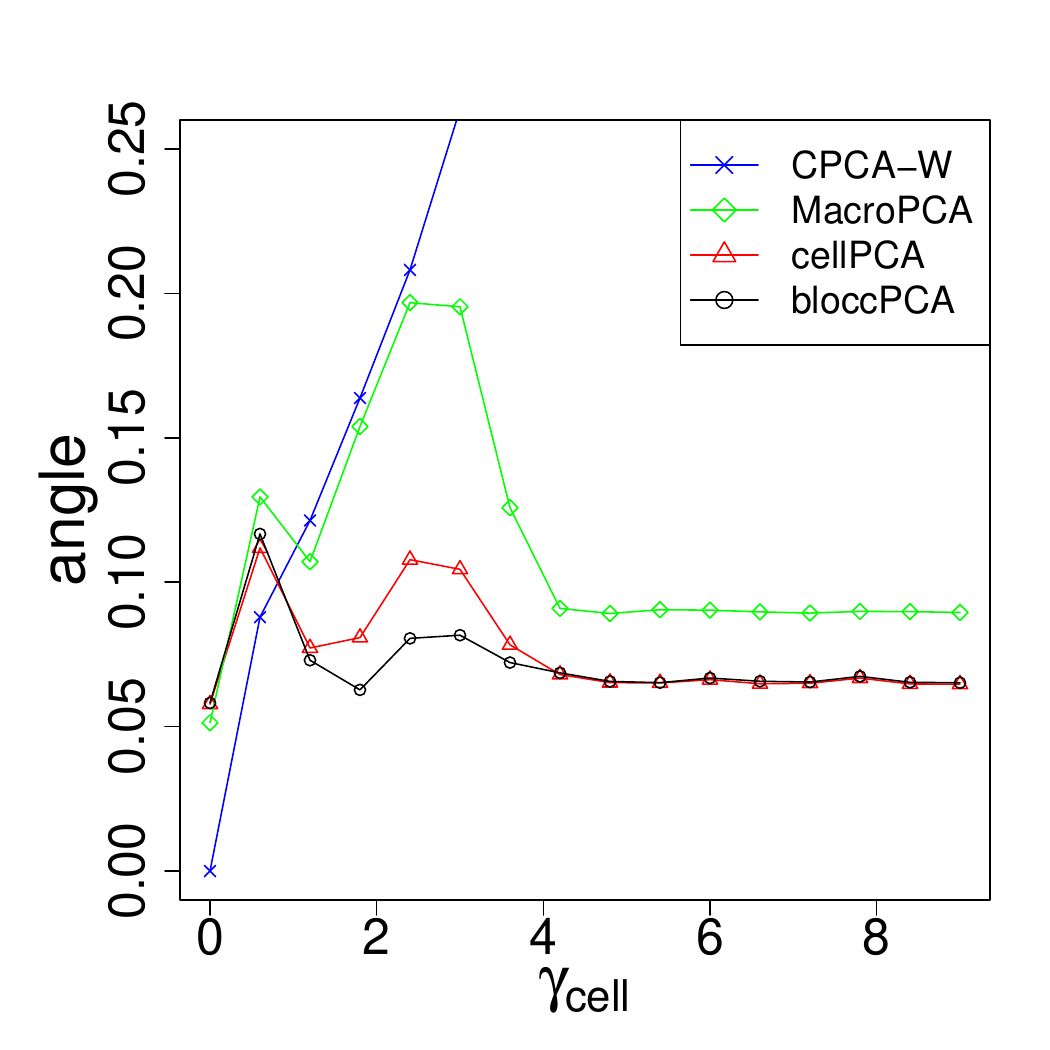} &
		\includegraphics[width=0.28\textwidth]{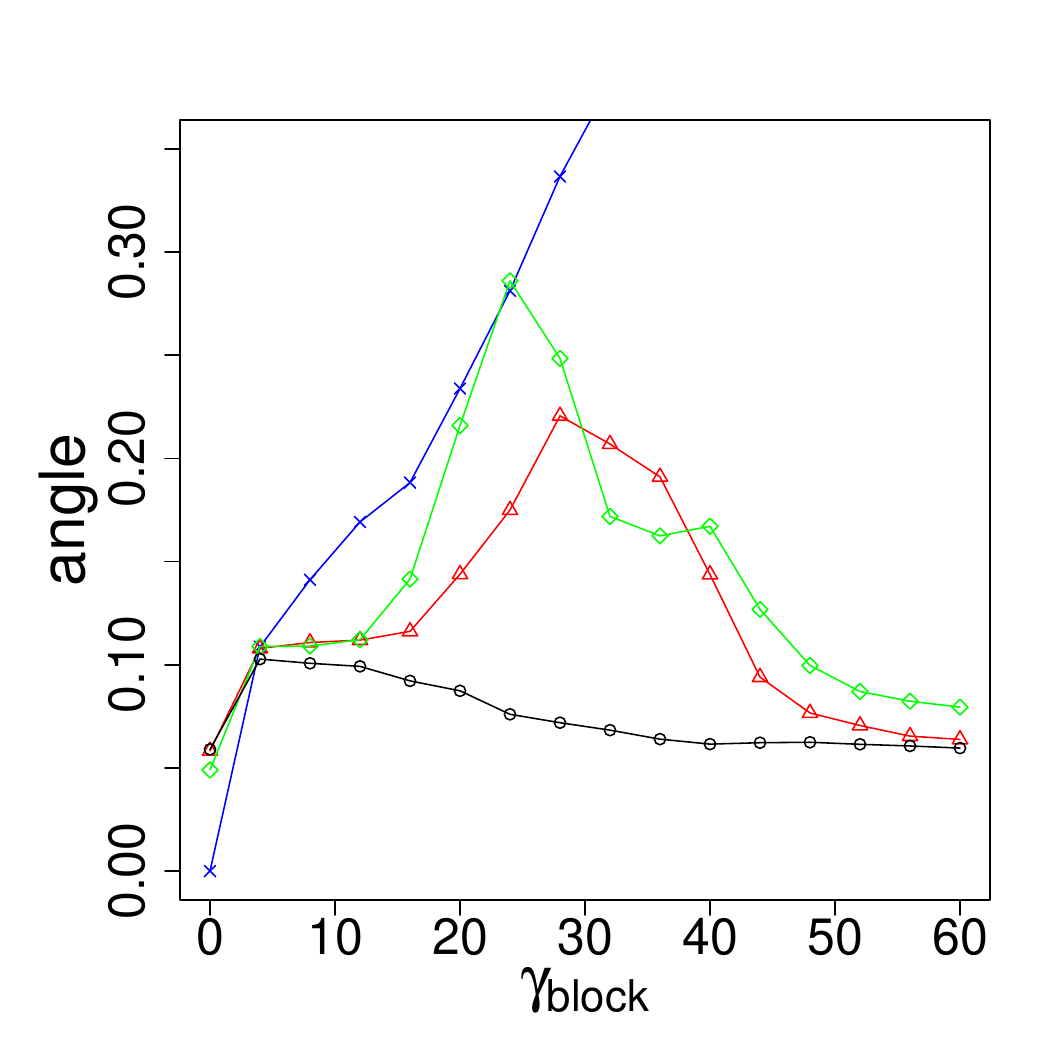} & 
        \includegraphics[width=0.28\textwidth]{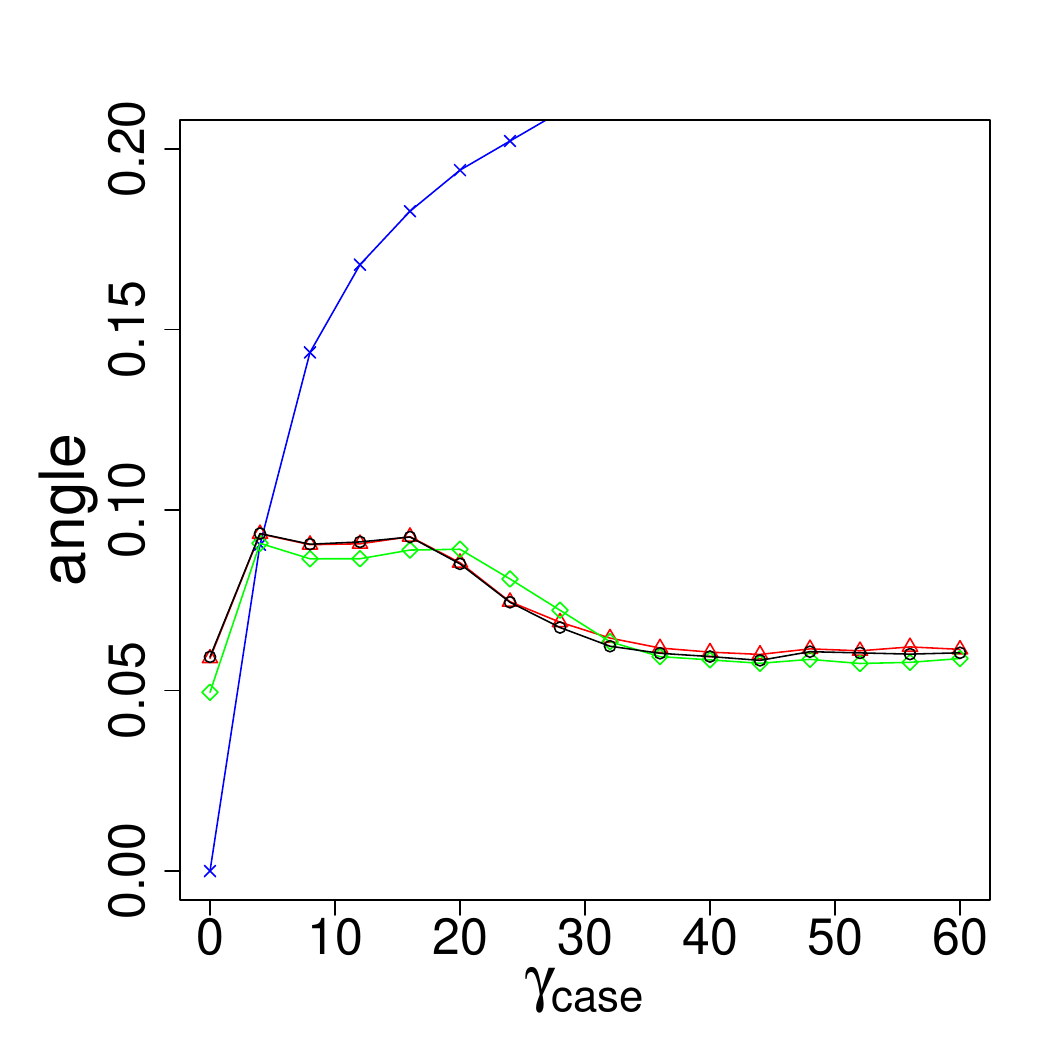}\\[6pt]
        
		\textbf{Cellwise-blockwise} &
        \textbf{Blockwise-casewise} & 
		\textbf{Cellwise-blockwise-casewise}\\[-12pt]

		\includegraphics[width=0.28\textwidth]{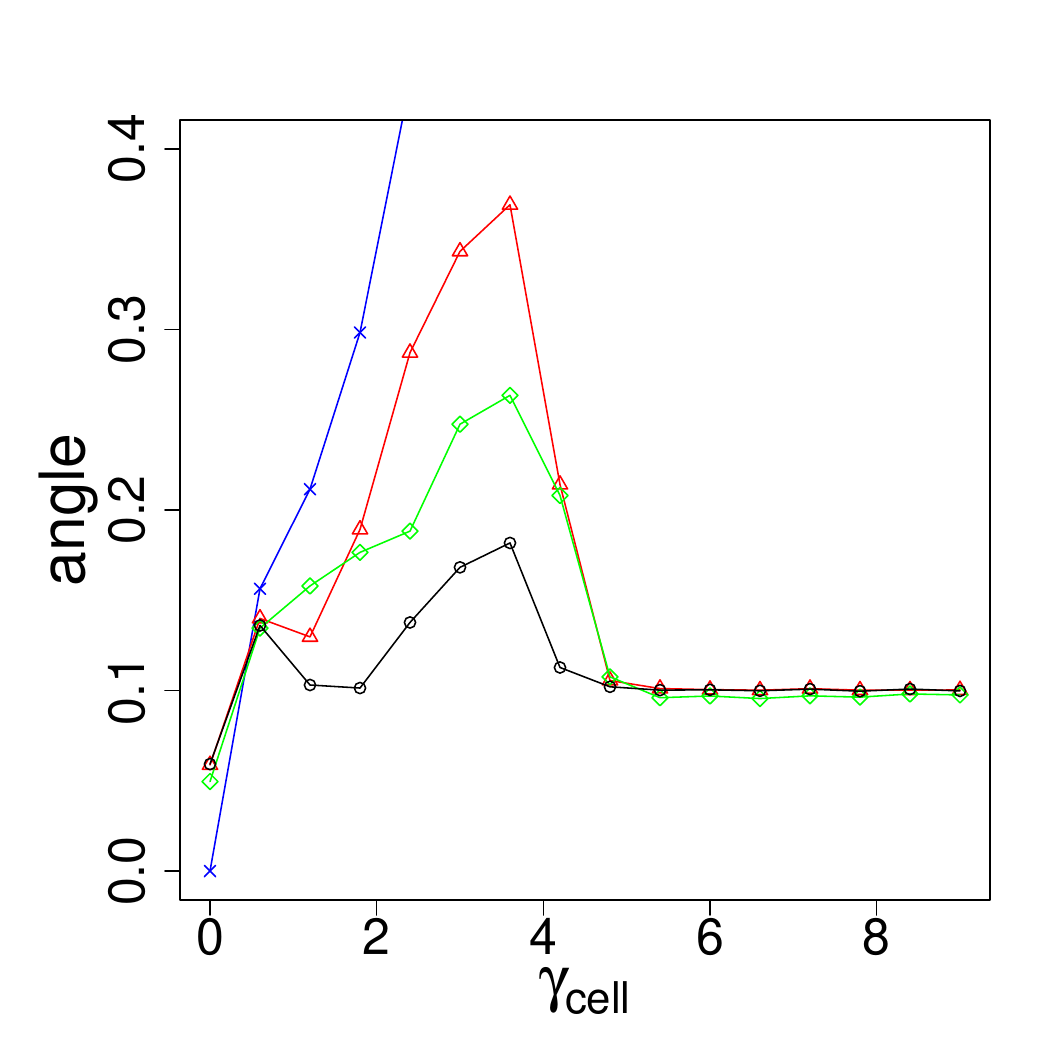}  &
		  \includegraphics[width=0.28\textwidth]{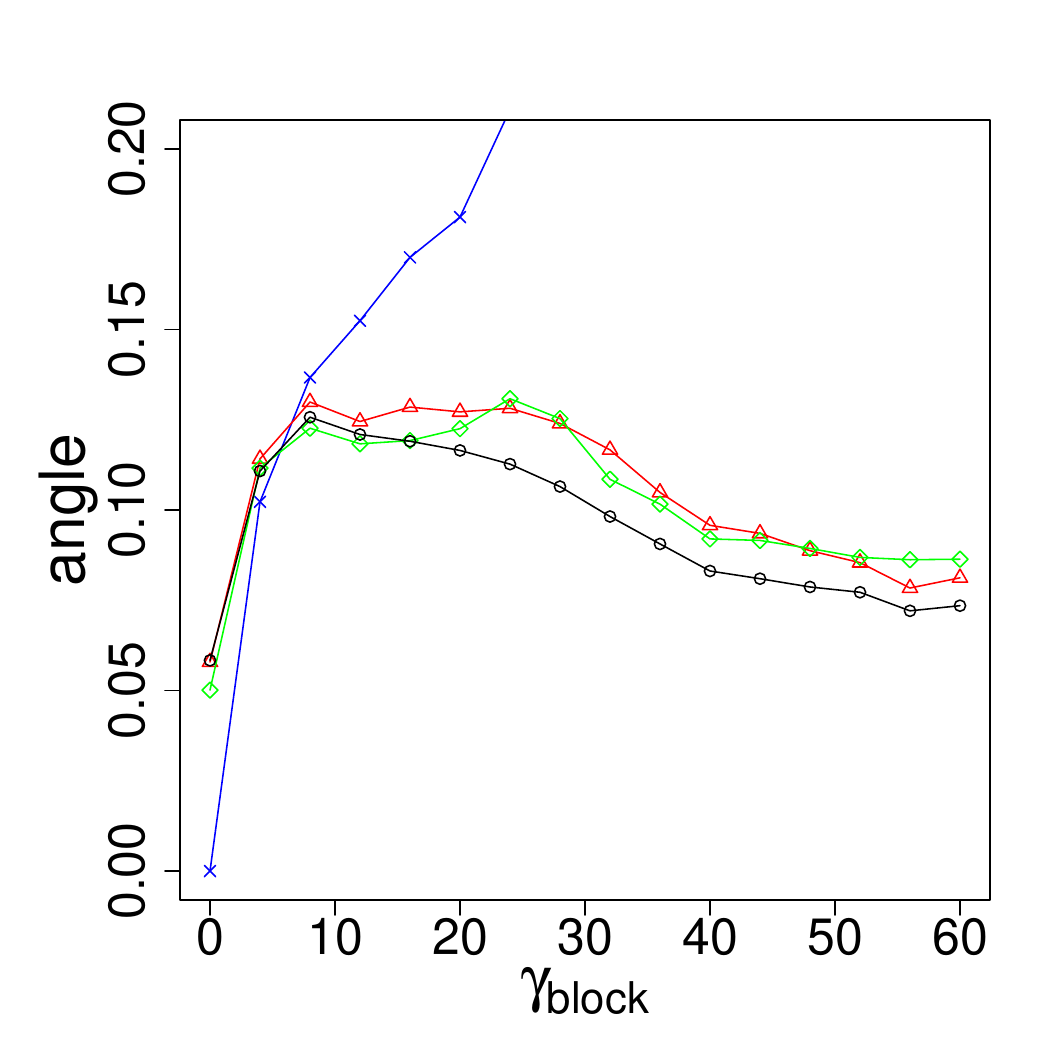} &
		\includegraphics[width=0.28\textwidth]{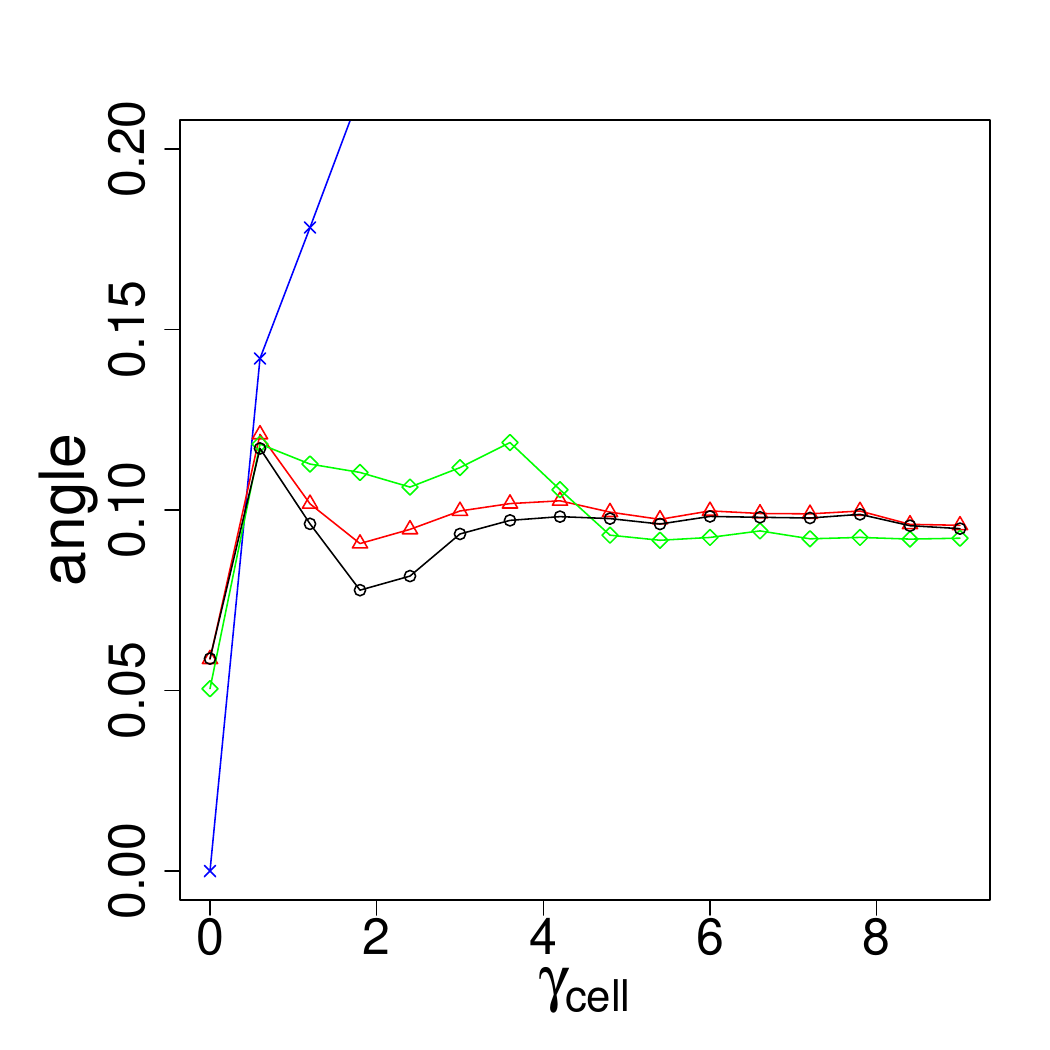}\\
	\end{tabular}
    \vspace{-0.5cm}
	\caption{Median angle attained by {CPCA-W}, MacroPCA, cellPCA, and bloccPCA under the different contamination scenarios for $n=100$, {$p=60$, $B=4$} without missing data.}
	\label{fig:sim_40_noNA}
\end{figure}

Figure~\ref{fig:sim_40_noNA} shows the median angle over 200 simulated
datasets obtained by {CPCA-W}, MacroPCA, cellPCA, and bloccPCA under the
different contamination scenarios for $n=100$, {$p=60$, and $B=4$}, without
missing data.

We first observe that when $\gamma_{\text{cell}} = 0$,
that is, when the data are uncontaminated, all methods have a similar performance
and {CPCA-W} reaches the {smallest angle}. This  was expected as CPCA-W is designed for clean data. However, in each contamination scenario, the method quickly breaks down. 

When the data are contaminated with blockwise outliers, bloccPCA significantly outperforms the competitors, with little to no impact of the blockwise outliers on the performance. 
In the scenario with cellwise outliers only, bloccPCA performs better than cellPCA at all contamination levels. This improvement is due to the additional blockwise robustness introduced through the $\rho_2$ function in the objective function. Indeed, a mildly contaminated cell may not be sufficiently downweighted by the cellwise function $\rho_1$. However, its contribution can still increase the corresponding blockwise deviation, causing the block to be identified as partially outlying and downweighted through $\rho_2$. As a result, the influence of the contaminated cell is further reduced.
In the casewise setting, bloccPCA, cellPCA and MacroPCA show a similar performance.
When multiple types of contamination are present, bloccPCA consistently provides
more accurate (or similar) estimates than the competitors. This
advantage is especially clear in the cellwise-blockwise 
contamination setting.

Overall, the simulation study shows that, in the presence of blockwise outliers,
bloccPCA provides substantial improvements over PCA methods that are not
blockwise robust, such as cellPCA and MacroPCA. Importantly, the additional
blockwise robustness does not lead to a loss of robustness against cellwise or
casewise outliers.

These findings are also confirmed by the additional simulation settings with
$p=200$, and with missing data, whose results are reported in
Section~\ref{supp:additional_simulation} of the Supplementary Material.

\section{Empirical Results}\label{sec:realdata}

\subsection{Global and blockwise structures of EU wellbeing}
\label{subsec:mbpca_results}
This subsection addresses research question \textbf{Q1}, namely how the
thematic areas of wellbeing contribute to the global structure of EU wellbeing
and how differences between countries can be traced back to specific blocks and
indicators. 
To answer this question, 
we apply bloccPCA to the EU wellbeing dataset, where the number of components is
selected using the Kneedle algorithm, which identifies $k=3$ as the elbow
point. The first three supercomponents explain $66\%$ of the variability in the
supermatrix. 

The first step in addressing \textbf{Q1} is to evaluate whether the selected
components provide a meaningful representation not only of the whole dataset,
but also of each thematic area. Although the rank is selected at the level of
the supermatrix, this does not guarantee that the corresponding blockcomponents
summarize each block equally well. Table~\ref{tab:expvar_block} shows the explained variance
within each block, using the criterion described in Subsection~\ref{subsec:block}.
\begin{table}[h]
    \centering
    \small
    \begin{tabular}{lc}
        \hline
        \textbf{Block} & \textbf{Explained variance} \\
        \hline
        GDP & 1.00 \\
        Wellbeing today & 0.55 \\
        Resources for the future & 0.69 \\
        Societal resilience & 0.57 \\
        Nature & 0.53 \\
        Inclusiveness & 1.00 \\
        Institutions & 0.97 \\
        \hline
    \end{tabular}
    \caption{Explained variance in each block of the EU wellbeing dataset.}
    \label{tab:expvar_block}
\end{table}
The first three components provide a
substantial summary of the supermatrix. 
At the block level, the explained
variance logically reaches $100\%$ in the blocks ``GDP" and ``Inclusiveness" since there are less variables than principal components
in these blocks. Moreover, it is very high for the block ``Institutions", and remains
large for ``Resources for the future". In contrast, ``Wellbeing today", 
``Societal resilience", and ``Nature" are summarized less completely by the first three
components. Still, 
they capture a meaningful proportion of the variability in these blocks.

\begin{figure}[!ht]
    \centering
    \begin{tabular}{ccc}
        \includegraphics[width=0.3\linewidth]{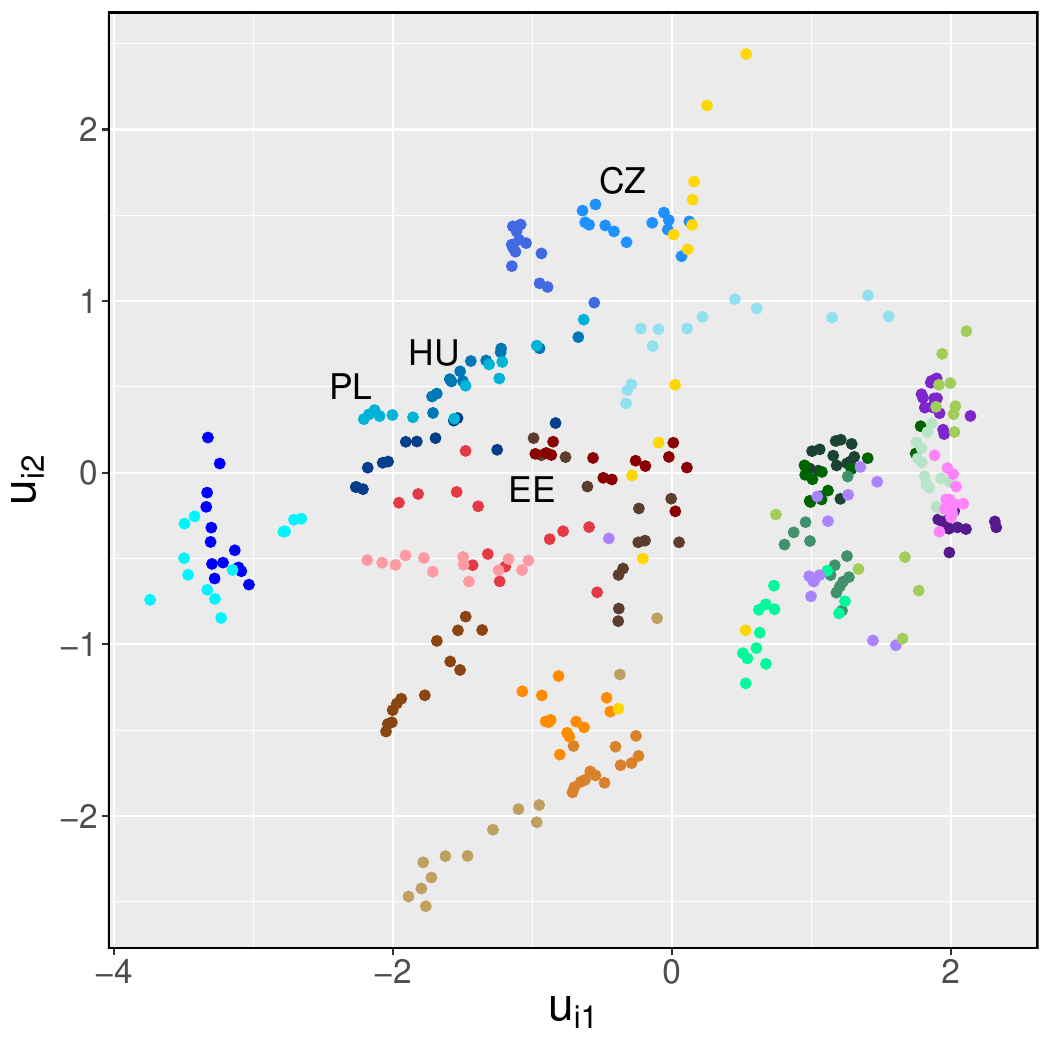}
        &
        \includegraphics[width=0.3\linewidth]{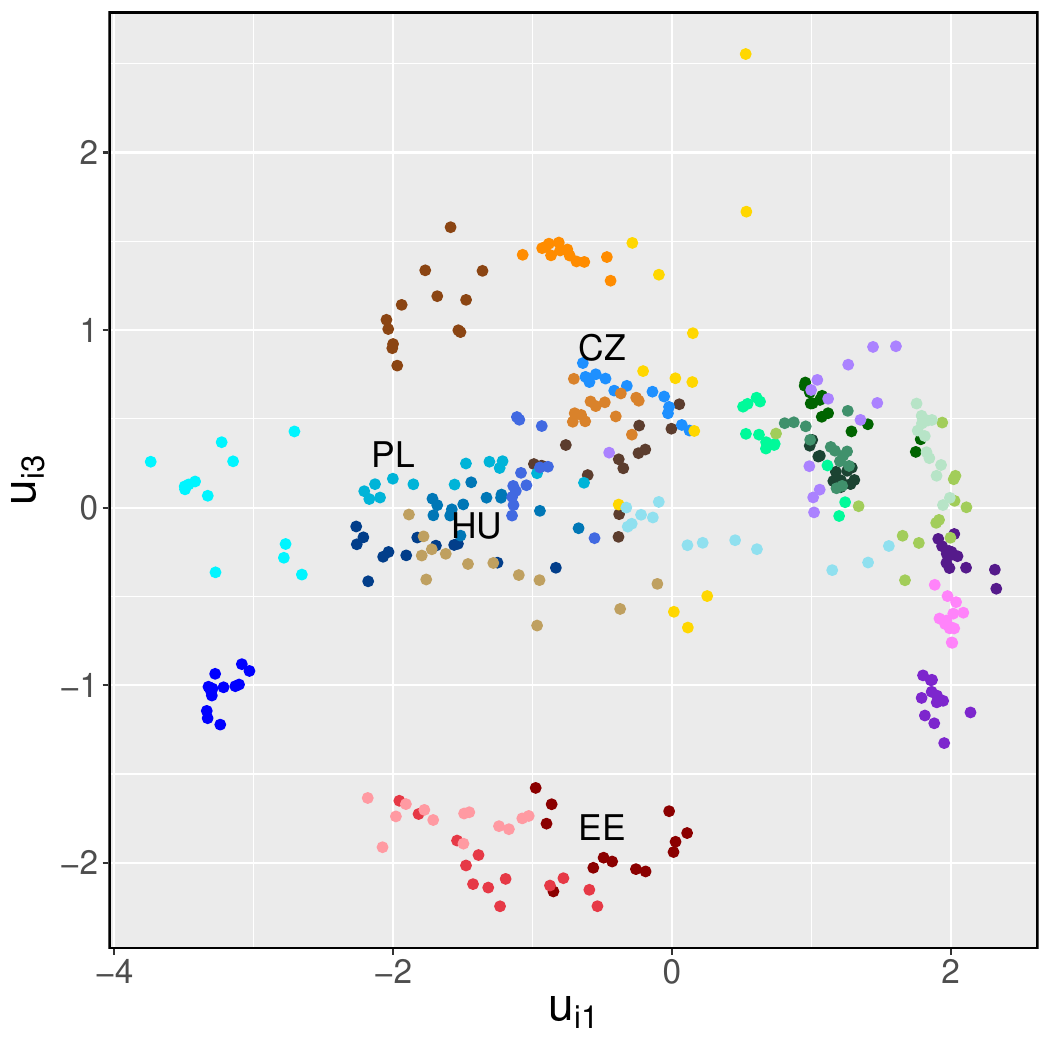}
        &
        \includegraphics[width=0.3\linewidth]{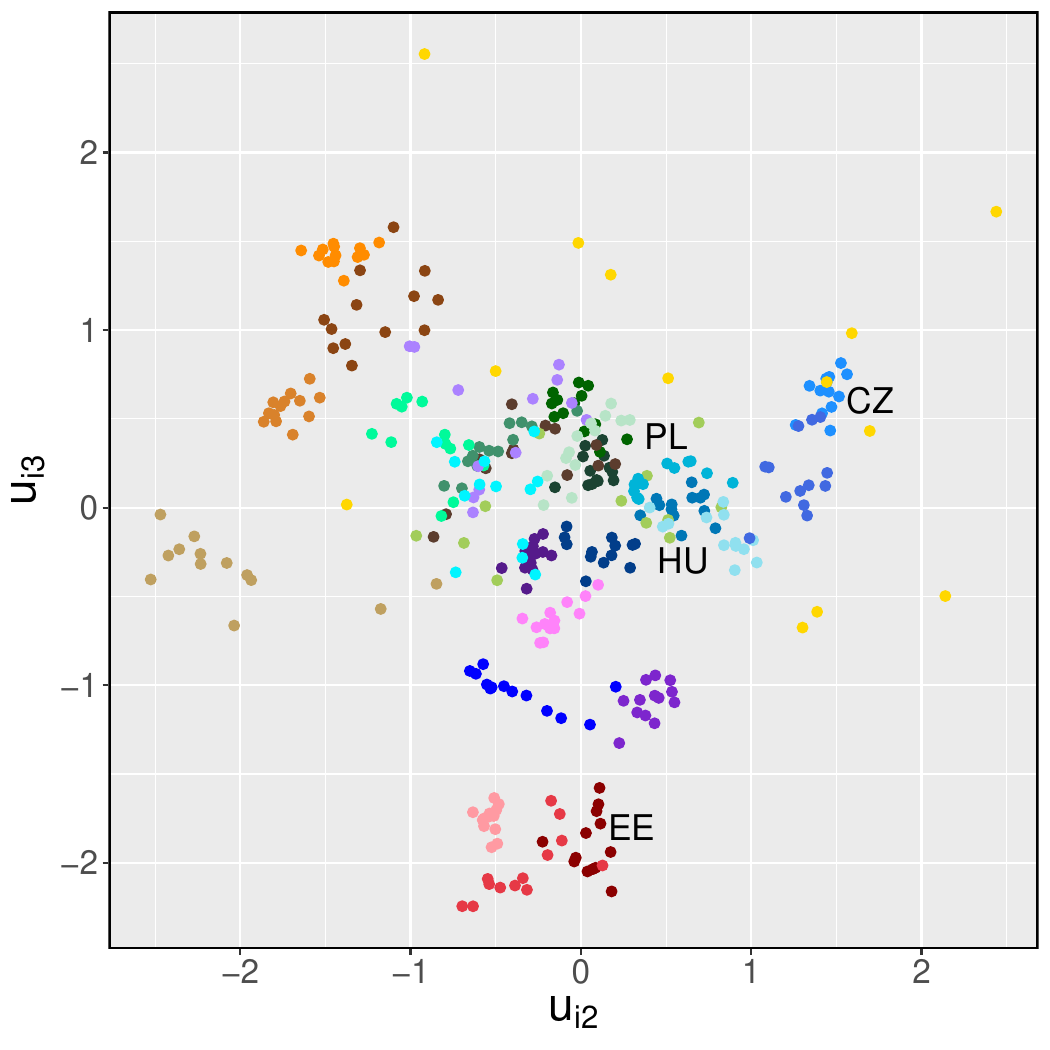}
    \end{tabular}
  \vspace{-0.5cm}  
    \includegraphics[width=1\linewidth]{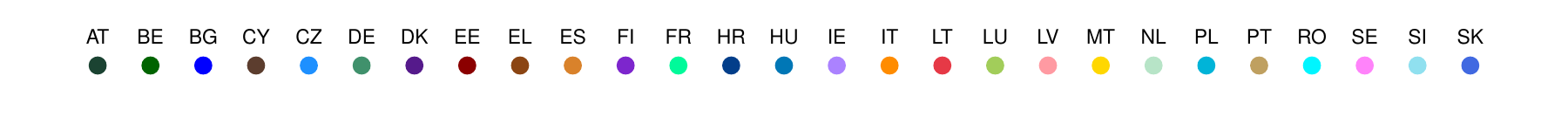}
    \vspace{-0.7cm}
    \caption{Scatter plots of the first three superscores of bloccPCA, colored by country:
components 1--2 (left), 1--3 (middle), and 2--3 (right).}
    \label{fig:superscores_123}
\end{figure}

 To describe the global structure, Figure~\ref{fig:superscores_123}
displays the scatter plots of the first three superscores, considering the
component pairs $(1,2)$, $(1,3)$, and $(2,3)$, where points
have been colored by country. The legend below associates each color with the corresponding country,
identified by its two-letter code. We follow the ISO 3166-1 alpha-2 codes,
except for Greece, which is denoted by EL following the usual EU convention.
Observations that are close in
these plots, for example Poland (PL) and Hungary (HU), have similar overall wellbeing profiles, whereas observations that
are far apart, such as Czech Republic (CZ) and Estonia (EE), differ in their global configuration of indicators.

The superscore plots
alone do not answer which thematic areas drive
the observed global patterns. To link the global structure back to the blocks,
Table~\ref{tab:block_contributions} shows  the contribution of block $b$ to component $s$, which is computed as
$100\|\bv^{(b)s}\|^2$. Since each superloading vector has unit norm, these
contributions sum to 100 over the blocks for each component. Large values
therefore indicate that the corresponding thematic area plays an important role
in defining that global direction of variation.
\begin{table}[h]
\centering
\small
\begin{tabular}{lccc}
\hline
\textbf{Block} & \textbf{Component 1} & \textbf{Component 2} & \textbf{Component 3} \\
\hline
GDP & 25 & 4 & 3 \\
Wellbeing today & 11 & 7 & 11 \\
Resources for the future & 12 & 19 & 7 \\
Societal resilience & 9 & 11 & 15 \\
Nature & 7 & 10 & 16 \\
Inclusiveness & 9 & 43 & 44 \\
Institutions & 27 & 6 & 4 \\
\hline
\end{tabular}
\caption{Block contributions to the first three supercomponents, computed as
$100\|\bv^{(b)s}\|^2$.}
\label{tab:block_contributions}
\end{table}

Thus, the first component is
mainly driven by ``Institutions" and ``GDP", suggesting a global direction associated
with institutional quality and economic performance. The second component is
largely driven by ``Inclusiveness", with a relevant contribution from ``Resources for
the future", while the third component is again dominated by ``Inclusiveness",
together with ``Nature" and ``Societal resilience". 

\begin{figure}[h]
    \centering
    \begin{tabular}{ccc}
        \includegraphics[width=0.31\linewidth]{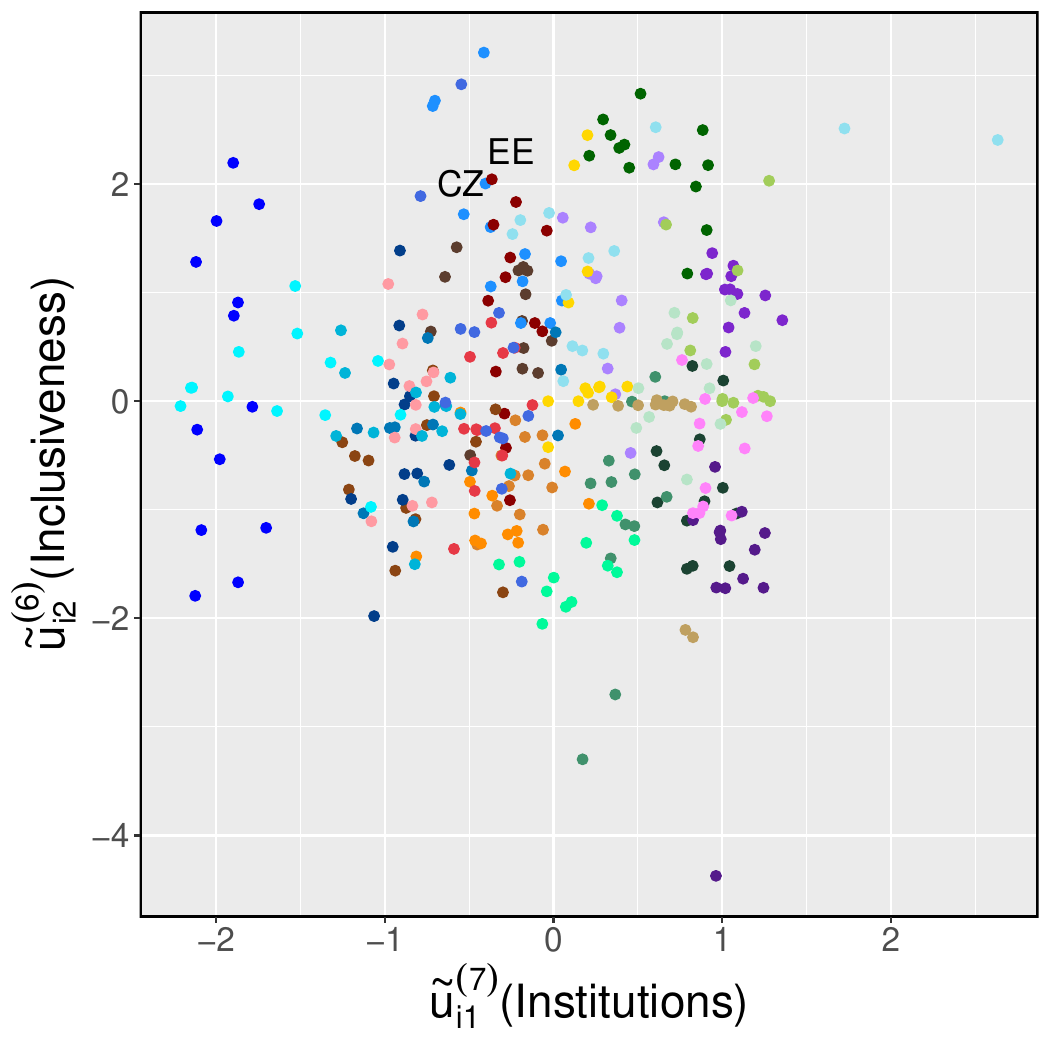}
        &
        \includegraphics[width=0.31\linewidth]{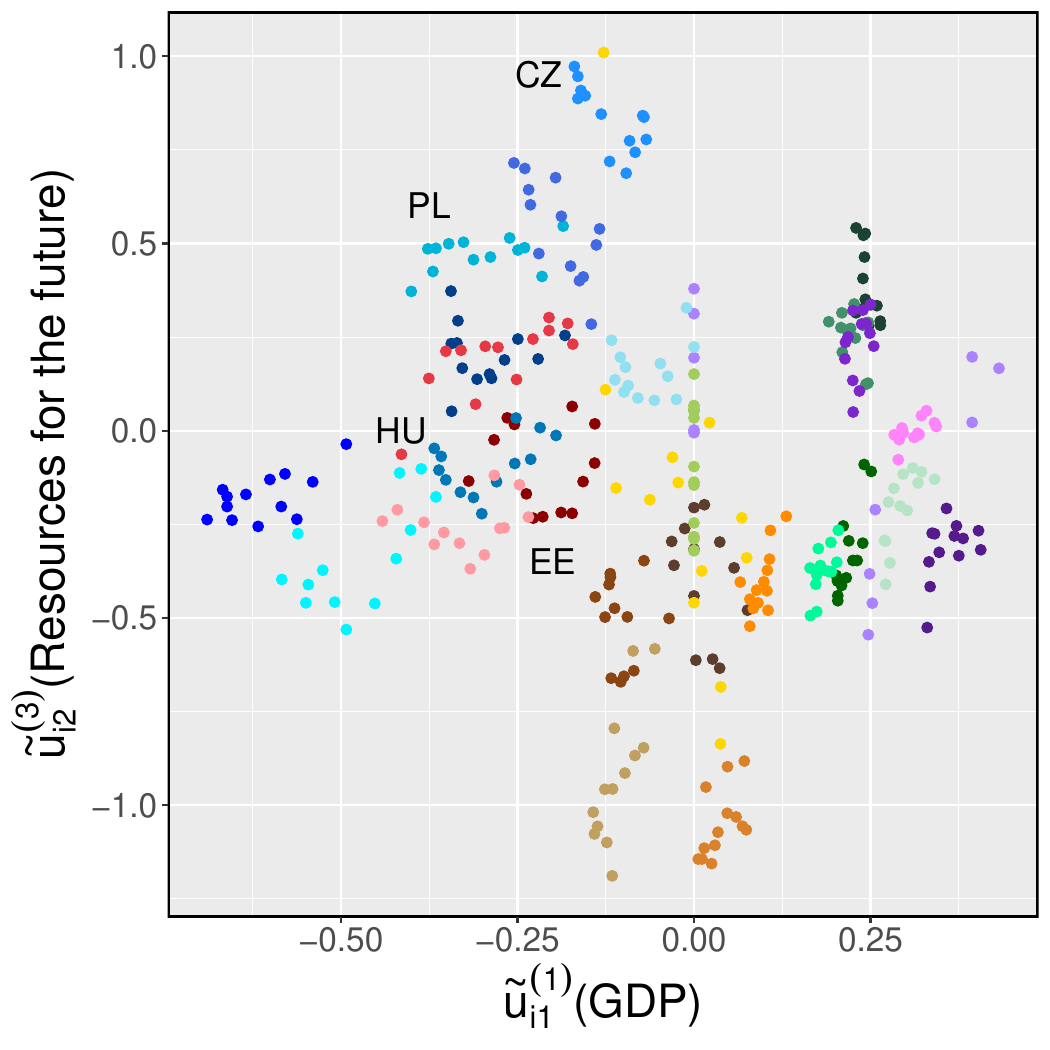}
       
    \end{tabular}
    \vspace{-0.5cm}
    
    \includegraphics[width=1\linewidth]{Figures_SIWB/legend.pdf}
     \vspace{-1.2cm}
    \caption{Blockscore plots for the two blocks contributing most to the first two
    principal components colored by country.}
    \label{fig:blockscores_main}
\end{figure}

To compare the global and blockwise representations,
Figure~\ref{fig:blockscores_main} reports the blockscore plots for the two blocks
that contribute most to the first two components from Table~\ref{tab:block_contributions}.
The comparison between superscores and blockscores makes it possible to
distinguish between global similarity and area-specific differences. Two
observations may be close in the superscore space, indicating similar overall
wellbeing profiles, but far apart in a blockscore plot, indicating that they
differ in a specific thematic area. 
For example, Hungary and Poland, close in terms of superscores, 
 differ along the second blockscore component of the block ``Resources for
the future'', as shown in the right panel of Figure~\ref{fig:blockscores_main}.

To investigate which indicators may contribute most to this difference,
Table~\ref{tab:blockload} reports the variables with the largest blockloading
components in absolute value for the selected blocks and principal components.
The blockloadings indicate which variables mainly define the corresponding
blockscore direction: variables with large positive or negative loadings are
those that contribute most to the separation of the observations in the
blockscore plot. This allows us to explain not only whether two countries
differ in a thematic area, but also which specific indicators are responsible
for this difference.
For Hungary (HU)--Poland (PL), the inspection of the blockloadings shows that the
difference in the block ``Resources for the future'' is mainly driven by the
variable ``Total fixed assets'', for which Hungary has higher values than Poland.
The two countries have similar values for ``Gross domestic expenditure on R\&D'' and ``Children aged less than 3 years in formal childcare''.

\begin{table}[h]
\centering
\footnotesize

\begin{tabular}{p{6.5cm}r}
\hline
\multicolumn{2}{l}{\textbf{Institutions}} \\
\hline
Voice and accountability & 0.43\\
Government effectiveness & 0.42\\
Worldwide Governance Index & 0.40 \\
Rule of Law & 0.40  \\
\hline
\end{tabular}

\vspace{0.3cm}

\begin{tabular}{p{6.5cm}r p{6.5cm}r}
\hline
\multicolumn{2}{l}{\textbf{Inclusiveness}} &
\multicolumn{2}{l}{\textbf{Resources for the future}} \\
\hline
Income share of the bottom 40\% of the population & 0.64
& Gross domestic expenditure on R\&D & 0.38 \\
Income quintile share ratio (S80/S20) & -0.61
& Total fixed assets & 0.37 \\
& & Children aged less then 3 years in formal childcare & 0.36 \\
\hline
\end{tabular}

\caption{Variables associated with the largest blockloading coefficients in
absolute value for the most contributing blocks of the first component
(Institutions, top) and the second component (Inclusiveness and Resources for
the future, bottom).}
\label{tab:blockload}
\end{table}

The pair Estonia (EE)--Czech Republic (CZ) illustrates observations that are
distant in terms of superscores but close in specific blockscore
representations. The two countries show similar behavior in the ``Institutions'', ``Inclusiveness'', and ``GDP'' blocks, whereas they differ in the
``Resources for the future'' block. This contrast appears to
drive their separation in the superscore space, suggesting that their global
difference is mainly associated with future-oriented
resources, rather than with institutional quality, inclusiveness, or GDP.
Specifically,
within the ``Resources for the future'' block, the Czech Republic has lower
values for ``Children aged less then 3 years in formal childcare'', and
higher values for ``Total fixed assets'' than Estonia.

\subsection{Casewise, blockwise, and cellwise outlyingness}\label{subsec:out_analysis}
To address
\textbf{Q2}, which concerns the identification and interpretation of anomalous
country-year observations,
 we introduce two graphical tools, i.e.\ the \textit{blockwise residual cellmap} and \textit{blockwise outlier map}, which combine different types
of outlyingness in a single display.  They allow us to distinguish between casewise, blockwise, and
cellwise outlyingness, thereby providing a more detailed interpretation of the
anomalous patterns detected by bloccPCA.

\subsubsection{Blockwise residual cellmap}

\begin{figure}[h]
        \includegraphics[width=\linewidth]{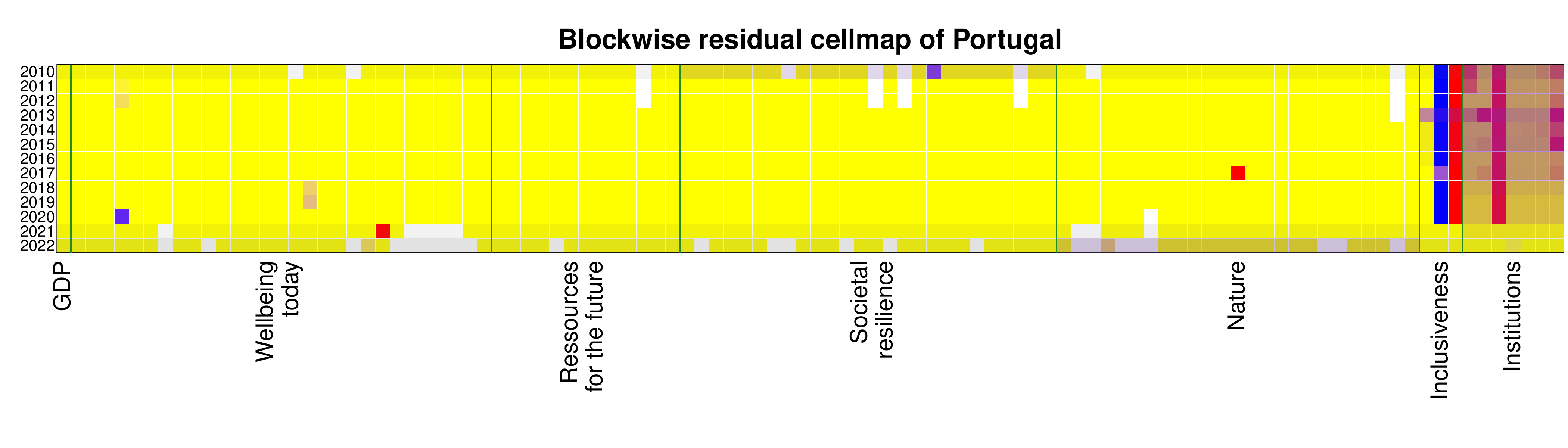}
         \vspace{-1.3cm}
        \caption{Blockwise residual cellmap of Portugal.}
        \label{fig:cellmap_pt}
\end{figure}

From the output of bloccPCA, we obtain the residual block matrices
$\bR^{(b)}=\bX^{(b)}-\bhX^{(b)}$, where
$\bhX^{(b)}=\bone_n(\bhmu^{(b)})^T+\bhU(\bhV^{(b)})^T$ for $b=1,\dots,B$. 
These residuals are then standardized columnwise using the corresponding
M-scale estimates as defined in \eqref{eq:mscale} to obtain $\btR^{(b)}$.
We can then visualize the concatenated $\btR^{(b)}$, or some of their rows 
and columns, by the blockwise residual cellmap as in 
Figure~\ref{fig:cellmap_pt}, which shows the observations corresponding to Portugal.

The cellmap represents the data in its table form, where
each row corresponds to a case, and each column is a variable.
Green vertical 
lines are added to delimit the blocks.
A data cell is colored yellow if its associated 
absolute standardized cellwise residual is small. Here we use $|\tr_{ij}^{(b)}| <
\sqrt{\chi^2_{1, 0.999}} \approx 3.3$. These represent 
the regular cells. Outlying cells
with a standardized residual larger than $3.3$
are highlighted in light orange up to dark red 
(for $\tr_{ij}^{(b)}  \geqslant 
6$). Analogously, cells with a large negative residual
are gradually colored by a light purple to dark blue color 
(for $\tr_{ij}^{(b)}  \leqslant -6$). 
Missing values are shown as white cells.

Blockwise and casewise information are added to the blockwise cellmap by
introducing additional shading at the block and row levels, extending the shaded cellmap of \cite{Hirari:cellGraphics}. From $\btR^{(b)}$, the blockwise deviations $d_i^{(b)}$ and the casewise total deviations $t_i$ can be computed as well as their M-scale, yielding the standardized blockwise deviations $\tilde d_i^{(b)}$ and the standardized casewise total deviations $\tilde{t}_i$.   A block is shaded
when $\tilde d_i^{(b)}$ exceeds the
cutoff $c_{\tilde d,0.99}^{(b)}$, and the intensity of the shading increases
with the magnitude of $\tilde d_i^{(b)}$. The strongest shading is assigned to
blocks for which $\tilde d_i^{(b)} \geqslant 1.5 c_{\tilde d,0.99}^{(b)}$. These shaded
blocks indicate blockwise outlyingness.

Similarly, a row is
shaded when $\tilde t_i$ exceeds
$c_{\tilde t,0.99}$, with the strongest gray shading assigned when
$\tilde t_i \geqslant 1.5 c_{\tilde t,0.99}$. In this way, observations that are
casewise outlying can be distinguished from observations that are outlying
only within specific blocks.

The cutoffs are obtained from a clean-data reference distribution. Specifically,
we use the superloadings and the estimated residual scales to simulate clean
data, and then apply bloccPCA to these simulated datasets. For each block $b$,
$c_{\tilde d,\alpha}^{(b)}$ is defined as the $\alpha$-quantile of the
simulated standardized blockwise deviations. Analogously, $c_{\tilde t,\alpha}$
is the $\alpha$-quantile of the simulated standardized casewise total
deviations. 

This display allows us to interpret the anomalous behavior of Portugal at different levels. The casewise outlyingness observed in 2010, 2021, and 2022 suggests that, in these years, the overall profile of Portugal deviates from the multiblock structure. The case of 2010 should be distinguished from those of 2021--2022. Indeed, in 2010, the casewise outlyingness is mainly driven by the blockwise deviations of the blocks ``Societal resilience", ``Inclusiveness", and ``Institutions". In particular, the block ``Societal resilience" contains one cell with an abnormally large negative residual. The value for the indicator ``Net lending/borrowing'' is tagged as a cellwise outlier, since it is $-8.9$ in that year, whereas in the other years it ranges between $-4.5$ and $3.3$. For the years 2021--2022, the casewise outlyingness is driven by the blockwise deviations of the blocks ``GDP" and ``Nature", while the blocks ``Inclusiveness" and ``Institutions" show lower deviations compared to previous years.

At the block level, the persistent shading of the ``Institutions" block indicates that Portugal differs systematically from the main pattern captured by the model in this thematic area. This block contains variables related to ``Institutional quality'', such as ``Control of corruption'', ``Government effectiveness'', and ``Political stability''. The blockwise outlyingness therefore suggests that the institutional profile of Portugal is not well explained by its overall position in the global wellbeing structure. In particular, the third variable, ``Political stability", has a very large positive residual for all years except 2021 and 2022. All other variables of the block show a similar pattern, with positive residuals until 2021. This year and the following one present a shift in the behavior of the outlyingness, as the ``Institutions" block is no longer outlying and the sign of the residuals is even reversed, becoming negative for most variables. Similarly to the casewise outlyingness, this shift is probably due to the COVID-19 crisis, which began in 2020 and had a significant impact.

The cellwise outliers in the ``Inclusiveness'' block provide an even more localized interpretation. They indicate that the anomalous behavior is not associated with the whole block, but rather with specific indicators. In this case, the last two variables of the ``Inclusiveness" block, namely ``Income quintile share ratio (S80/S20)'' and ``Income share of the bottom 40\% of the population'', show large residuals for several years. The co-occurring outlyingness of these two variables is explained by their similarity in definition, as they both measure income inequality but with opposite signs. This means that Portugal shows lower income inequality than expected and, therefore, higher inclusiveness.

Overall, the cellmap shows that Portugal's anomalous behavior is multi-layered: some years deviate at the case level, the ``Institutions block" shows a systematic blockwise difference, and the ``Inclusiveness block" highlights localized cellwise deviations. This illustrates the usefulness of the blockwise cellmap for addressing \textbf{Q2}, as it clarifies whether anomalies arise at the case, block, or cell level. An additional cellmap (of Sweden) and its interpretation are provided in Section~\ref{supp:add_results} of the Supplementary Material.

\subsubsection{Blockwise outlier map}\label{subsec:blockoutmap}
The blockwise residual cellmap mainly represents outlyingness through the distance of an observation from its fitted value in the estimated subspace. In particular, it shows which cells, blocks, or entire cases have large residuals with respect to the fitted value. However, the orthogonal projection of an observation onto the estimated subspace may also be unusual, even if its residual from the subspace is not large. To capture this second type of outlyingness as well, we use the enhanced outlier map introduced by \citet{Centofanti2026cellPCA}. This plot displays, for each observation, the norm of its standardized residual against its score distance. The former measures how far the observation is from the estimated subspace, whereas the latter measures how far the projected observation lies within the fitted subspace.
In the multiblock PCA setting, this diagnostic tool can be computed at both the global and block levels. This yields one global outlier map and $B$ blockwise outlier maps.

\begin{figure}[!ht]
   
        \centering
        \includegraphics[width=0.07\linewidth]{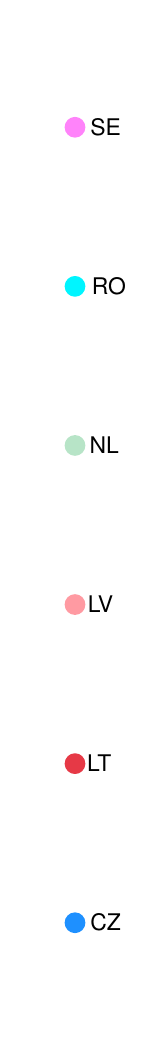}
        \includegraphics[width=0.47\linewidth]{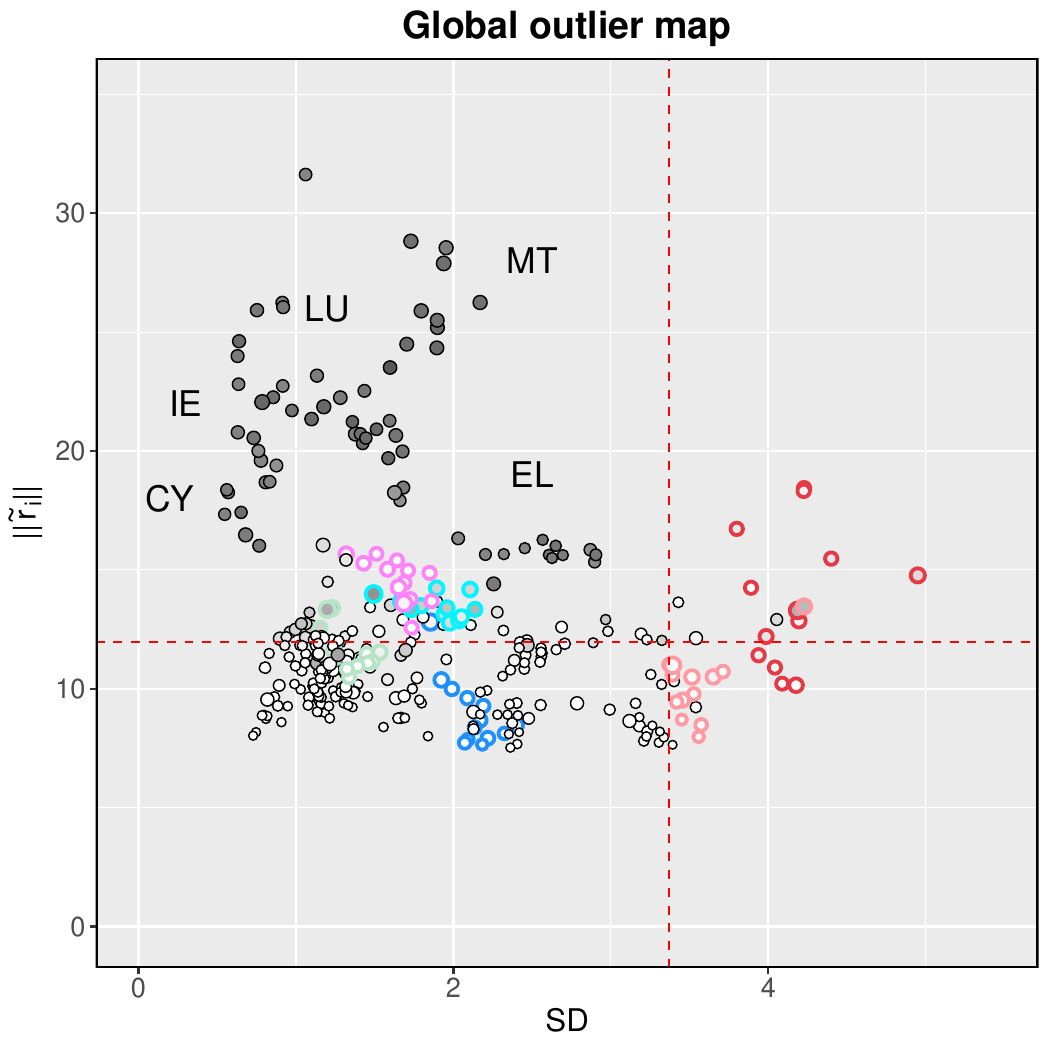}\\
        \includegraphics[width=0.47\linewidth]{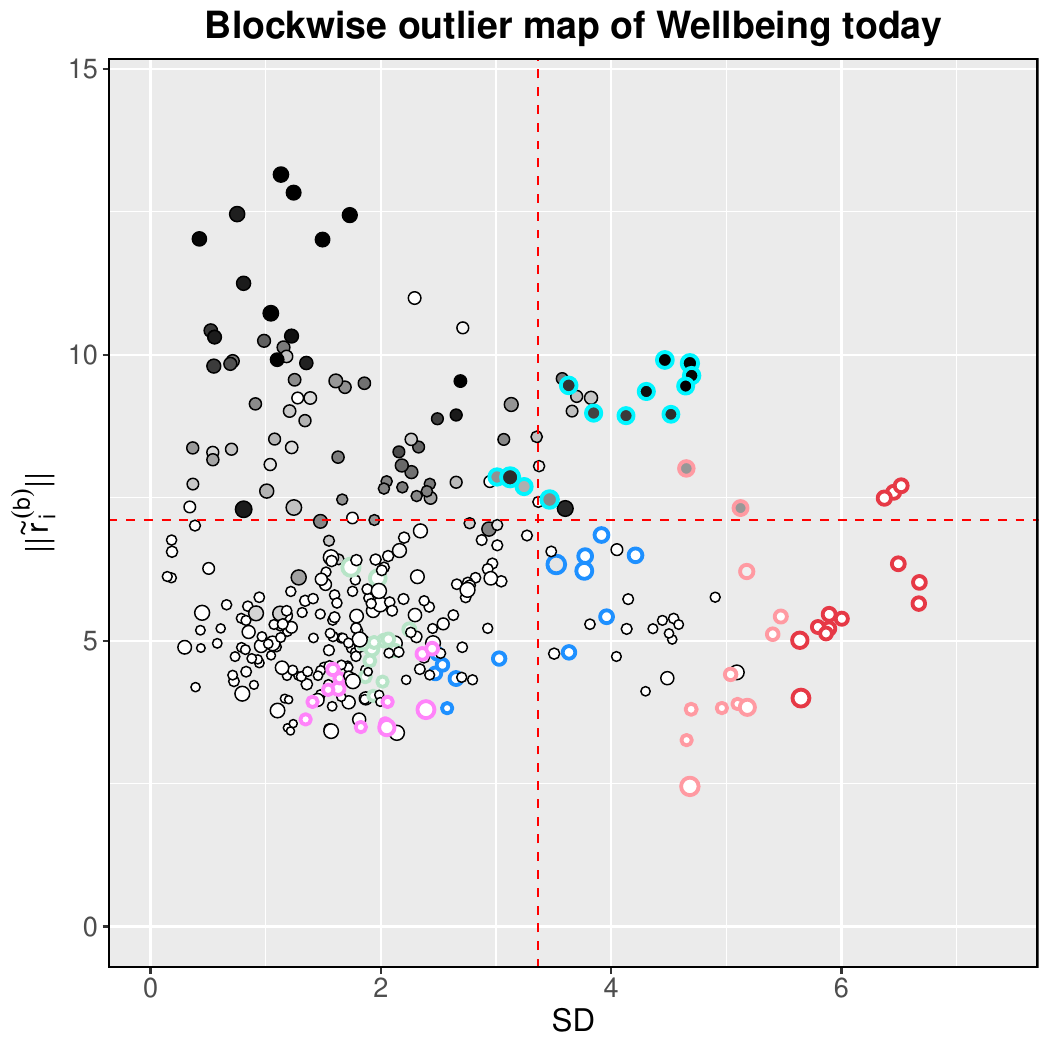}
        \includegraphics[width=0.47\linewidth]{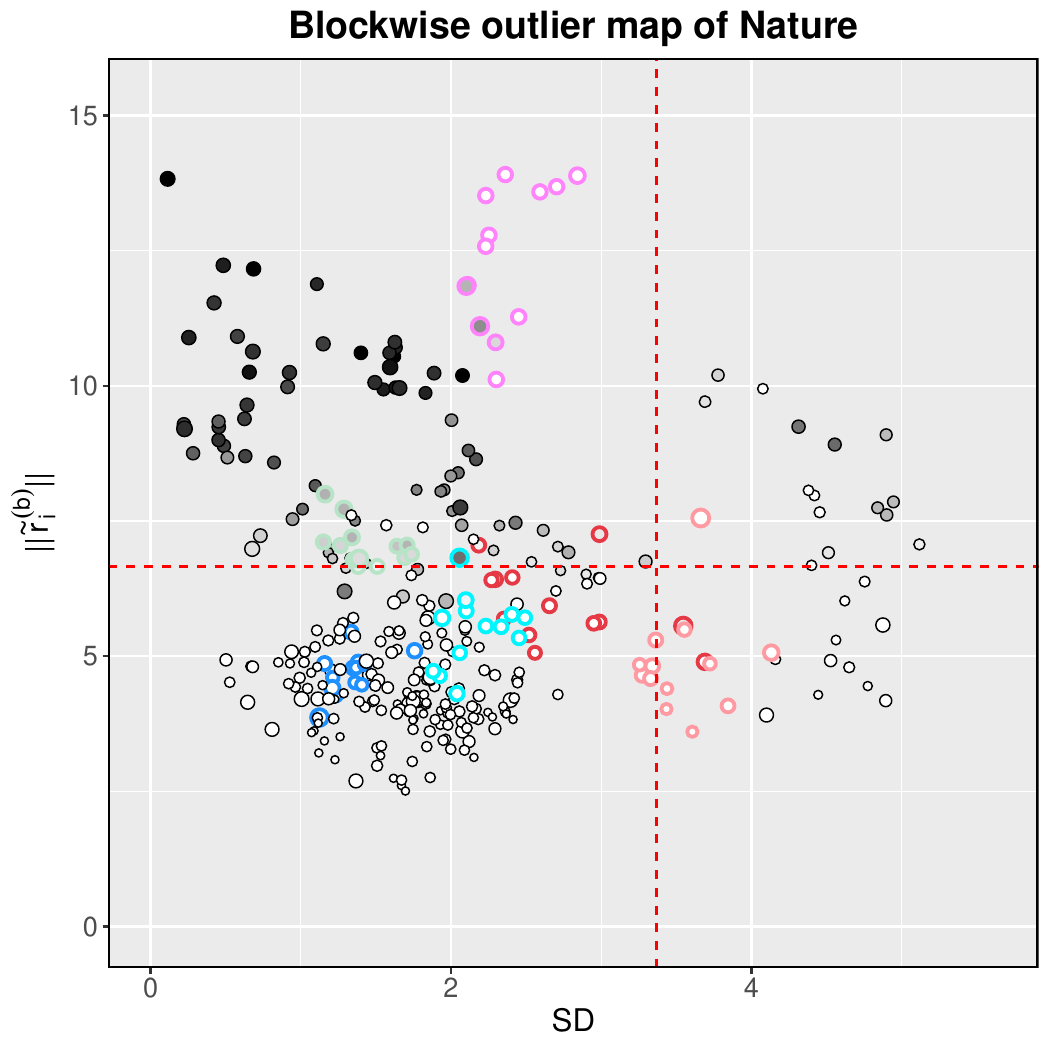}
         \vspace{-0.3cm}
    \caption{Global outlier map and blockwise outlier maps for the blocks ``Wellbeing today''
and ``Nature''.}
    \label{fig:outmap}
\end{figure}

The top row of Figure~\ref{fig:outmap} shows the global outlier map, where each observation is represented by a point whose vertical coordinate is
the norm of its standardized global residual, i.e., 
$\|\btr_i\|=\sqrt{\sum_{b=1}^B\|\btr_i^{(b)}\|^2}$. 
The horizontal coordinate is the global score distance $\mathrm{SD}_i$, defined
as the {Mahalanobis distance of the projected observation, i.e.\ the} norm of
$\big(\bhV^T(\bx_i-\bhmu)\big)\odot
(1/\sqrt{\hlambda_1},\ldots,1/\sqrt{\hlambda_k})^T$. 
An observation is considered outlying in residual distance, score distance,
or both, when it exceeds the corresponding cutoff represented by the red dashed
lines. The cutoff for the norm of the standardized residuals is obtained using
the same simulation-based strategy adopted for the blockwise residual cellmap. That is, clean data are
generated from the fitted model, the corresponding standardized residual norms
are computed, and their $0.99$-quantile is used as threshold. For the score
distances, the cutoff is given by $\sqrt{\chi^2_{k,0.99}}$, which is appropriate when we assume normality of the scores of outlier-free observations.
Additional information is encoded through the size and color of the points. The
size of each point is proportional to
$1-\frac{1}{p}\sum_{b=1}^{B}\sum_{j=1}^{p_b}
o_{ij}^{(b)}w_{ij}^{(b)\mathrm{cell}}$, so that larger points correspond to
observations containing more missing or cellwise downweighted entries. Finally,
the color intensity reflects the casewise outlyingness of the observation, that is, darker points correspond to larger casewise total deviations.

The bottom rows of Figure~\ref{fig:outmap} show the blockwise outlier maps for the
 blocks ``Wellbeing today'' and ``Nature''.  For each block $b$, the
vertical coordinate is the norm of the standardized block residual
$\|\btr_i^{(b)}\|$. The horizontal coordinate is the corresponding
block score distance,
computed as the Mahalanobis distance of
$(\btV^{(b)})^T(\bx_i^{(b)}-\bhmu^{(b)})$, based on MCD estimates of
location and covariance.
The cutoff for the vertical coordinate is obtained as the $0.99$-quantile of
the norm of the standardized block residuals simulated under the clean-data
reference model. The cutoff for the block score distance is
$\sqrt{\chi^2_{\text{min}(p_b,k),0.99}}$. 
The size of each point is proportional to
$1-\frac{1}{p_b}\sum_{j=1}^{p_b}o_{ij}^{(b)}
w_{ij}^{(b)\mathrm{cell}}$, so that larger points correspond to observations
with more missing or cellwise downweighted entries in block $b$. The darkness of
the point is proportional to the blockwise outlyingness of the observation,
with darker points corresponding to larger blockwise deviations.

The global outlier map shows that some countries have large residual distances, e.g., Sweden (SE), and Romania (RO). 
Some countries (Cyprus (CY), Greece (EL), Ireland (IE), Luxembourg (LU), and Malta (MT))
show even higher residual distances along with high casewise total 
deviations, as seen by their darkness.  Notice that this is not
the case for Sweden and Romania shown in white because their outlyingness
is caused by a few cellwise and blockwise outliers whose influences were
downweighted by the functions $\rho_1$ and $\rho_2$ in the computation of the casewise total deviations.

A comparison between the global outlier map and the blockwise outlier maps helps
identify the thematic areas that contribute to the detected outlyingness. For
instance, Sweden (SE) is mildly anomalous in terms of residual distance at
the global level. Inspection of the blockwise outlier maps shows that this
outlyingness mainly originates from the block ``Nature''. In particular, Sweden
shows unusually low values for the indicators ``GHG emissions from land use,
land use change and forestry'' and ``Net greenhouse gas emissions''. This is
consistent with the analysis of the blockwise residual cellmap of Sweden
provided in Section~\ref{supp:add_results} of the Supplementary Material.
Moreover, at the global scale, Romania also displays a slight outlyingness in
residual distance. This deviation is principally driven by the block ``Wellbeing today'', where Romania records very high values for
``Inactive population due to caring responsibilities by sex'' and ``Share of people with good or very good perceived health''.

The global outlier map also reveals that Latvia (LV) and
Lithuania (LT) have outlying score distances while having
non-outlying residual distances for Latvia and several mildly outlying residual distances for Lithuania.  This means that these observations are mostly well
reconstructed by the model, since they lie close to the fitted subspace, but {they}
are extreme within  {the estimated} subspace. 
For these countries, the large score
distance appears mainly associated with the blocks ``Wellbeing today'' and
``Nature''.  In the
``Wellbeing today'' block, these countries show high values of
``Long working hours in main job'' and low values of ``Share of
people with good or very good perceived health''. In the ``Nature'' block,  Latvia show high values for
``Air pollution pm10'' and ``Air pollution pm2.5''.

Finally, the blockwise outlier maps show that some observations that are not
outlying at the global level may become outlying within a specific thematic
area. For example, the Czech Republic (CZ) appears regular in the global map,
but for some years displays outlying behavior in the ``Wellbeing today'' block. This
block-specific outlyingness is associated with unusually low values of the
indicators ``In-work at-risk-of-poverty rate'' and ``At risk of poverty or
social exclusion rate (AROPE)''. Similarly, the Netherlands (NL) shows
block-specific outlyingness in the ``Nature'' block, mainly due to its very low
value for ``Consumption footprint per capita''.
\section{Conclusion}\label{sec:conc}
We have introduced bloccPCA, a new robust multiblock PCA method designed for data whose variables are organized into meaningful blocks and may contain casewise, blockwise, and cellwise outliers, as well as missing values. The method is motivated by a real problem arising in the analysis of sustainable and inclusive wellbeing indicators, where the goal is  to understand how different thematic areas contribute to the global wellbeing structure. In this sense, bloccPCA provides a direct answer to \textbf{Q1}, by extracting robust global components while preserving the interpretation of the block-specific contributions.
The method also addresses \textbf{Q2}, as it provides diagnostic tools to identify whether anomalous country--year observations arise at the case, block, or cell level. This is achieved through the combination of robust estimation, blockwise and cellwise weights, residual cellmaps, and global and blockwise outlier maps. These tools make it possible to distinguish observations that deviate as a whole from those that are anomalous only within specific thematic areas or individual indicators. The application to the EU wellbeing dataset shows that bloccPCA can reveal interpretable global and area-specific patterns that would be hidden by a purely global analysis.

This work opens several directions for further research. One possible extension is the development of robust multiblock regression or classification methods for data with cellwise, blockwise, and casewise contamination. 

\bigskip

\noindent{\bf Software availability.} The \textsf{R} code and the EU wellbeing dataset are publicly available from the webpage 
\if0\blind 
{
\texttt{https://wis.kuleuven.be/statdatascience/robust/software}.
} \fi
\if1\blind 
{\url{xxx}.
} \fi

\bibliographystyle{chicago}
\setlength{\bibsep}{5pt plus 0.2ex}
{\small
\spacingset{1}
\bibliography{reference}
}

\end{document}


\def\spacingset#1{\renewcommand{\baselinestretch}
{#1}\small\normalsize} \spacingset{1}

\newcommand{\blind}{0}


\if0\blind
{

\title{\bf Supplementary Material to: \\\vskip10mm Cellwise, Blockwise, and Casewise Robust Multiblock PCA 
for Sustainable and Inclusive Wellbeing in the EU}
\author[1,2]{Arthur Daisomont}
\author[1]{Fabio Centofanti}
\author[1]{Mia Hubert}

\affil[1]{Section of Statistics and Data Science, Department 
           of Mathematics, KU Leuven, Belgium}
\affil[2]{Joint Research Center, European Commission, Ispra, Italy}

\setcounter{Maxaffil}{0}
\renewcommand\Affilfont{\itshape\small}
\date{October 2, 2026} 
  \maketitle
} \fi

\if1\blind
{
  \phantom{abc}\\
  \vspace{20mm}
  
  \begin{center}
     {\LARGE \bf Supplementary Material to: \\\vskip10mm Cellwise, Blockwise, and Casewise Robust \\Multiblock PCA 
for Sustainable and\\ 
\vskip4mm Inclusive Wellbeing in the EU}\\

    \vskip12mm
    \large{October 1, 2026} 
\end{center}
  \medskip
} \fi


\pagenumbering{arabic}

\appendix

\setcounter{equation}{0} 
\renewcommand{\theequation}
  {A.\arabic{equation}} 

\spacingset{1.45} 

\section{Cellwise, blockwise, and casewise contamination models}\label{subsec:blockoutmodel}

The concept of outliers originally referred to outlying cases, that is,
observations that differ from the majority due to their deviation
in a large number of components of the data. It was formally described by
the Tukey-Huber model \citep{tukey-1962,huber-1964}, 
and is also
known as the \textit{casewise contamination model} \citep{Centofanti2026cellPCA}.
It assumes
that the supermatrix $\bX$ originates
from a contaminated $p$-dimensional random variable $X_{\varepsilon}$ distributed as 
$(1-\varepsilon^{\text{case}})H_0 + \varepsilon^{\text{case}}H_Z$, where
$0 \leqslant \varepsilon^{\text{case}} < 0.5$ represents the proportion of contaminated
cases on average. The distributions $H_0$ and $H_Z$ correspond
to the clean data and the outlier generating distributions, respectively. The random variable
$X_{\varepsilon}$ can also be expressed as
\begin{equation}\label{eq:cont_mod_case}
    X_{\varepsilon} = A^{\text{case}}\odot X + (\bone_p - A^{\text{case}})\odot Z,
\end{equation}
where $\bone_p$ is a column vector of size $p$ with all components
equal to 1, and $\odot$ is the Hadamard product, which is the 
element-wise multiplication between vectors or matrices. The random
variables $X$ and $Z$ follow the distributions $H_0$
and $H_Z$, respectively. Finally, $A^{\text{case}}$ is a random vector of dimension
$p$ such that each of its components $A^{\text{case}}_j$ is Bernoulli 
distributed with a probability of success equal to $1-
\varepsilon^{\text{case}}$ for $j=1,\ldots,p$ and
the marginals $A^{\text{case}}_j$ are jointly fully dependent,
that is, $P(A^{\text{case}}_1=\ldots=A^{\text{case}}_p) = 1$.
If $X$, $A^{\text{case}}$, and $Z$ are independent of
each other, the model described by \eqref{eq:cont_mod_case} corresponds
to the $\varepsilon$-contamination model from \cite{huber-1981} or, equivalently,
to the \textit{fully dependent contamination model} (FDCM) from \cite{alqallaf-2009}.

By the condition $0 \leqslant \varepsilon^{\text{case}} < 0.5$, the model
assumes that at most half of the observations are contaminated.
While this is often verified in low-dimensional settings,
it is rarely the case in higher dimensions
where each new variable carries an additional risk of being contaminated.
Casewise-only robust methods
would delete or downweight an entire row or column even if they
have a partial contamination rate. In order to avoid removing
clean cells within partially contaminated observations, the concept
of \textit{cellwise outliers} was introduced.
Cellwise outliers are defined as cells generated from a contaminated distribution
and can affect any entry in the data matrix. The \textit{cellwise contamination model}
\citep{Centofanti2026cellPCA}
describes the process of generating cellwise contaminated data as
\begin{equation}\label{eq:cont_mod_cell}
    X_{\varepsilon} = A^{\text{cell}}\odot X + (\bone_p - A^{\text{cell}})\odot Z,
\end{equation}
where $X$ and $Z$ are the same as in \eqref{eq:cont_mod_case}. The difference
with its casewise counterpart lies in the components $A_j^{\text{cell}}$ of the 
$p$-variate variable $A^{\text{cell}}$. While they are still Bernoulli distributed with a
success probability of $1-\varepsilon_j^{\text{cell}}$, we do not impose their dependence.
This implies that contamination can affect some cells individually and 
not necessarily simultaneously.
Furthermore, if the components $A_j^{\text{cell}}$ are independent,
the model becomes equivalent to the \textit{fully independent contamination model}
(FICM) of \cite{alqallaf-2009}.

In our framework, we assume that the observed data matrix $\bX$ is the concatenation 
of $B$ block matrices $\bX^{(b)}$. In such a context, we define a new type of outliers
called \textit{blockwise outliers} as
all cells within a block generated by a contaminated distribution,
as illustrated in the middle panel of Figure \ref{tab:con_models}.
Blockwise outliers differ conceptually from the casewise and cellwise outliers
in the sense that they are not casewise outliers
because they only partially affect  an observation
and therefore should not lead
to the down-weighting or removal of the entire observation. 
Moreover, unlike cellwise outliers, they are defined by the co-occuring 
contamination of several cells.
We introduce the \textit{blockwise contamination model}, which is defined as
\begin{equation}\label{eq:cont_mod_block}
    X_{\varepsilon} = A^{\text{block}}\odot X + (\bone_p - A^{\text{block}})\odot Z,
\end{equation}
where $X$ and $Z$ are still the same as in \eqref{eq:cont_mod_case}. 
The $p$-variate variable \sloppy
$A^{\text{block}} =
\left(
(A^{(1)})^T,\ldots,
(A^{(B)})^T
\right)^T$
is defined blockwise, where $A^{(b)}$ is a $p_b$-variate
random vector with Bernoulli marginals $A_j^{(b)}$ having success
probability $1-\varepsilon_j^{(b)}$. Moreover, within each block~$b$, the marginals of $A^{(b)}$ are fully dependent, that is,
$P\left(A_1^{(b)}=\cdots=A_{p_b}^{(b)}\right)=1$.
The contamination model \eqref{eq:cont_mod_block} can also be written
as
\begin{equation}\label{eq:cont_mod_block_adv}
    X_{\varepsilon}^{(b)}
    =
    A^{(b)}\odot X^{(b)}
    +
    \left(\bone_{p_b}-A^{(b)}\right)\odot Z^{(b)},
\end{equation}
where $X_{\varepsilon}^{(b)}$, $X^{(b)}$, and $Z^{(b)}$ denote the restrictions
of $X_{\varepsilon}$, $X$, and $Z$ to block $b$, respectively.

In practice, it is common to encounter all three types of contamination
along with missing values in the same dataset.
For this reason, we combine models \eqref{eq:cont_mod_case},
\eqref{eq:cont_mod_cell}, and \eqref{eq:cont_mod_block} to form 
the \textit{fully mixed contaminated and partially observed contamination model} as
\begin{equation}\label{eq:cont_mod_full}
    X_{\varepsilon} = A\odot X + (\bone_p - A)\odot Z,
\end{equation}
where $A = A^{\text{case}} \odot A^{\text{cell}} \odot A^{\text{block}} \odot A^{\text{obs}}$.
The variables $A^{\text{case}}$, $A^{\text{cell}}$ and $A^{\text{block}}$ are defined
identically as in \eqref{eq:cont_mod_case},
\eqref{eq:cont_mod_cell} and \eqref{eq:cont_mod_block}, respectively.
Moreover, $A^{\text{obs}}$ is a random vector with marginals $A^{\text{obs}}_j$ which are
binary variables with possible outcomes 1 and NA such that
$P(A^{\text{obs}}_j = 1) = 1-\varepsilon_j^{\text{obs}}$.

\section{First-order conditions}
\label{app:firstorder}
In this section, the first-order
necessary conditions of \eqref{eq:objP} are derived.
Recall that the objective function \eqref{eq:objP} is given by
\begin{multline*}
		L_{\rho_1,\rho_2,\rho_3}(\bX,\bV,\bU,\bmu) = \\
		\frac{\hsigma_3^2}{o}\sum_{i=1}^{n}
		o_i\rho_3\!\left(\frac{1}{\hsigma_{3}}\sqrt{\frac{1}{o_i}
			\sum_{b=1}^{B} o^{(b)}_{i} 
			\left(\hsigma_{2}^{(b)}\right)^2\rho_2\! 
			\left(\frac{1}{\hsigma_{2}^{(b)}}\sqrt{\frac{1}
				{o^{(b)}_{i}}
				\sum_{j=1}^ {p_b} o^{(b)}_{ij}\, 
				\left(\hsigma_{1,j}^{(b)}\right)^2\,
				\rho_1\!\left(\frac{r_{ij}^{(b)}}
				{\hsigma_{1,j}^{(b)}}\right)}\, \right)}\right).
\end{multline*}
We first compute the gradient of $L_{\rho_1,\rho_2,\rho_3}$ 
with respect to $\bv_{j}^{(b)}$,
\begin{align*} 
\bzero = \frac{\partial}{\partial \bv_{j}^{(b)}}
    L_{\rho_1,\rho_2,\rho_3} 
  &= \frac{\hsigma_3^2}{o} \sum_{i=1}^{n}
     o_i \rho'_3(\rt_i/\hsigma_3) 
     \frac{1}{\hsigma_3}
     \frac{\partial}{\partial \bv_{j}^{(b)}} \rt_i
     \nonumber\\
  &= \frac{1}{o} \sum_{i=1}^{n} o_i
     \frac{\psi_3(\rt_i/\hsigma_3)}
     {\rt_i/\hsigma_3}\, \frac{1}{2o_i}
     \frac{\partial}{\partial \bv_{j}^{(b)}}\left(\sum_{b=1}^{B} 
     o_{i}^{(b)} (\hsigma_{2}^{(b)})^2\rho_2\left(\frac{d_{i}^{(b)}}
     {\hsigma_{2}^{(b)}}\right)\right)\nonumber\\
  &= \frac{1}{2o} \sum_{i=1}^{n}
     \:w_i^{\text{case}}\, o_{i}^{(b)} (\hsigma_{2}^{(b)})^2
     \rho'_2\left(\frac{d_{i}^{(b)}}
     {\hsigma_{2}^{(b)}}\right)\frac{1}{\hsigma_{2}^{(b)}}
     \frac{\partial}{\partial \bv_{j}^{(b)}}d_{i}^{(b)}\nonumber\\
  &= \frac{1}{2o} \sum_{i=1}^{n}
     \:w_i^{\text{case}}\,o_{i}^{(b)} 
     \frac{\psi_2(d_{i}^{(b)}/\hsigma_{2}^{(b)})}
     {d_{i}^{(b)}/\hsigma_{2}^{(b)}}\frac{1}{2o_{i}^{(b)}}
     \frac{\partial}{\partial \bv_{j}^{(b)}}
     \left(\sum_{j=1}^{p_b} o_{ij}^{(b)}(\hsigma_{1,j}^{(b)})^2 
     \rho_1\left( \frac{r_{ij}^{(b)}}
     {\hsigma_{1,j}^{(b)}}\right) \right)\nonumber\\
  &= \frac{1}{4o} \sum_{i=1}^{n}
     \:w_i^{\text{case}}\:w_{i}^{\text{block}(b)}
     o_{ij}^{(b)}(\hsigma_{1,j}^{(b)})^2 
     \rho'_1\left( \frac{r_{ij}^{(b)}}
     {\hsigma_{1,j}^{(b)}}\right)
     \frac{1}{\hsigma_{1,j}^{(b)}}
     \frac{\partial}{\partial \bv_{j}^{(b)}}
     (x_{ij}^{(b)} - \mu_{j}^{(b)} - 
     \bu_i^T\bv_{j}^{(b)})\nonumber\\
  &= \frac{-1}{4o} \sum_{i=1}^{n}
     \:w_i^{\text{case}}\:w_{i}^{\text{block}(b)}
     o_{ij}^{(b)}(\hsigma_{1,j}^{(b)})^2 
     \psi_1\left( \frac{r_{ij}^{(b)}}
     {\hsigma_{1,j}^{(b)}}\right)
     \frac{1}{\hsigma_{1,j}^{(b)}}\bu_i\nonumber\\
  &= \frac{-1}{4o}\sum_{i=1}^{n}\:w_i^{\text{case}}\:
     w_{i}^{\text{block}(b)}\, \frac{\psi_1(r_{ij}^{(b)}/
     \hsigma_{1,j}^{(b)})}{r_{ij}^{(b)}/\hsigma_{1,j}^{(b)}}
     o_{ij}^{(b)}r_{ij}^{(b)}\bu_i\nonumber\\
  &= \frac{-1}{4o}\sum_{i=1}^{n} 
     \bu_i\:w_i^{\text{case}}\:w_{i}^{\text{block}(b)}\:w_{ij}^{\text{cell}(b)} o_{ij}^{(b)} 
     (x_{ij}^{(b)}-\mu_{j}^{(b)}-\bu_i^T\bv_{j}^{(b)})\nonumber\\
  &= \frac{-1}{4o}\sum_{i=1}^{n} 
     \bu_i w_{ij}^{(b)} 
     (x_{ij}^{(b)}-\mu_{j}^{(b)}-\bu_i^T\bv_{j}^{(b)})\nonumber\\
  &= \frac{-1}{4o}
     \left(\bU^T\bW^{(b)j}(\bx^{(b)j}-\mu_{j}^{(b)} \bone_n)-
     \bU^T\bW^{(b)j}\bU\bv_j^{(b)}\right).
\end{align*}
The gradient of $L_{\rho_1,\rho_2, \rho_3}$ with respect to $\bu_i$ is 
\begin{align*} 
\bzero = \frac{\partial}{\partial \bu_i}
    L_{\rho_1,\rho_2,\rho_3} 
  &= \frac{\hsigma_3^2}{o}
     o_i \rho'_3(\rt_i/\hsigma_3) 
     \frac{1}{\hsigma_3}
     \frac{\partial}{\partial \bu_i} \rt_i
     \nonumber\\
  &= \frac{1}{o} o_i
     \frac{\psi_3(\rt_i/\hsigma_3)}
     {\rt_i/\hsigma_3}\, \frac{1}{2o_i}
     \frac{\partial}{\partial \bu_i}\left(\sum_{b=1}^{B} 
     o_{i}^{(b)} (\hsigma_{2}^{(b)})^2\rho_2\left(\frac{d_{i}^{(b)}}
     {\hsigma_{2}^{(b)}}\right)\right)\nonumber\\
  &= \frac{1}{2o} \sum_{b=1}^{B}
     \:w_i^{\text{case}}\, o_{i}^{(b)} (\hsigma_{2}^{(b)})^2
     \rho'_2\left(\frac{d_{i}^{(b)}}
     {\hsigma_{2}^{(b)}}\right)\frac{1}{\hsigma_{2}^{(b)}}
     \frac{\partial}{\partial \bu_i}d_{i}^{(b)}\nonumber\\
  &= \frac{1}{2o} \sum_{b=1}^{B}
     \:w_i^{\text{case}}\,o_{i}^{(b)} 
     \frac{\psi_2(d_{i}^{(b)}/\hsigma_{2}^{(b)})}
     {d_{i}^{(b)}/\hsigma_{2}^{(b)}}\frac{1}{2o_{i}^{(b)}}
     \frac{\partial}{\partial \bu_i}
     \left(\sum_{j=1}^{p_b} o_{ij}^{(b)}(\hsigma_{1,j}^{(b)})^2 
     \rho_1\left( \frac{r_{ij}^{(b)}}
     {\hsigma_{1,j}^{(b)}}\right) \right)\nonumber\\
  &= \frac{1}{4o} \sum_{k=1}^{B}\sum_{j=1}^{p_b}
     \:w_i^{\text{case}}\:w_{i}^{\text{block}(b)}
     o_{ij}^{(b)}(\hsigma_{1,j}^{(b)})^2 
     \rho'_1\left( \frac{r_{ij}^{(b)}}
     {\hsigma_{1,j}^{(b)}}\right)
     \frac{1}{\hsigma_{1,j}^{(b)}}
     \frac{\partial}{\partial \bu_i}
     \left(x_{ij}^{(b)} - \mu_{j}^{(b)} - 
     \bu_i^T\bv_{j}^{(b)}\right)\nonumber\\
  &= \frac{-1}{4o} \sum_{b=1}^{B}\sum_{j=1}^{p_b}
     \:w_i^{\text{case}}\:w_{i}^{\text{block}(b)}
     o_{ij}^{(b)}(\hsigma_{1,j}^{(b)})^2 
     \psi_1\left( \frac{r_{ij}^{(b)}}
     {\hsigma_{1,j}^{(b)}}\right)
     \frac{1}{\hsigma_{1,j}^{(b)}}\bv_{j}^{(b)}\nonumber\\
  &= \frac{-1}{4o}\sum_{b=1}^{B}\sum_{j=1}^{p_b}
     \:w_i^{\text{case}}\:w_{i}^{\text{block}(b)}\, \frac{\psi_1(r_{ij}^{(b)}/
     \hsigma_{1,j}^{(b)})}{r_{ij}^{(b)}/\hsigma_{1,j}^{(b)}}
     o_{ij}^{(b)}r_{ij}^{(b)}\bv_{j}^{(b)}\nonumber\\
  &= \frac{-1}{4o}\sum_{b=1}^{B}\sum_{j=1}^{p_b} 
     \bv_{j}^{(b)} \:w_i^{\text{case}} \:w_{i}^{\text{block}(b)} \:w_{ij}^{\text{cell}(b)} 
     o_{ij}^{(b)} 
     \left(x_{ij}^{(b)}-\mu_{j}^{(b)}-
     (\bv_{j}^{(b)})^T\bu_i\right)\nonumber\\
  &= \frac{-1}{4o}\sum_{b=1}^{B}\sum_{j=1}^{p_b}
     \bv_{j}^{(b)} w_{ij}^{(b)} 
     \left(x_{ij}^{(b)}-\mu_{j}^{(b)}-
     (\bv_{j}^{(b)})^T\bu_i \right)\nonumber\\
  &= \frac{-1}{4o}\sum_{b=1}^{B}(\bV^{(b)})^T\bW^{(b)}_i 
     \left(\bx^{(b)}_i -\bmu^{(b)}-\bV^{(b)}\bu_i \right).
\end{align*}
Finally, the derivative of $L_{\rho_1,\rho_2,\rho_3}$ with 
respect to $\mu_{j}^{(b)}$ is 
\begin{align*} 
\bzero = \frac{\partial}{\partial \mu_{j}^{(b)}}
    L_{\rho_1,\rho_2,\rho_3} 
  &= \frac{\hsigma_3^2}{o} \sum_{i=1}^{n}
     o_i \rho'_3(\rt_i/\hsigma_3) 
     \frac{1}{\hsigma_3}
     \frac{\partial}{\partial \mu_{j}^{(b)}} \rt_i
     \nonumber\\
  &= \frac{1}{o} \sum_{i=1}^{n} o_i
     \frac{\psi_3(\rt_i/\hsigma_3)}
     {\rt_i/\hsigma_3}\, \frac{1}{2o_i}
     \frac{\partial}{\partial \mu_{j}^{(b)}}\left(\sum_{b=1}^{B} 
     o_{i}^{(b)} (\hsigma_{2}^{(b)})^2\rho_2\left(\frac{d_{i}^{(b)}}
     {\hsigma_{2}^{(b)}}\right)\right)\nonumber\\
  &= \frac{1}{2o} \sum_{i=1}^{n}
     \:w_i^{\text{case}}\, o_{i}^{(b)} (\hsigma_{2}^{(b)})^2
     \rho'_2\left(\frac{d_{i}^{(b)}}
     {\hsigma_{2}^{(b)}}\right)\frac{1}{\hsigma_{2}^{(b)}}
     \frac{\partial}{\partial \mu_{j}^{(b)}}d_{i}^{(b)}\nonumber\\
  &= \frac{1}{2o} \sum_{i=1}^{n}
     \:w_i^{\text{case}}\,o_{i}^{(b)} 
     \frac{\psi_2(d_{i}^{(b)}/\hsigma_{2}^{(b)})}
     {d_{i}^{(b)}/\hsigma_{2}^{(b)}}\frac{1}{2o_{i}^{(b)}}
     \frac{\partial}{\partial \mu_{j}^{(b)}}
     \left(\sum_{j=1}^{p_b} o_{ij}^{(b)}(\hsigma_{1,j}^{(b)})^2 
     \rho_1\left( \frac{r_{ij}^{(b)}}
     {\hsigma_{1,j}^{(b)}}\right) \right)\nonumber\\
  &= \frac{1}{4o} \sum_{i=1}^{n}
     \:w_i^{\text{case}}\:w_{i}^{\text{block}(b)}
     o_{ij}^{(b)}(\hsigma_{1,j}^{(b)})^2 
     \rho'_1\left( \frac{r_{ij}^{(b)}}
     {\hsigma_{1,j}^{(b)}}\right)
     \frac{1}{\hsigma_{1,j}^{(b)}}
     \frac{\partial}{\partial \mu_{j}^{(b)}}
     \left(x_{ij}^{(b)} - \mu_{j}^{(b)} - 
     \bu_i^T\bv_{j}^{(b)}\right)\nonumber\\
  &= \frac{-1}{4o} \sum_{i=1}^{n}
     \:w_i^{\text{case}}\,w_{i}^{\text{block}(b)}
     o_{ij}^{(b)}(\hsigma_{1,j}^{(b)})^2 
     \psi_1\left( \frac{r_{ij}^{(b)}}
     {\hsigma_{1,j}^{(b)}}\right)
     \frac{1}{\hsigma_{1,j}^{(b)}}\nonumber\\
  &= \frac{-1}{4o}\sum_{i=1}^{n} \:w_i^{\text{case}}\,
     \:w_{i}^{\text{block}(b)}\, \frac{\psi_1(r_{ij}^{(b)}/
     \hsigma_{1,j}^{(b)})}{r_{ij}^{(b)}/\hsigma_{1,j}^{(b)}}
     o_{ij}^{(b)}r_{ij}^{(b)}\nonumber\\
  &= \frac{-1}{4o}\sum_{i=1}^{n} 
     \:w_i^{\text{case}} \:w_{i}^{\text{block}(b)} \:w_{ij}^{\text{cell}(b)} o_{ij}^{(b)} 
     \left(x_{ij}^{(b)}-\mu_{j}^{(b)}-
     \bu_i^T\bv_{j}^{(b)}\right),\nonumber
\end{align*}
so the gradient with respect to the vector $\bmu^{(b)}$ becomes
\begin{align*} 
\hspace{-0.7cm}
\bzero = \frac{\partial}{\partial \bmu^{(b)}}L_{\rho_1,\rho_2, \rho_3} 
  &= \sum_{i=1}^n \bW^{(b)}_i\, \left( \bx^{(b)}_i - \bmu^{(b)}
  - \bV^{(b)} \bu_i \right).
\end{align*}
\vskip6mm

We also derive the first order conditions of the 
weighted PCA objective function
\begin{equation*} 
  \tL = 
  \sum_{i=1}^n\sum_{b=1}^{B}\sum_{j=1}^{p_b} w_{ij}^{(b)}
    \left(x_{ij}^{(b)}-\mu_j^{(b)}-\bu_i^T\bv_j^{(b)}\right)^2\;.
\end{equation*}
We start with the  gradient of $\tL$ 
with respect to $\bv_j^{(b)}$ which is 
\begin{align*} 
\bzero = \frac{\partial}{\partial \bv_j^{(b)}} \tL 
  &= 
  \sum_{i=1}^{n} w_{ij}^{(b)}\,
  \frac{\partial }{\partial \bv_j^{(b)}}
  \left(x_{ij}^{(b)}-\mu_j^{(b)}-\bu_i^T\bv_j^{(b)}\right)^2\nonumber\\
  &= 
  -2\sum_{i=1}^{n} \bu_i w_{ij}^{(b)} 
   (x_{ij}^{(b)}-\mu_j^{(b)}-\bu_i^T\bv_j^{(b)})\nonumber\\
  &= 
  -2\left(\bU^T\bW^{(b)j}(\bx^{(b)j}-\mu_j^{(b)} \bone_n)-
     \bU^T\bW^{(b)j}\bU\bv_j^{(b)}\right).
\end{align*}
The gradient of $\tL$ with respect to $\bu_i$ is 
\begin{align*} 
\bzero = \frac{\partial}{\partial \bu_i} \tL 
  &= 
  \sum_{b=1}^{B}\sum_{j=1}^{p_b} w_{ij}^{(b)}\,
  \frac{\partial}{\partial \bu_i}
  \left(x_{ij}^{(b)}-\mu_j^{(b)}-\bu_i^T\bv_j^{(b)}\right)^2\nonumber\\
  &= 
  -2\sum_{b=1}^{B}\sum_{j=1}^{p_b} \bv_j^{(b)} w_{ij}^{(b)}
   \left(x_{ij}^{(b)}-\mu_j^{(b)}-
   \left(\bv_j^{(b)}\right)^T\bu_i\right)\nonumber\\
  &= 
  -2\sum_{b=1}^{B}(\bV^{(b)})^T\bW^{(b)}_i 
     \left(\bx^{(b)}_i -\bmu^{(b)}-\bV^{(b)}\bu_i \right).
\end{align*}
Finally, the derivative of $\tL$ with 
with respect to $\mu_j^{(b)}$ is 
\begin{align*}
  \frac{\partial}{\partial \mu_j^{(b)}} \tL 
  &= 
  \sum_{i=1}^{n} w_{ij}^{(b)}\,
  \frac{\partial}{\partial \mu_j^{(b)}}
  \left(x_{ij}^{(b)}-\mu_j^{(b)}-\bu_i^T
  \bv_j^{(b)}\right)^2\nonumber\\
  &= 
  -2\sum_{i=1}^{n} w_{ij}^{(b)} 
   \left(x_{ij}^{(b)}-\mu_j^{(b)}-
   \left(\bv_j^{(b)}\right)^T\bu_i\right)\,,
\end{align*}
so the gradient with respect to the vector $\bmu^{(b)}$ becomes
\begin{align*}
\bzero = \frac{\partial}{\partial \bmu^{(b)}}\tL &= 
  -2\sum_{i=1}^n \bW^{(b)}_i\, \left( \bx^{(b)}_i - \bmu^{(b)}
  - \bV^{(b)} \bu_i \right)\,.
\end{align*}

\section{Description of the algorithm}
\label{app:algo}

In this section, we provide a detailed description of the IRLS algorithm in Section~\ref{subsec:algo}.
In order to simplify the 
problem, we introduce a new objective function
\begin{equation} \label{eq:WPCA}
	\sum_{i=1}^n\sum_{b=1}^B\sum_{j=1}^{p_b} 
	w_{ij}^{(b)}\left(x_{ij}^{(b)}
	- \mu_j^{(b)} - 
	\left(\bU\left(\bV^{(b)}\right)^T\right)_{ij}\right)^2
\end{equation}
where $\left(\bU\left(\bV^{(b)}\right)^T\right)_{ij}
= \bu_i^T\bv_j^{(b)} =
\sum_{\ell=1}^k u_{i\ell}v_{j\ell}^{(b)}$ and the 
weight matrices $\bW^{(b)}$ are assumed fixed for now.
The objective function \eqref{eq:WPCA} has the particularity of
leading to the same first order conditions of optimality as \eqref{eq:objP}.
The proof is available in Section \ref{app:firstorder} of the Supplementary 
Material. 
To achieve minimization of \eqref{eq:WPCA}, \cite{gabriel1978}
showed that an alternating least squares type of procedure can be followed. At each step,  the minimization of the objective function \eqref{eq:WPCA} is achieved
by iteratively updating each component 
by following the first-order optimality conditions
\eqref{eq:condi1}-\eqref{eq:condi3}. This stepwise procedure
is performed for fixed weights. Thereafter, 
the weights are updated as well.
The initial estimates $\bV_{(0)}$, $\bU_{(0)}$,  and
$\bmu_{(0)}$ are obtained from cellPCA
\citep{Centofanti2026cellPCA}. The initial weight matrices $\bW^{(b)}_{(0)}$ are derived from \eqref{eq:cellweight}, \eqref{eq:blockweight}, \eqref{eq:caseweight}. 
At each iteration, for $\ell=0,1,\ldots$, the
estimates are updated from
$\bV_{(\ell)}=\left[
\big(\bV^{(1)}_{(\ell)}\big)^T,\dots,
\big(\bV^{(B)}_{(\ell)}\big)^T
\right]^T$,  
$\bU_{(\ell)}$, and
$\bmu_{(\ell)}= \left(
\big(\bmu^{(1)}_{(\ell)}\big)^T,\dots,
\big(\bmu^{(B)}_{(\ell)}\big)^T
\right)^T $ to
$\bV_{(\ell+1)}$, $\bU_{(\ell+1)}$, and 
$\bmu_{(\ell+1)}$ through the following four-step
procedure.

\vspace{3mm}
{\bf (a) Minimize~\eqref{eq:WPCA} with 
respect to $\bV$} by applying \eqref{eq:condi1} 
with $\bU_{(\ell)}$\,, $\bmu^{(b)}_{(\ell)}$\,,
and the weight matrices $\bW^{(b)}_{(\ell)}$\,. 
The minimization can be done component by component
because, for fixed $b$ and $j$, we can isolate the term of the  
objective function~\eqref{eq:WPCA} associated
with the column vector $\bv_j^{(b)}$,
\begin{align}\label{eq:normWUV}
 \sum_{i=1}^n 
  \left(\bW^{(b)j}_{(\ell)}\right)_{ii}\left(x_{ij}^{(b)}
  - \mu^{(b)}_{j,(\ell)} - \hx_{ij}^{(b)}\right)^2
 &= \sum_{i=1}^n 
  \left(\bW^{(b)j}_{(\ell)}\right)_{ii}\left(x_{ij}^{(b)}
  - \mu^{(b)}_{j,(\ell)} - \left(\bU\left(\bV^{(b)}\right)^T\right)_{ij}\right)^2\\
 &= \big|\big|
  \big(\bW^{(b)j}_{(\ell)}\big)^{1/2}\big(\bx^{(b)j}-\mu^{(b)}_{j,(\ell)} \bone_n\big)
  - (\bW^{(b)j}_{(\ell)})^{1/2}\bU_{(\ell)} \bv^{(b)}_j
  \big|\big|^2. \notag
\end{align}
The minimization of this term with respect to $\bv_j^{(b)}$ is
equivalent to the least squares regression of 
$\big(\bW^{(b)j}_{(\ell)}\big)^{1/2}\big(\bx^{(b)j}-\mu^{(b)}_{j,(\ell)} \bone_n\big)$
on $(\bW^{(b)j}_{(\ell)})^{1/2}\bU_{(\ell)}$\,.
The solution is given by 
\begin{align}\label{eq:WUdagger}
\bv^{(b)}_j&=\Big(\big[(\bW^{(b)j}_{(\ell)})^{1/2}\bU_{(\ell)}\big]^T
  (\bW^{(b)j}_{(\ell)})^{1/2}\bU_{(\ell)}\Big)^{\!\dagger}\,
  \big[(\bW^{(b)j}_{(l)})^{1/2}\bU_{(\ell)}\big]^T
  \big(\bW^{(b)j}_{(\ell)}\big)^{1/2}\big(\bx^{(b)j}-\mu^{(b)}_{j,(\ell)} \bone_n\big)\nonumber \\
  &=\big(\bU_{(\ell)}^T\bW_{(\ell)}^{(b)j}\bU_{(\ell)}\big)^\dagger
  \bU_{(\ell)}^T\bW_{(\ell)}^{(b)j}
  \left(\bx^j-(\bmu_{(\ell)})_j\bone_n\right)
\end{align}
where $^\dagger$ denotes the Moore-Penrose
generalized inverse.
Repeating this for all $b = 1,\ldots,B$ and all $j=1,\ldots,p_b$ yields
the new matrices $\bV_{(\ell+1)}^{(b)}$.

\vspace{3mm}
\textbf{(b) Minimize~\eqref{eq:WPCA} with 
respect to $\bU$} by \eqref{eq:condi2} with 
the new $\bV_{(\ell+1)}^{(b)}$
and the old $\bmu^{(b)}_{(\ell)}$\,
and weight matrices $\bW^{(b)}_{(\ell)}$.
This is achieved in a similar way as in step (a). 
For a fixed $i$ the new $\bu_i$ 
must minimize its corresponding term of the objective function, that is,
\begin{equation} \label{eq:normWVU}
 \big|\big|
  \big(\bW_{i,(\ell)}\big)^{1/2}\big(\bx_i-\bmu_{(\ell)}\big)
  - (\bW_{i,(\ell)})^{1/2}\bV_{(\ell+1)} \bu_i
  \big|\big|^2 
\end{equation}
where the diagonal entries of the $p \times p$ diagonal matrix $\bW_i$
are equal to the $i$th row of $\left[\bW^{(1)}, \dots, \bW^{(B)}\right]$.
This is equivalent to minimize
\begin{equation} \label{eq:normWVUcellwise}
 \big|\big|
  \big(\btW_{i,(\ell)}\big)^{1/2}\big(\bx_i-\bmu_{(\ell)}\big)
  - (\btW_{i,(\ell)})^{1/2}\bV_{(\ell+1)} \bu_i
  \big|\big|^2 
\end{equation}
with $\btW_{i,(\ell)}$ a diagonal matrix whose elements are equal to the $i$th row of  $\left[\bW_{\text{cell}}^{(1)} \odot \bW_{\text{block}}^{(1)} \odot \bO^{(1)}, \dots, \bW_{\text{cell}}^{(B)} \odot \bW_{\text{block}}^{(B)} \odot \bO^{(B)}\right]$.
The reason behind this is that \eqref{eq:normWVU} is equal
to \eqref{eq:normWVUcellwise} multiplied by the casewise
weight of the observation $\bx_i$. If this weight is strictly positive,
then the formulations are equivalent. In the scenario of a zero
casewise weight, \eqref{eq:normWVU} achieves its lower bound
of zero anyway. The minimization of \eqref{eq:normWVUcellwise} can be seen
as the least squares regression of
$\big(\btW_{i,(\ell)}\big)^{1/2}\big(\bx_i-\bmu_{(\ell)}\big)$ on
$(\btW_{i,(\ell)})^{1/2}\bV_{(\ell+1)}$ whose solution is given by
\begin{align} \label{eq:WVdagger}
\bu_{i,(\ell+1)}&=
  \Big( \big[(\btW_{i,(\ell)})^{1/2}\bV_{(\ell+1)}\big]^T 
  (\btW_{i,(\ell)})^{1/2}\bV_{(\ell+1)} \Big)^{\!\dagger}\,
  \big[(\btW_{i,(\ell)})^{1/2}\bV_{(\ell+1)}\big]^T
  (\btW_{i,(\ell)})^{1/2}
  \left(\bx_i-\bmu_{(\ell)}\right)\nonumber \\
  &=\Big(\bV_{(\ell+1)}^T(\btW_{i,(\ell)})
  \bV_{(\ell+1)}\Big)^{\!\dagger}
  \bV_{(\ell+1)}^T(\btW_{i,(\ell)})
  \left(\bx_i-\bmu_{(\ell)}\right)\\
  &=\left(\sum_{b=1}^{B}
  (\bV^{(b)}_{(\ell+1)})^T\btW^{(b)}_{i,(\ell)}
  \bV^{(b)}_{(\ell+1)}\right)^{\dagger}\;
  \left(\sum_{b=1}^{B}(\bV^{(b)}_{(\ell+1)})^T\btW^{(b)}_{i,
  (\ell)}\big(\bx_i^{(b)}-\bmu_{(\ell)}^{(b)}\big)\right)\nonumber.
\end{align}

\vspace{3mm}
\textbf{(c) Minimize~\eqref{eq:WPCA} with 
respect to $\bmu$} with the new $\bV_{(\ell+1)}^{(b)}$
and $\bU_{(\ell+1)}$\,
and the old weight matrices $\bW^{(b)}_{(\ell)}$.
Identically to step (a), for each $b=1,\ldots,B$ and each $j=1,\ldots,p_b$,
the isolated term of the objective 
function~\eqref{eq:WPCA} that contains $\mu^{(b)}_j$ is 
\begin{multline} \label{eq:columndist}
  \big|\big|
  \big(\bW^{(b)j}_{(\ell)}\big)^{1/2}\left(\bx^{(b)j}-\mu^{(b)}_{j,(\ell)} \bone_n
  - \left(\bU_{(\ell+1)}\big(\bV_{(\ell+1)}^{(b)}\big)^T\right)^j\right)
  \big|\big|^2  \\
  =\sum_{i=1}^n \bW_{ij,(\ell)}^{(b)} \left(
  x_{ij}^{(b)}-\mu_{j,(\ell+1)}^{(b)} -\left(\bU_{(\ell+1)}\big(\bV_{(\ell+1)}^{(b)}\big)^T
  \right)_{ij}\right)^2
\end{multline}
where $\bx^{(b)j}$ is the $j$th column of $\bX^{(b)}$,
which is minimized by the weighted mean 
\begin{equation}\label{eq:tmu}
 \mu_{j,(\ell+1)}^{(b)} =  \left(\sum_{i=1}^n 
		\bW^{(b)}_{ij,(\ell)}\right)^{-1}
		\sum_{i=1}^n \bW^{(b)}_{ij,(\ell)}
		\left(x_{ij}^{(b)} -\left(\bU_{(\ell+1)}\big(\bV_{(\ell+1)}^{(b)}\big)^T
        \right)_{ij}\right)\;.
\end{equation}

\vspace{3mm}
\textbf{(d) Update $\bW_{(\ell)}^{(b)}$} for all $b$ according 
to \eqref{eq:updateW}, \eqref{eq:cellweight},
\eqref{eq:blockweight},
and \eqref{eq:caseweight} with the new components
$\bV_{(\ell+1)}^{(b)}$,
$\bU_{(\ell+1)}$ and $\bmu_{(\ell+1)}^{(b)}$.

We run this iterative process until the following convergence criterion is met
\[ \dfrac{\dfrac{1}{np}\left|\left|\sum_{b=1}^B\bU_{(\ell+1)}
\big( \bV^{(b)}_{(\ell+1)}\big)^T
- \sum_{b=1}^B\bU_{(\ell)} \big( \bV^{(b)}_{(\ell)}\big)^T\right|\right|_F^2}{\dfrac{1}
{np}\left|\left|\sum_{b=1}^B\bU_{(\ell)} \big( \bV^{(b)}_{(\ell)}\big)^T\right|\right|_F^2} 
\leqslant 10^{-12}.\]

The quality of the approximation depends on the conditioning of the matrices whose generalized inverses are computed. To ensure well-conditioned matrices, we add the
requirement that each column of  $\bW^{(b)}$ should have
at most $50\%$ (by default) of values equal to $0$. This is equivalent to requiring
that every variable contains a maximum amount of contaminated data, either by outliers
or missing values. This ensures that the weight matrices $\bW^{(b)j}$ and $\btW_i^{(b)}$
are not filled with too many zeros. In the case where this condition should be broken,
we return the estimates from the previous step and advise investigating the 
variable(s) concerned.

\section{Proof of proposition \ref{prop:mono}}
\label{app:proofmono}
In this section, we prove the monotonicity property of the algorithm stated
by Proposition~\ref{prop:mono}. The proof relies on the following three
lemmas.

\begin{lemma}
    For fixed weight matrices $\bW^{(b)}$, the steps
    (a), (b), and (c) of the algorithm described in Section \ref{app:algo} of the
    Supplementary Material decrease the objective function \eqref{eq:WPCA}.
\end{lemma}
\begin{proof}
By construction, each step successively finds new estimates that minimize
the weighted objective function for fixed weights.
\end{proof}
For the second lemma, similarly to \cite{Centofanti2026cellPCA}, we first introduce a
new notation for a potential fit by $\btheta=\bhX=[\bhX^{(1)},\dots,\bhX^{(B)}]$, with $\bhX^{(b)}=\bU(\bV^{(b)})^T+\bone_n(\bmu^{(b)})^T$.
Then, we denote as $\bof(\btheta)$ the matrix $(\bX-\btheta)\odot(\bX-\btheta)$ transformed
into a $np$-dimensional vector. The objective function \eqref{eq:objP} can therefore
be rewritten as $L(\bof(\btheta)) := L_{\rho_1,\rho_2,\rho_3}(\bX,\bV,\bU,\bmu)$.

\begin{lemma}
    The function $\bof \rightarrow L(\bof)$ is concave.
\end{lemma}

\begin{proof}
    From Lemma 2 of Section F from the Supplementary Material of \cite{Centofanti2026cellPCA},
    we know that the functions $h_1: \mathbb R_{+} \rightarrow \mathbb R_{+}:z \mapsto 
 \rho_{1}(\sqrt{z})$, $h_2: \mathbb R_{+} \rightarrow \mathbb R_{+}:
 z \mapsto \rho_{2}(\sqrt{z})$, and $h_3: \mathbb R_{+} \rightarrow \mathbb R_{+}:
 z \mapsto \rho_{3}(\sqrt{z})$ are concave, when $\rho_1$, $\rho_2$ and $\rho_3$ are selected as the hyperbolic tangent function $\rho_{b,c}$.

    By the definition of concavity of a multivariate function,
    we need to show that for any column vectors $\bof, \bg \in \mathbb{R}_+^{np}$ and
    any $\lambda \in (0,1)$ we have $L(\lambda \bof + (1-\lambda)\bg) \geqslant
    \lambda L(\bof) + (1-\lambda)L(\bg)$.
        \begin{align*}
            L(\lambda \bof + (1-\lambda)\bg) = & \;  \frac{\hsigma_3^2}{o}\sum_{i=1}^{n}
		o_i h_3\!\left(\frac{1}{o_i \hsigma_{3}^2}
			\sum_{b=1}^{B} o^{(b)}_{i} 
			\big(\hsigma_{2}^{(b)}\big)^2 h_2\! \left(\frac{1}{o^{(b)}_{i}\big(\hsigma_{2}^{(b)}\big)^2}
				\sum_{j=1}^{p_b} o^{(b)}_{ij}\, 
				\big(\hsigma_{1,j}^{(b)}\big)^2\,\right. \right.\\
            & \; \quad \quad \quad \quad \left. \left.
				h_1\!\left(\frac{\lambda f_{ij}^{(b)}+(1-\lambda)g^{(b)}_{ij}}{\big(\hsigma_{1,j}^{(b)}\big)^2}\right) \right) \right) \\
                \geqslant & \; \frac{\hsigma_3^2}{o} \sum_{i=1}^{n} o_i h_3 
                \left(\frac{1}{o_i \hsigma_{3}^2} \sum_{b=1}^{B} o^{(b)}_{i} \big(\hsigma_{2}^{(b)}\big)^2 h_2\left(\frac{1}{o^{(b)}_{i}\big(\hsigma_{2}^{(b)}\big)^2} 
                 \sum_{j=1}^{p_b} o^{(b)}_{ij} \big(\hsigma_{1,j}^{(b)}\big)^2  \right.\right.\\
                 & \; \quad \quad \quad \quad\left. \left. \left[\lambda h_1(f_{ij}^{(b)}/\big(\hsigma_{1,j}^{(b)}\big)^2)
                 + (1-\lambda)h_1(g^{(b)}_{ij}/\big(\hsigma_{1,j}^{(b)}\big)^2)
                 \right]\right)\right)\\
                 = & \; \frac{\hsigma_3^2}{o} \sum_{i=1}^{n} o_i h_3 
                \left(\frac{1}{o_i \hsigma_{3}^2} \sum_{b=1}^{B} o^{(b)}_{i} \big(\hsigma_{2}^{(b)}\big)^2\right.\\
                & \; \quad \quad \quad \quad h_2\left(\lambda \frac{1}{o^{(b)}_{i}\big(\hsigma_{2}^{(b)}\big)^2} 
                 \sum_{j=1}^{p_b} o^{(b)}_{ij} \big(\hsigma_{1,j}^{(b)}\big)^2 
                 h_1(f_{ij}^{(b)}/\big(\hsigma_{1,j}^{(b)}\big)^2)\right.\\
                 & \; \quad \quad \quad \quad \left. +(1-\lambda)\frac{1}{o^{(b)}_{i}\big(\hsigma_{2}^{(b)}\big)^2}
                 \sum_{j=1}^{p_b} o^{(b)}_{ij} \big(\hsigma_{1,j}^{(b)}\big)^2 
                 h_1(g^{(b)}_{ij}/\big(\hsigma_{1,j}^{(b)}\big)^2 \right)\\
                 \geqslant & \; \frac{\hsigma_3^2}{o} \sum_{i=1}^{n} o_i  
                 h_3 \left(\frac{1}{o_i \hsigma_{3}^2} \sum_{b=1}^{B} o^{(b)}_{i} 
                 \big(\hsigma_{2}^{(b)}\big)^2 \right.\\
                 & \; \quad \quad \quad \quad \left[ \lambda h_2\left( \frac{1}{o^{(b)}_{i}\big(\hsigma_{2}^{(b)}\big)^2} 
                 \sum_{j=1}^{p_b} o^{(b)}_{ij} \big(\hsigma_{1,j}^{(b)}\big)^2 
                 h_1(f_{ij}^{(b)}/\big(\hsigma_{1,j}^{(b)}\big)^2)\right) \right.\\
                 & \;\quad \quad \quad \quad \left. \left. +(1-\lambda) h_2 \left( \frac{1}{o^{(b)}_{i}\big(\hsigma_{2}^{(b)}\big)^2}
                 \sum_{j=1}^{p_b} o^{(b)}_{ij} \big(\hsigma_{1,j}^{(b)}\big)^2 
                 h_1(g^{(b)}_{ij}/\big(\hsigma_{1,j}^{(b)}\big)^2\right) \right]\right)\\
                 = & \; \frac{\hsigma_3^2}{o} \sum_{i=1}^{n} o_i h_3 
                \left(\lambda \frac{1}{o_i \hsigma_{3}^2} \sum_{b=1}^{B} o^{(b)}_{i} \big(\hsigma_{2}^{(b)}\big)^2 h_2\left( \frac{1}{o^{(b)}_{i}\big(\hsigma_{2}^{(b)}\big)^2} 
                 \sum_{j=1}^{p_b} o^{(b)}_{ij} \big(\hsigma_{1,j}^{(b)}\big)^2 
                 \right .\right.\\
                 & \; \quad \quad \quad \quad \left. h_1(f_{ij}^{(b)}/\big(\hsigma_{1,j}^{(b)}\big)^2)\right)
                 +(1-\lambda) \frac{1}{o_i \hsigma_{3}^2} \sum_{b=1}^{B} o^{(b)}_{i} \big(\hsigma_{2}^{(b)}\big)^2\\
                 & \; \quad \quad \quad \quad \left. h_2\left( \frac{1}{o^{(b)}_{i}\big(\hsigma_{2}^{(b)}\big)^2} 
                 \sum_{j=1}^{p_b} o^{(b)}_{ij} \big(\hsigma_{1,j}^{(b)}\big)^2 
                 h_1(g_{ij}^{(b)}/\big(\hsigma_{1,j}^{(b)}\big)^2)\right) \right)\\
                 \geqslant & \; \lambda \, \frac{\hsigma_3^2}{o} \sum_{i=1}^{n} o_i
                 h_3 \left(\frac{1}{o_i \hsigma_{3}^2} \sum_{b=1}^{B}
                 o^{(b)}_{i} \big(\hsigma_{2}^{(b)}\big)^2 h_2\left( \frac{1}
                 {o^{(b)}_{i}\big(\hsigma_{2}^{(b)}\big)^2} 
                 \sum_{j=1}^{p_b} o^{(b)}_{ij} \big(\hsigma_{1,j}^{(b)}\big)^2 
                 \right .\right.\\
                 & \; \quad \quad \quad \quad \left. h_1(f_{ij}^{(b)}/\big(\hsigma_{1,j}^{(b)}\big)^2)\right)
                 +(1-\lambda) \frac{\hsigma_3^2}{o} \sum_{i=1}^{n}o_i 
                 h_3 \left(\frac{1}{o_i \hsigma_{3}^2} \sum_{b=1}^{B}
                 o^{(b)}_{i} \big(\hsigma_{2}^{(b)}\big)^2 \right.\\
                 & \; \quad \quad \quad \quad \left. h_2\left( \frac{1}{o^{(b)}_{i}\big(\hsigma_{2}^{(b)}\big)^2} 
                 \sum_{j=1}^{p_b} o^{(b)}_{ij} \big(\hsigma_{1,j}^{(b)}\big)^2 
                 h_1(g_{ij}^{(b)}/\big(\hsigma_{1,j}^{(b)}\big)^2)\right) \right)\\
                 = & \; \lambda L(\bof) + ( 1 -\lambda)L(\bg).
        \end{align*}
    The first inequality derives from the concavity of $h_1$
    and the fact that $h_2$ and $h_3$ are nondecreasing. The second comes from the
    concavity of $h_2$ and the fact that $h_2$ is nondecreasing. Finally, the third inequality 
    is due to the concavity of $h_3$. The concavity of $L$ is verified.
\end{proof}
For the third lemma, similarly to Lemma 2, we write the weighted
objective function~\eqref{eq:WPCA} as a function of $\bof$. The new 
notation will be $L_{\bW}(\bof):=\big(\text{vec}(\bW)\big)^T \bof$,
where $\text{vec}(\bW)$ denotes
the vectorized form of the weight matrix $\bW$ in a column vector of size $np$.
The last step is to demonstrate the link between the original objective function $L$
and the weighted PCA objective $L_{\bW}$.

\begin{lemma}
    If two column vectors $\bof, \bg \in \mathbb{R}_+^{np}$ verify $L_{\bW}(\bg) \leqslant 
    L_{\bW}(\bg)$, then $L(\bof) \leqslant L(\bg)$.
\end{lemma}
\begin{proof}
    Lemma 2 states the concavity of $L(\bof)$. Moreover, because $h_1$, $h_2$ and
    $h_3$ are differentiable, $L(\bof)$ is also differentiable. Therefore, it verifies the following
    inequality
    \[L(\bof) \leqslant L(\bg) + \big(\nabla L (\bg)\big)^T (\bof - \bg),\]
    where $\nabla L (\bg)$ is the gradient of $L$ taken in $\bg$.
    As shown in Section \ref{app:firstorder} of the Supplementary Material, 
    $\nabla L(\bg)$ is proportional to $\nabla L_{\bW}(\bg)$, and the latter is equal
    to $\text{vec}(\bW)$ by construction. Therefore, the quantity $\big(\nabla L (\bg)\big)^T (\bof - \bg)$ is proportional to 
    $\big(\text{vec}(\bW)\big)^T (\bof - \bg)=L_{\bW}(\bof)-L_{\bW}(\bg)$ 
    which is nonpositive from our hypothesis. This yields $L(\bof) \leqslant L(\bg)$.
\end{proof}

Proposition \ref{prop:mono} follows directly from the lemmas.
\begin{proof}[Proof of Proposition 1]
    When we go from $\btheta_{(\ell)}$ to $\btheta_{(\ell+1)}$, 
Lemma~1 ensures that\linebreak
$L_{\bW_{(\ell)}}(\bof(\btheta_{(\ell+1)})) \leqslant 
L_{\bW_{(\ell)}}(\bof(\btheta_{(\ell)}))$. Applying   
Lemma~3 then gives
$L(\bof(\btheta_{(\ell+1)})) \leqslant 
L(\bof(\btheta_{(\ell)}))$.
\end{proof}

\section{Imputation and prediction with bloccPCA} \label{sec:pracext}
\subsection{Imputation} \label{subsec:imputation}
When one or more cells of an observation $\bx_i$ receive weights below 1, the
method considers those cells as potentially contaminated. In this situation, it
may be useful to construct a cleaned version of the observation, in which
suspicious cells are shrunk towards their fitted values and missing cells are
filled in, while clean cells are left unchanged. Thus, we aim to obtain an
imputed point $\bimpx_i$, with block components $\bx_i^{(b)\text{imp}}$, whose cells
coincide with the original data when they are considered clean and observed,
and are modified otherwise.

Let $\bhx_i^{(b)}=\bhV^{(b)}\bhu_i+\bhmu^{(b)}$ denote the fitted value of
observation $i$ in block $b$, and let $\bhx_i=\bhV\bhu_i+\bhmu$ be the concatenation of these
fitted block values. 
We construct the imputed block observation as
\begin{equation}\label{eq:imp}
	\bx_i^{(b)\text{imp}}
	:=
	\bhx_i^{(b)}
	+
	\btW_i^{(b)}
	\left(\bx_i^{(b)}-\bhx_i^{(b)}\right),
	\qquad b=1,\ldots,B .
\end{equation}
with $\btW_i^{(b)}$ a diagonal matrix whose elements are equal to the $i$th row of $\bW_{\text{cell}}^{(b)} \odot \bW_{\text{block}}^{(b)} \odot \bO^{(b)}$.
The fully imputed observation is obtained by concatenating the imputed blocks,
that is, $\bimpx_i=\left((\bx_i^{(1)\text{imp}})^T,\ldots,
(\bx_i^{(B)\text{imp}})^T\right)^T$. From the following lemma, its orthogonal projection onto
the estimated subspace is equal to $\bhx_i$.

\begin{lemma}\label{lem:orth_imp}
Let
$\btW_i$ be the $p\times p$ diagonal matrix whose diagonal is obtained by
concatenating the diagonals of the block matrices $\btW_i^{(b)}$,
$b=1,\ldots,B$. 
Then,
$\left(\btW_i(\bx_i-\bhx_i)\right)^T\bhV=\bzero$.
\end{lemma}

\begin{proof}
Since
$\bhx_i=\left((\bhx_i^{(1)})^T,\ldots,(\bhx_i^{(B)})^T\right)^T$,
\begin{align*}
\left(\btW_i(\bx_i-\bhx_i)\right)^T\bhV
&=
\sum_{b=1}^{B}
\left(\btW_i^{(b)}
\left(\bx_i^{(b)}-\bhx_i^{(b)}\right)\right)^T
\bhV^{(b)}  \\
&=
\sum_{b=1}^{B}
\left(\btW_i^{(b)}
\left(\bx_i^{(b)}-\bhmu^{(b)}-\bhV^{(b)}\bhu_i\right)\right)^T
\bhV^{(b)} \\
&=\sum_{b=1}^{B}
\left(\bhV^{(b)}\right)^T
\btW_i^{(b)}
\left(\bx_i^{(b)}-\bhmu^{(b)}-\bhV^{(b)}\bhu_i\right)
=
\bzero
\end{align*}
where the last equality follows from \eqref{eq:condi2}.
\end{proof}

From \eqref{eq:imp}, each imputed cell is given by
\begin{equation*} 
x_{ij}^{(b)\text{imp}}
=
\hx_{ij}^{(b)}
+
w_{ij}^{(b)\mathrm{cell}}
w_{i}^{(b)\mathrm{block}}
o_{ij}^{(b)}
\left(x_{ij}^{(b)}-\hx_{ij}^{(b)}\right).
\end{equation*}
If $w_{ij}^{(b)\mathrm{cell}}w_i^{(b)\mathrm{block}}o_{ij}^{(b)}=1$, the cell
is kept unchanged and $x_{ij}^{(b)\text{imp}}=x_{ij}^{(b)}$. If this product is equal
to zero, the imputed cell coincides with the fitted value
$\hx_{ij}^{(b)}$. For intermediate values, the imputed cell lies between the
original and fitted value, thereby moving suspicious cells closer to the
estimated subspace.

\subsection{Out-of-sample prediction}
\label{sec:newx}
The application of bloccPCA can be seen as a training stage, where the robust
center and loading matrices are estimated from the available data. After this
stage, a new out-of-sample observation $\bx^*$ may become available, and one may
want to compute its fitted value $\bhx^*=\bhV\bhu^*+\bhmu$ without refitting the model.
In classical PCA, this is done by orthogonally projecting $\bx^*$ onto the
estimated principal subspace. However, this projection is not robust and may be
strongly affected if the new observation contains outlying cells, outlying
blocks, or missing values. Since the center and loading matrices are now fixed,
the problem reduces to estimating the score vector of the new observation.

For this purpose, we minimize the inner part of the objective function
\eqref{eq:objP} for the single observation $\bx^*$, omitting the summation over
the training observations. More precisely, the superscore $\bu^*$ is obtained by
minimizing, with respect to $\bu$,
\begin{equation*}
    \sum_{b=1}^{B} o^{*(b)}_{i} 
			\left(\hsigma_{2}^{(b)}\right)^2\rho_2\! 
			\left(\frac{1}{\hsigma_{2}^{(b)}}\sqrt{\frac{1}
				{o^{*(b)}_{i}}
				\sum_{j=1}^ {p_b} o^{*(b)}_{ij}\, 
				\left(\hsigma_{1,j}^{(b)}\right)^2\,
				\rho_1\!\left(\frac{x^{*(b)}_{ij}-\hmu_j^{(b)}-\bu^T \bhv_j^{(b)}}
				{\hsigma_{1,j}^{(b)}}\right)}\, \right)
\end{equation*}
where $o^{*(b)}_{ij}$ is equal to $1$ if $x^{*(b)}_{ij}$ is attributed and $0$ otherwise,
and $o^{*(b)}_{i} = \sum_{j=1}^{p_b} o^{*(b)}_{ij}$.
Differentiating this objective function yields the same first-order condition as
\eqref{eq:condi2}, with the superloadings and center kept fixed.
However, the condition remains nonlinear because the cellwise and blockwise
weights depend on the superscore. We therefore use the same IRLS strategy as in
the training stage, alternating between updating the superscore through
\eqref{eq:updateUblock} and updating the weights through
\eqref{eq:cellweight} and \eqref{eq:blockweight}. As in the training stage, the
initial value of $\bu$ is obtained from the prediction procedure of cellPCA
\citep{Centofanti2026cellPCA}.
Moreover, blockscores can be estimated as in Section~\ref{subsec:block}.

\clearpage
\section{Additional simulation results}
\label{supp:additional_simulation}
\begin{figure} [h]
	\centering
	\begin{tabular}{ccc}
		\textbf{Cellwise}  & \textbf{Blockwise} & \textbf{Casewise}\\[-12pt]
        
		\includegraphics[width=0.28\textwidth]{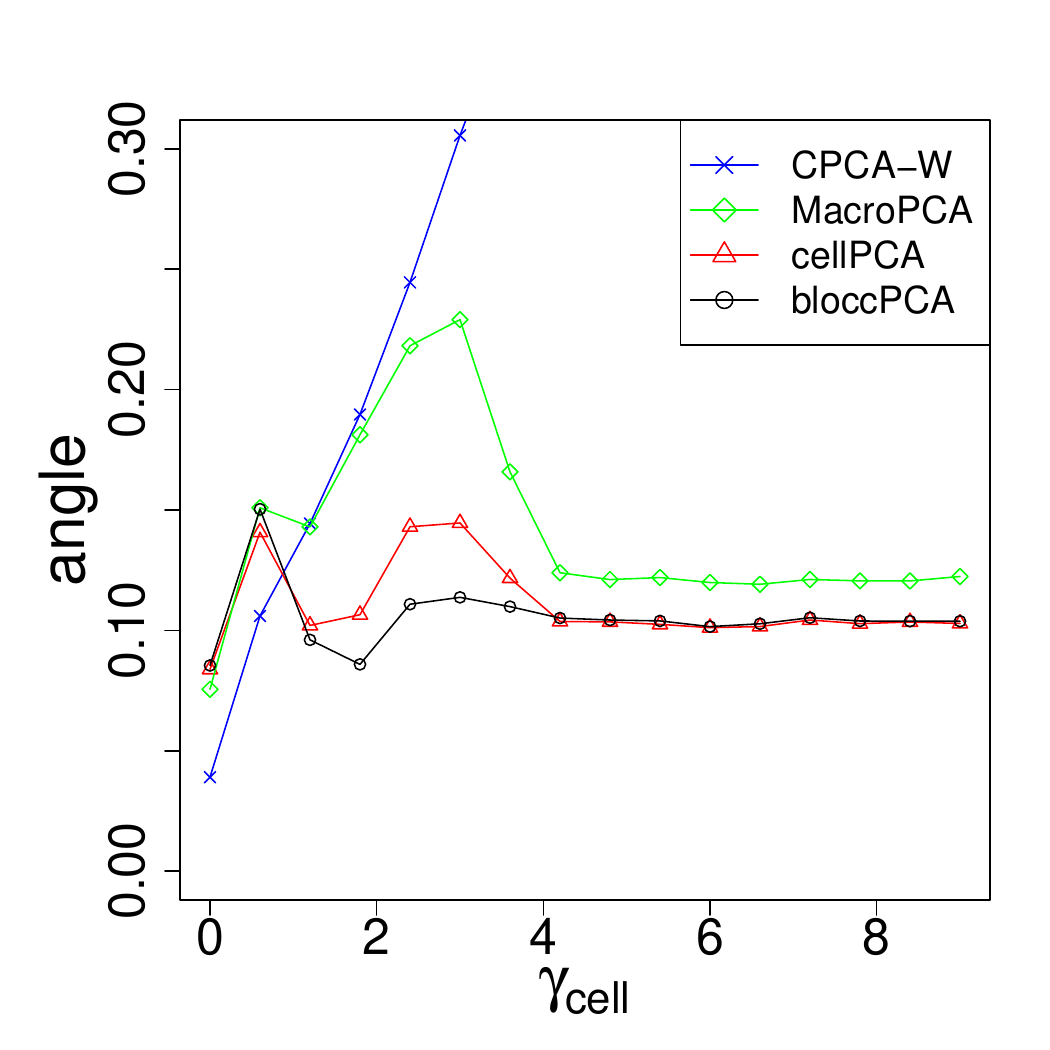} &
		\includegraphics[width=0.28\textwidth]{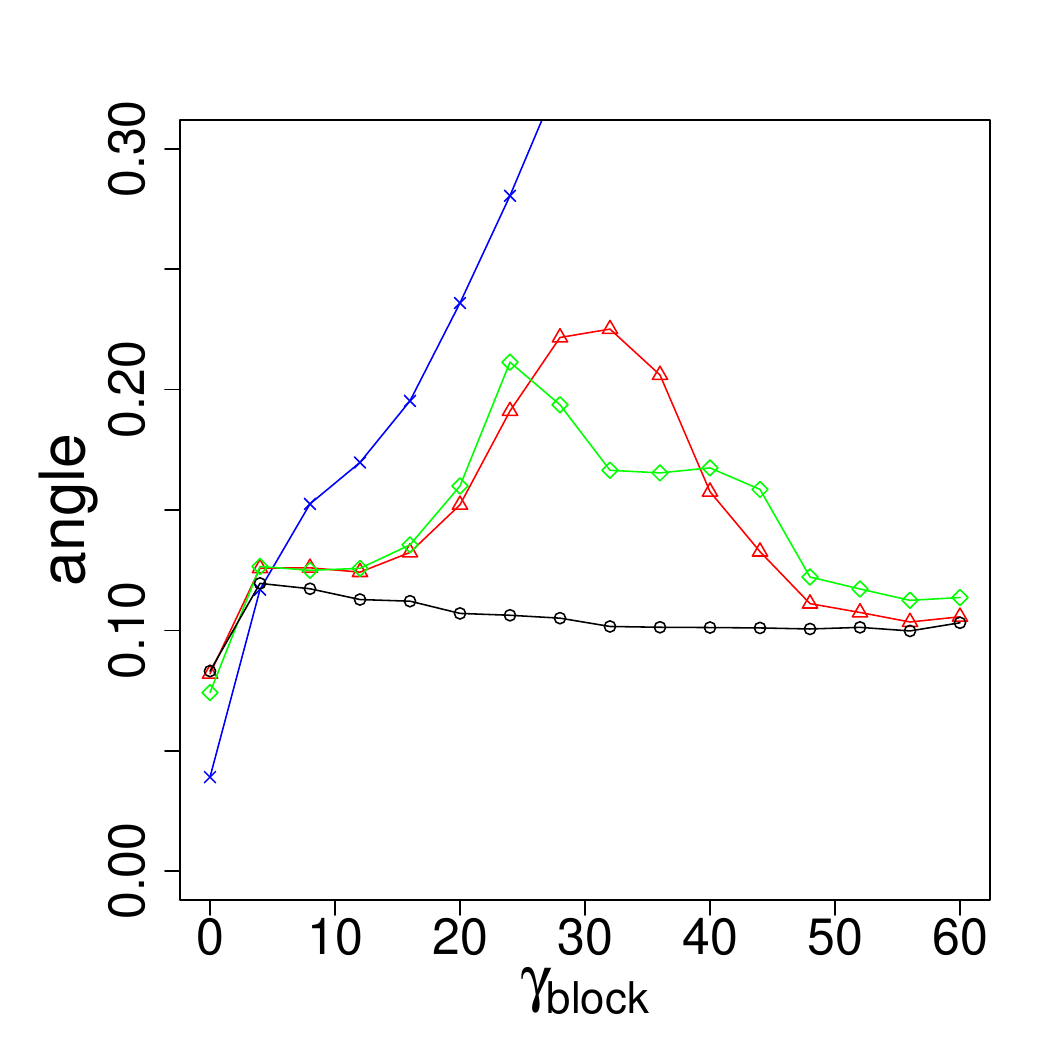} & 
        \includegraphics[width=0.28\textwidth]{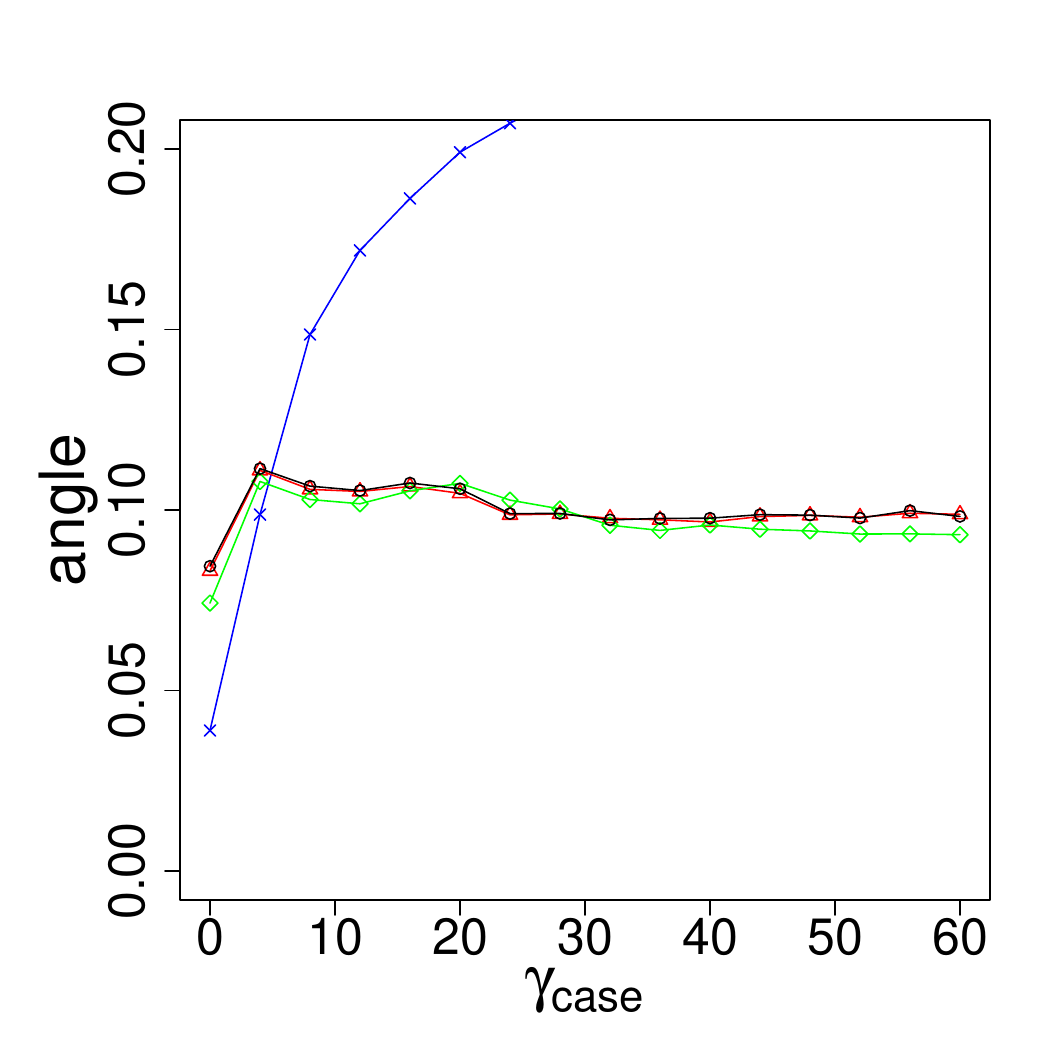}\\[6pt]
        
		\textbf{Cellwise-blockwise} &
        \textbf{Blockwise-casewise} & 
		\textbf{Cellwise-blockwise-casewise}\\[-12pt]

		\includegraphics[width=0.28\textwidth]{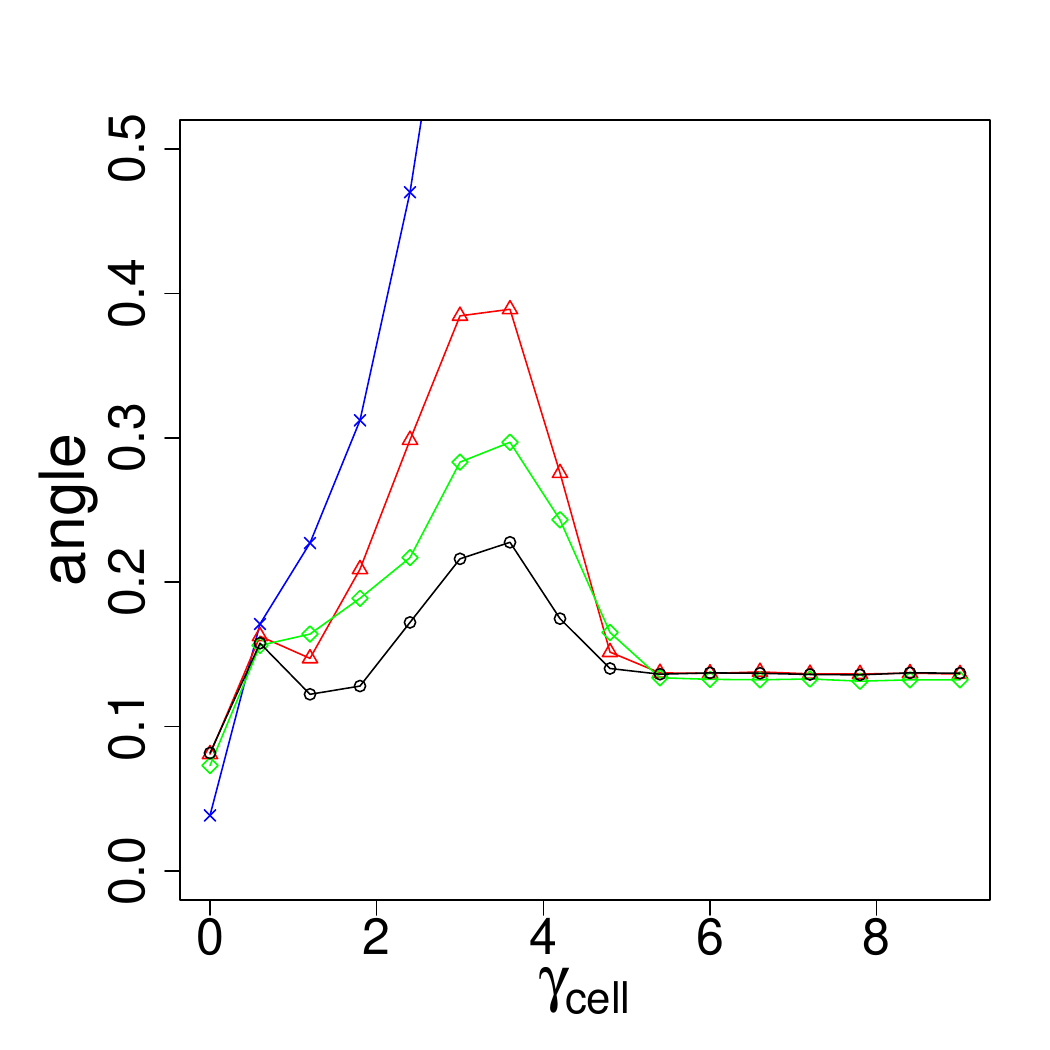}  &
		  \includegraphics[width=0.28\textwidth]{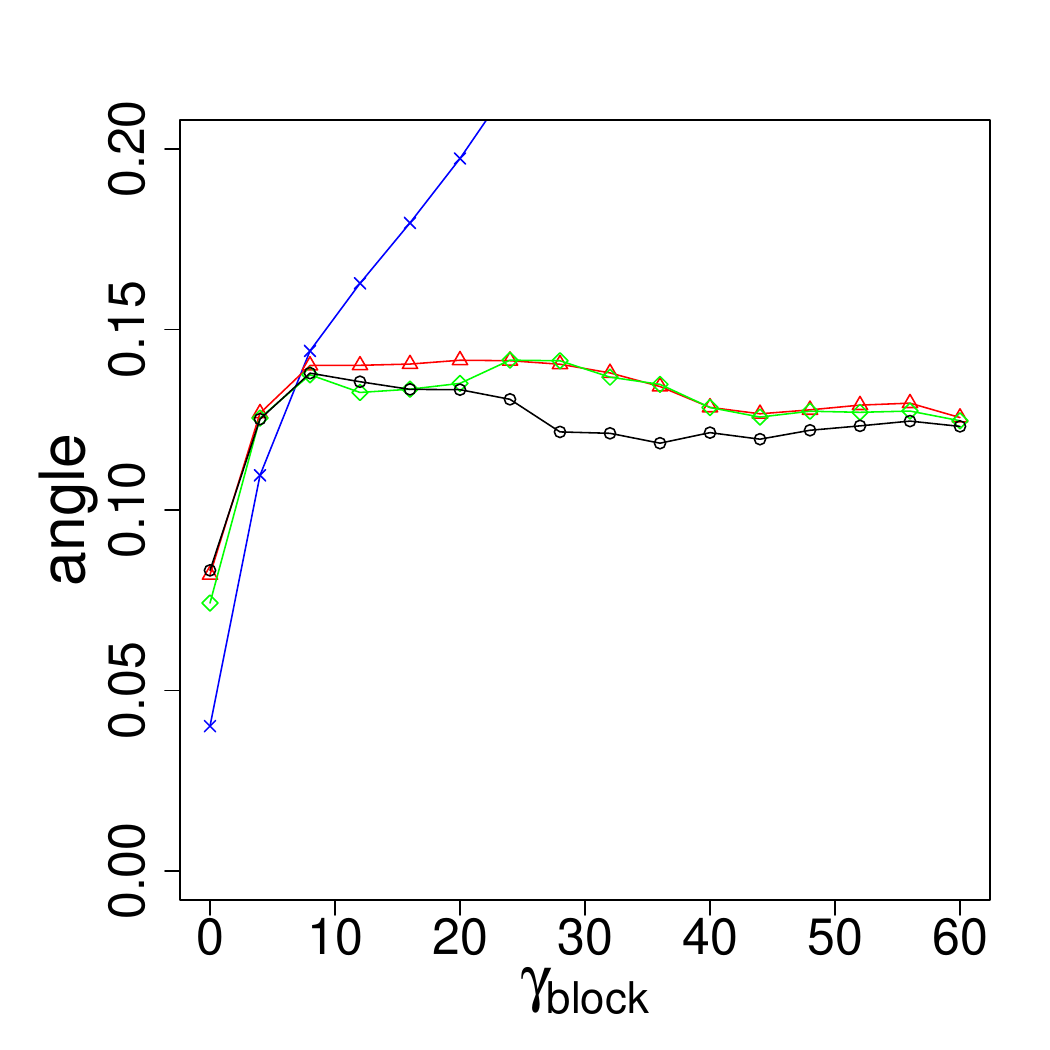} &
		\includegraphics[width=0.28\textwidth]{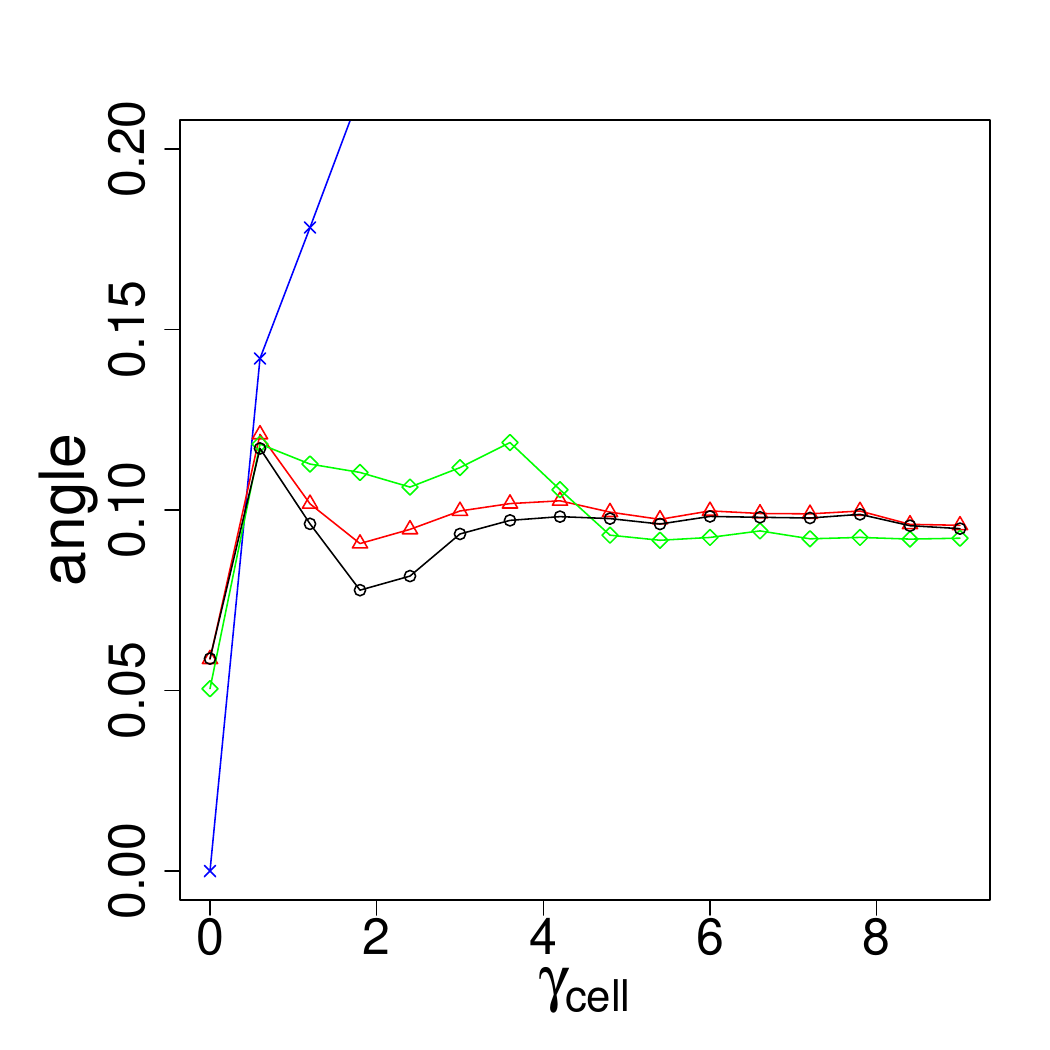}\\
	\end{tabular}
	\caption{Median angle attained by CPCA-W, MacroPCA, cellPCA, and bloccPCA under the different contamination scenarios for $n=100$, {$p=60$, $B=4$} with missing data.}
	\label{fig:sim_40_NA}
\end{figure}

\begin{figure}[!ht]
	\centering
	\begin{tabular}{ccc}
		\textbf{Cellwise}  & \textbf{Blockwise} & \textbf{Casewise}\\[-12pt]
        
		\includegraphics[width=0.28\textwidth]{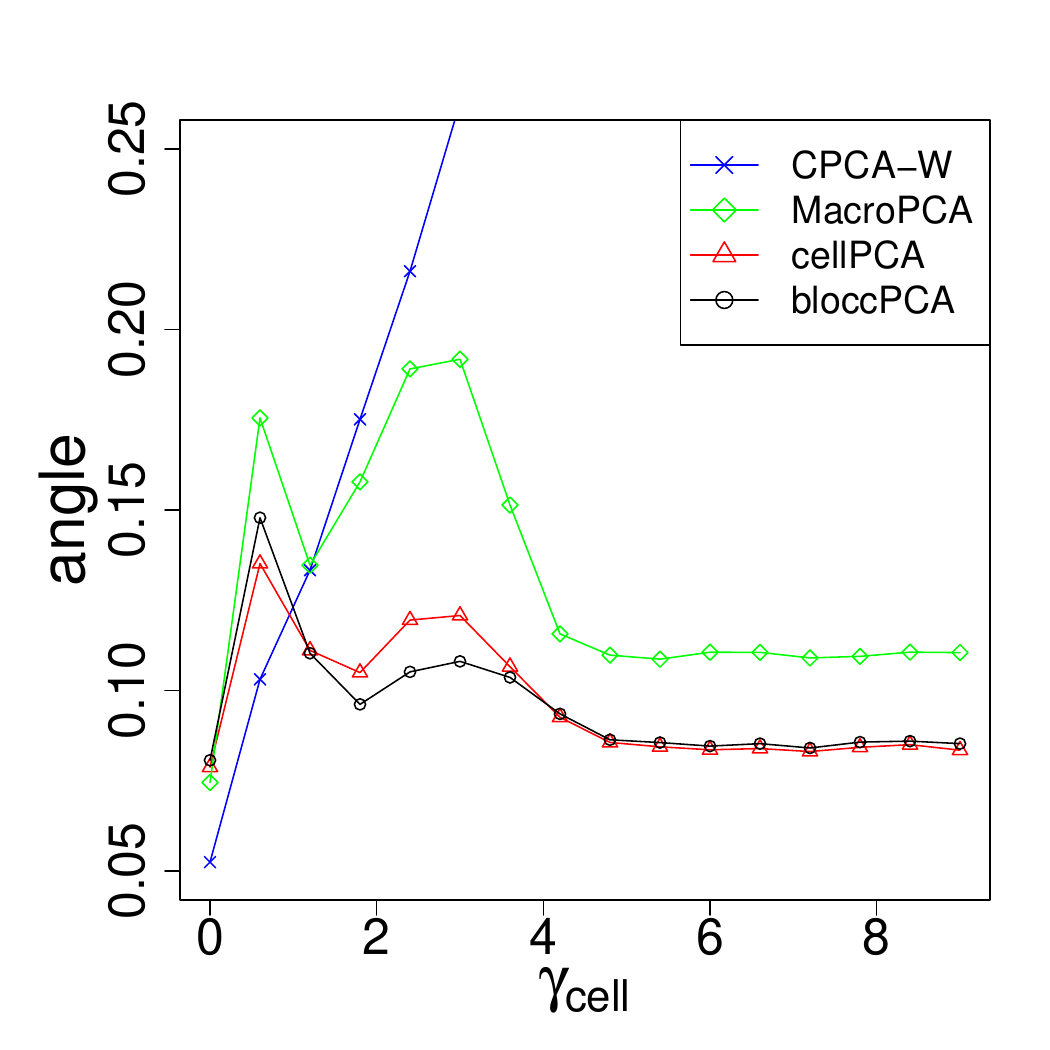} &
		\includegraphics[width=0.28\textwidth]{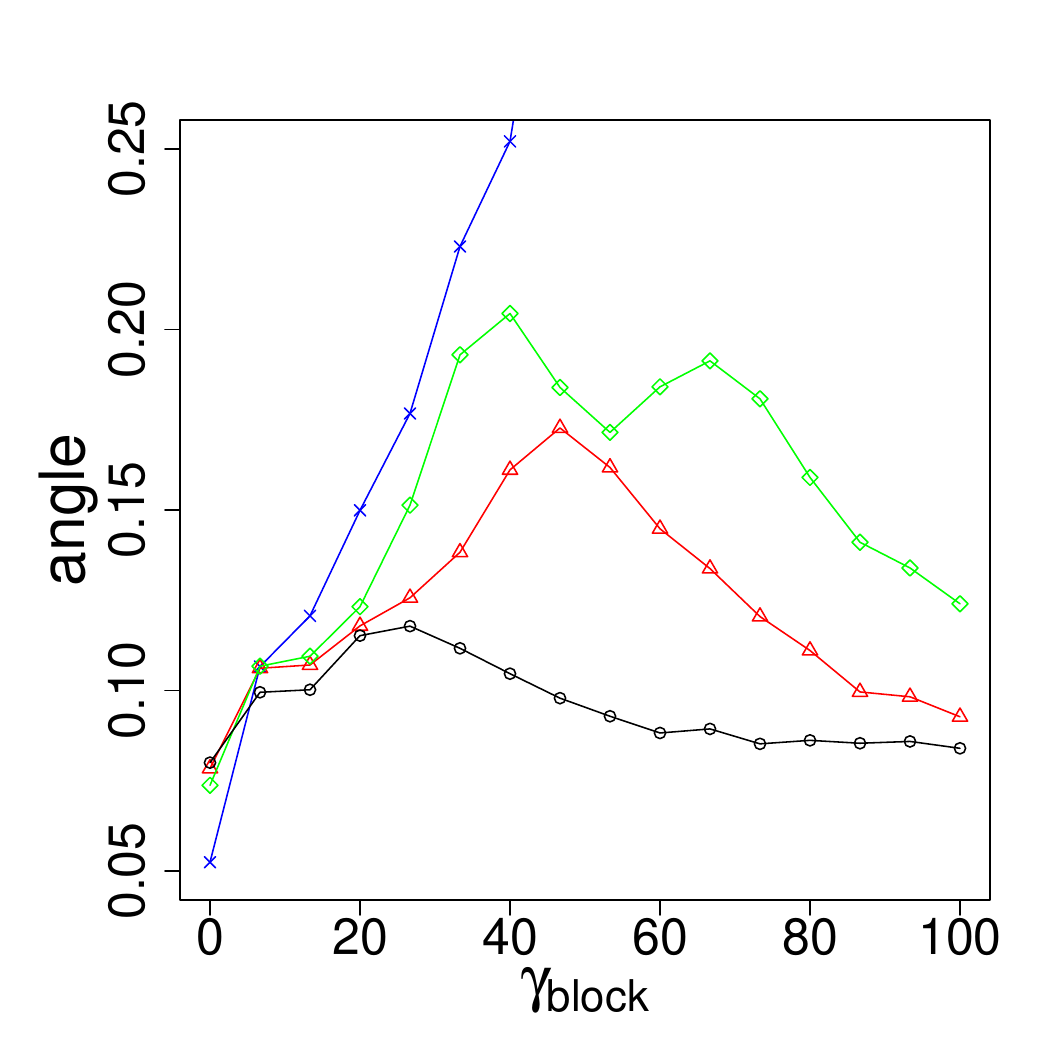} & 
        \includegraphics[width=0.28\textwidth]{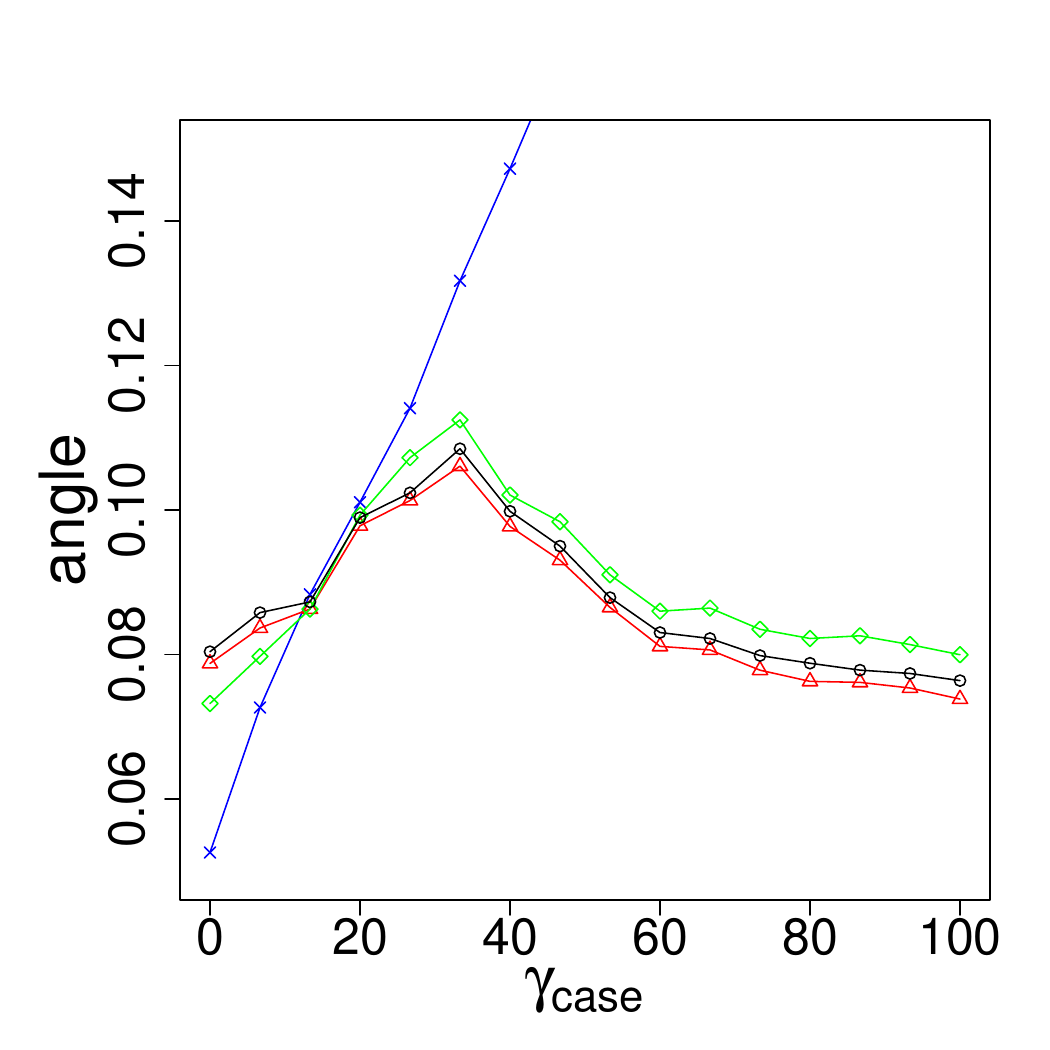}\\[6pt]
        
		\textbf{Cellwise-blockwise} &
        \textbf{Blockwise-casewise} & 
		\textbf{Cellwise-blockwise-casewise}\\[-12pt]

		\includegraphics[width=0.28\textwidth]{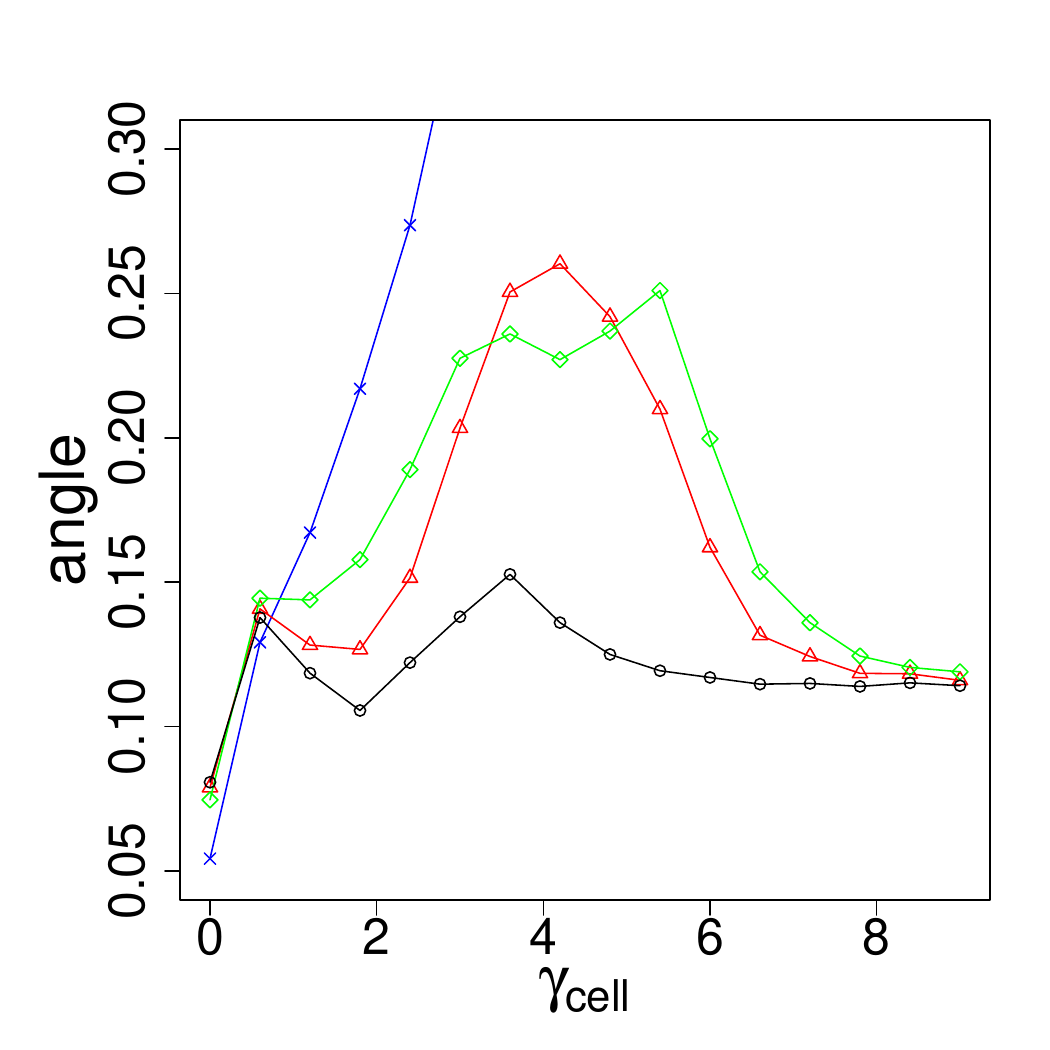}  &
		  \includegraphics[width=0.28\textwidth]{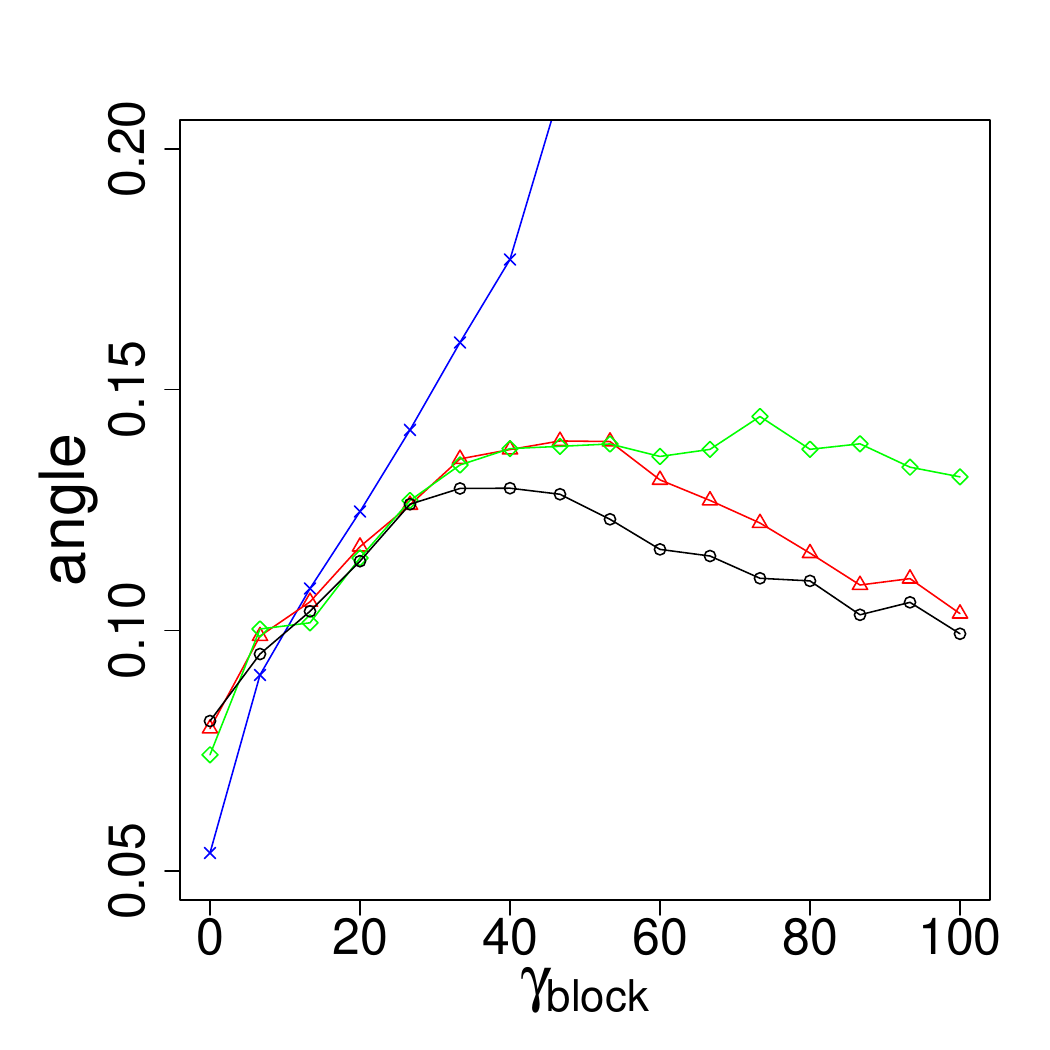} &
		\includegraphics[width=0.28\textwidth]{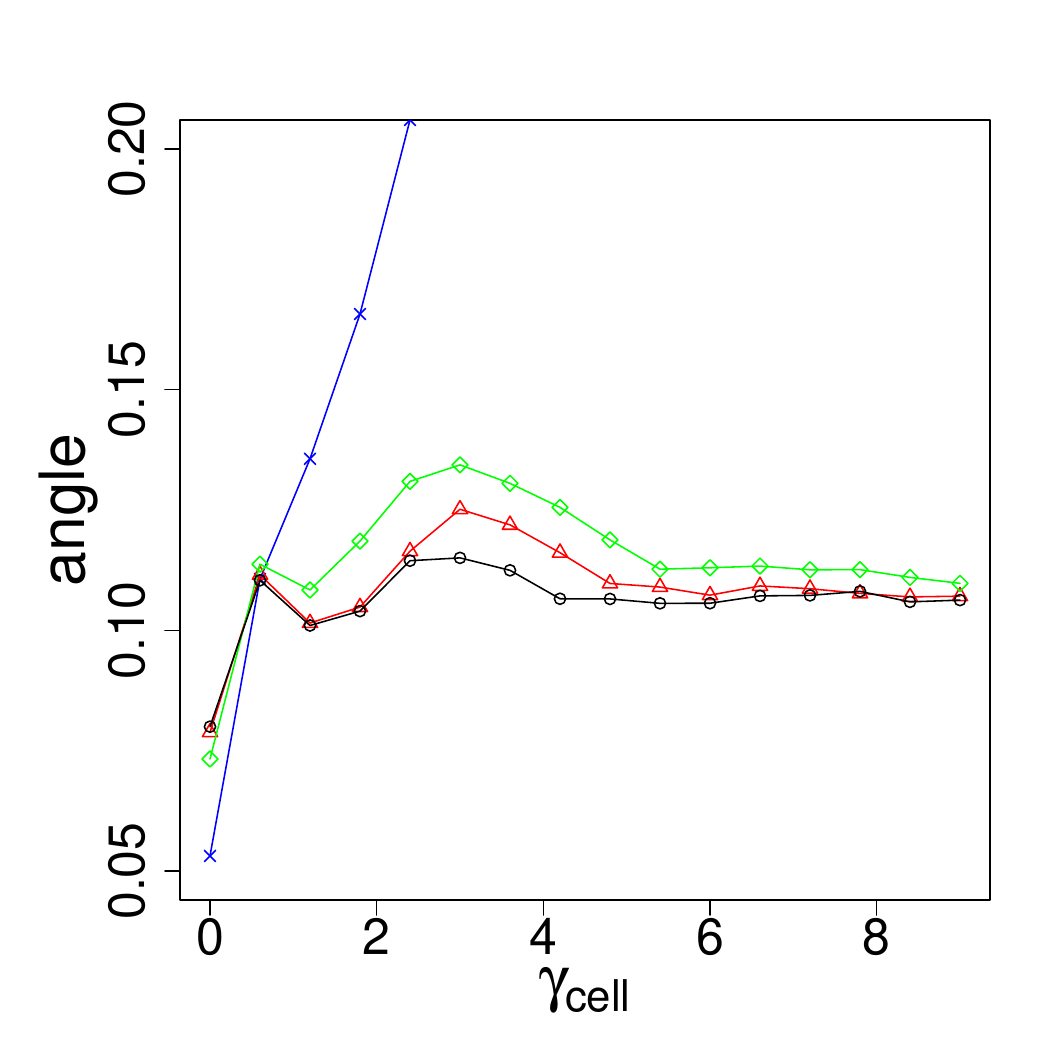}\\
	\end{tabular}
	\caption{Median angle attained by CPCA-W, MacroPCA, cellPCA, and bloccPCA under the different contamination scenarios for $n=100$, $p=200$, {$B=5$} without missing data.}
	\label{fig:sim_200_noNA}
\end{figure}

\begin{figure}[!ht]
	\centering
	\begin{tabular}{ccc}
		\textbf{Cellwise}  & \textbf{Blockwise} & \textbf{Casewise}\\[-12pt]
        
		\includegraphics[width=0.28\textwidth]{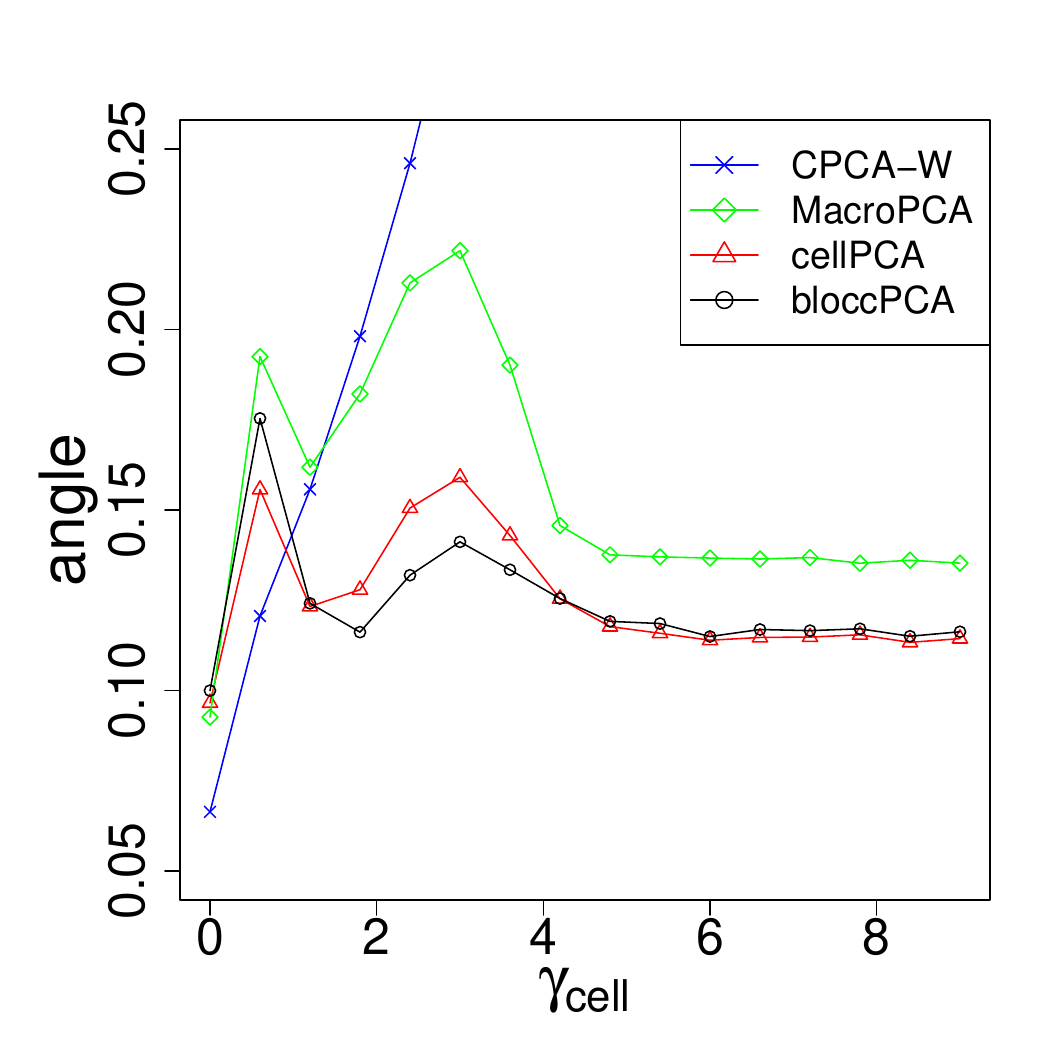} &
		\includegraphics[width=0.28\textwidth]{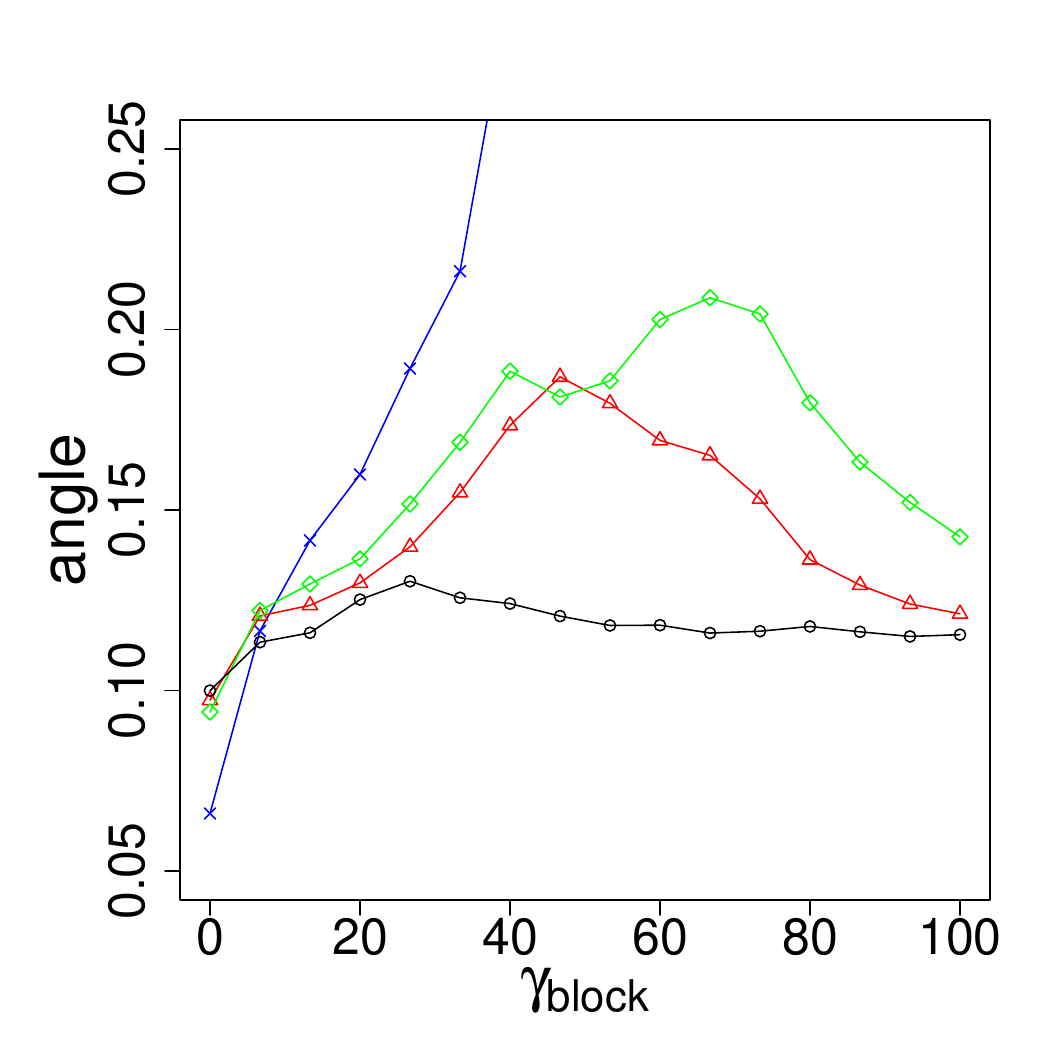} & 
        \includegraphics[width=0.28\textwidth]{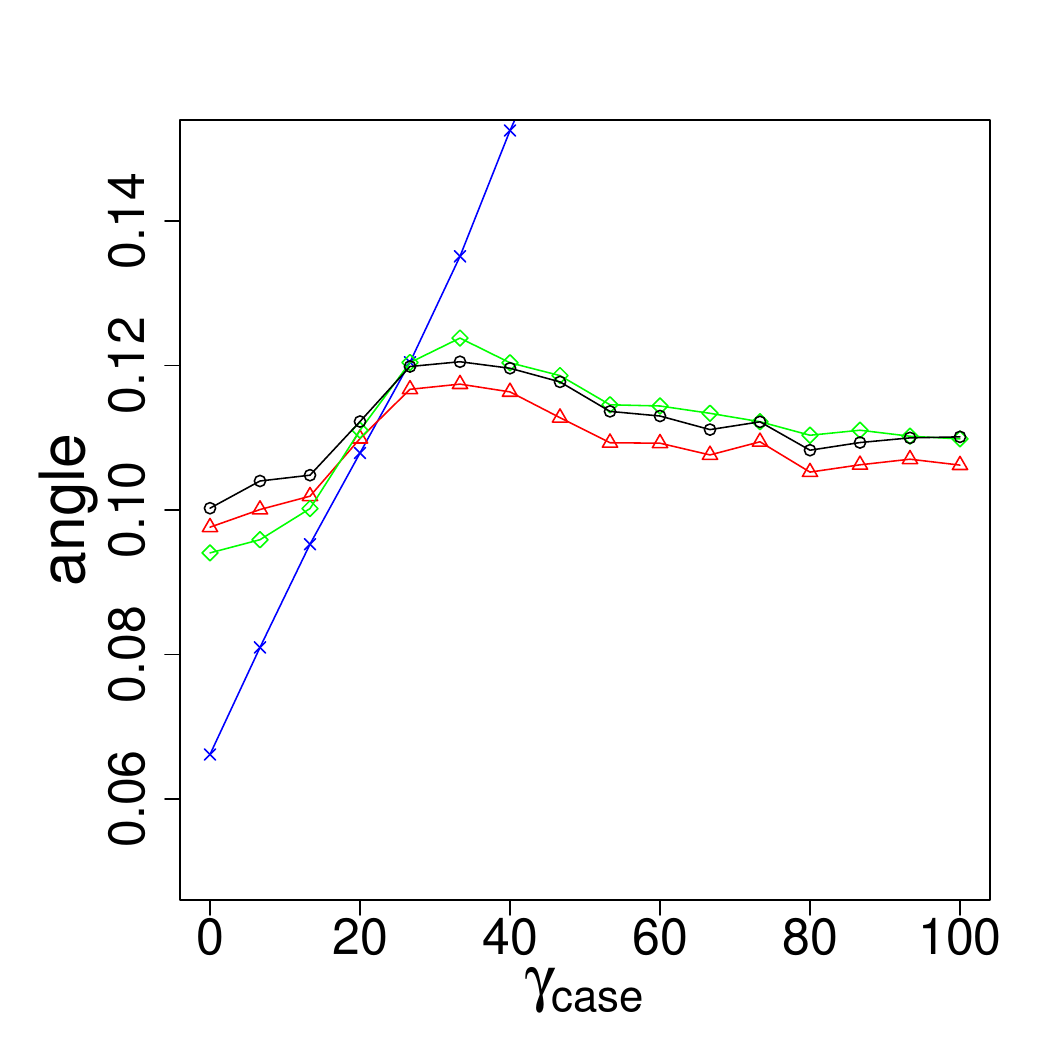}\\[6pt]
        
		\textbf{Cellwise-blockwise} &
        \textbf{Blockwise-casewise} & 
		\textbf{Cellwise-blockwise-casewise}\\[-12pt]

		\includegraphics[width=0.28\textwidth]{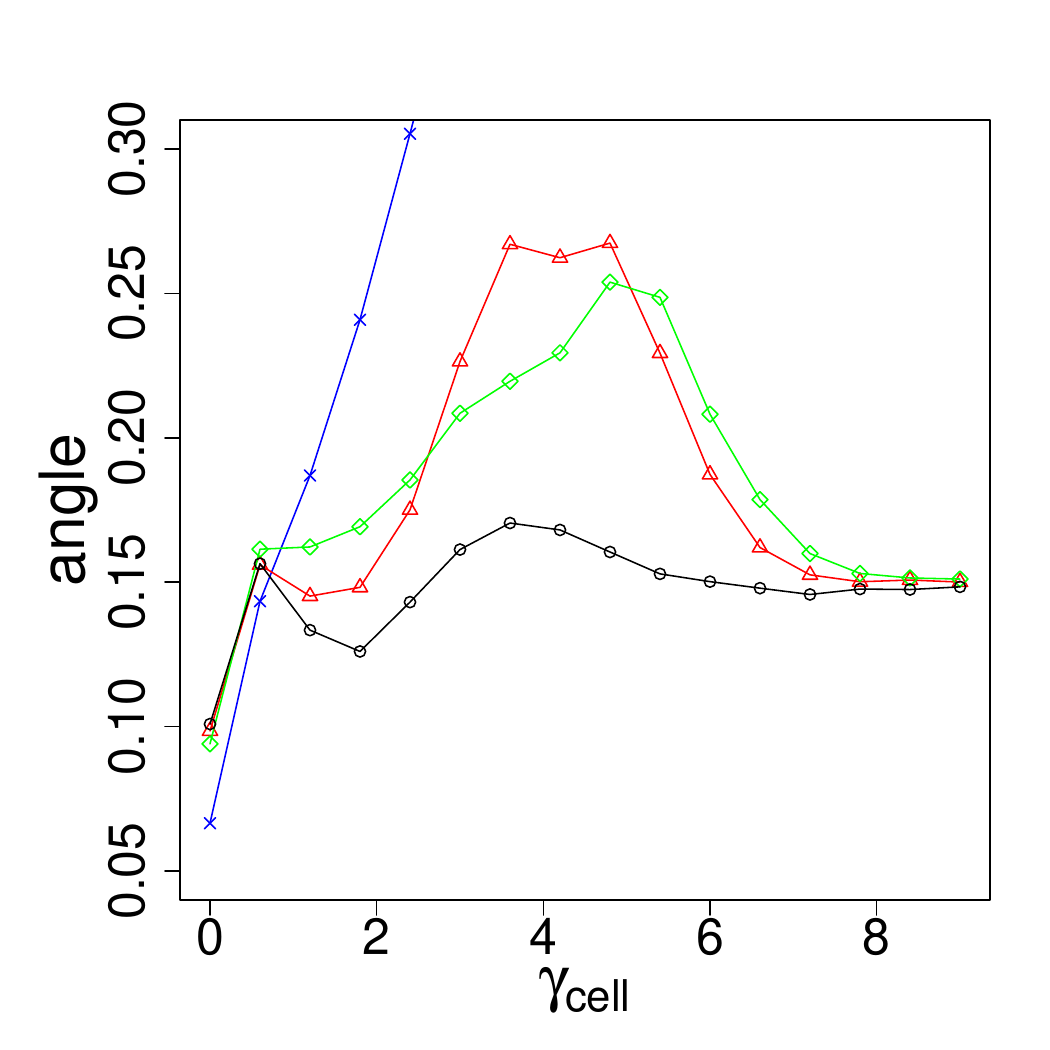}  &
		  \includegraphics[width=0.28\textwidth]{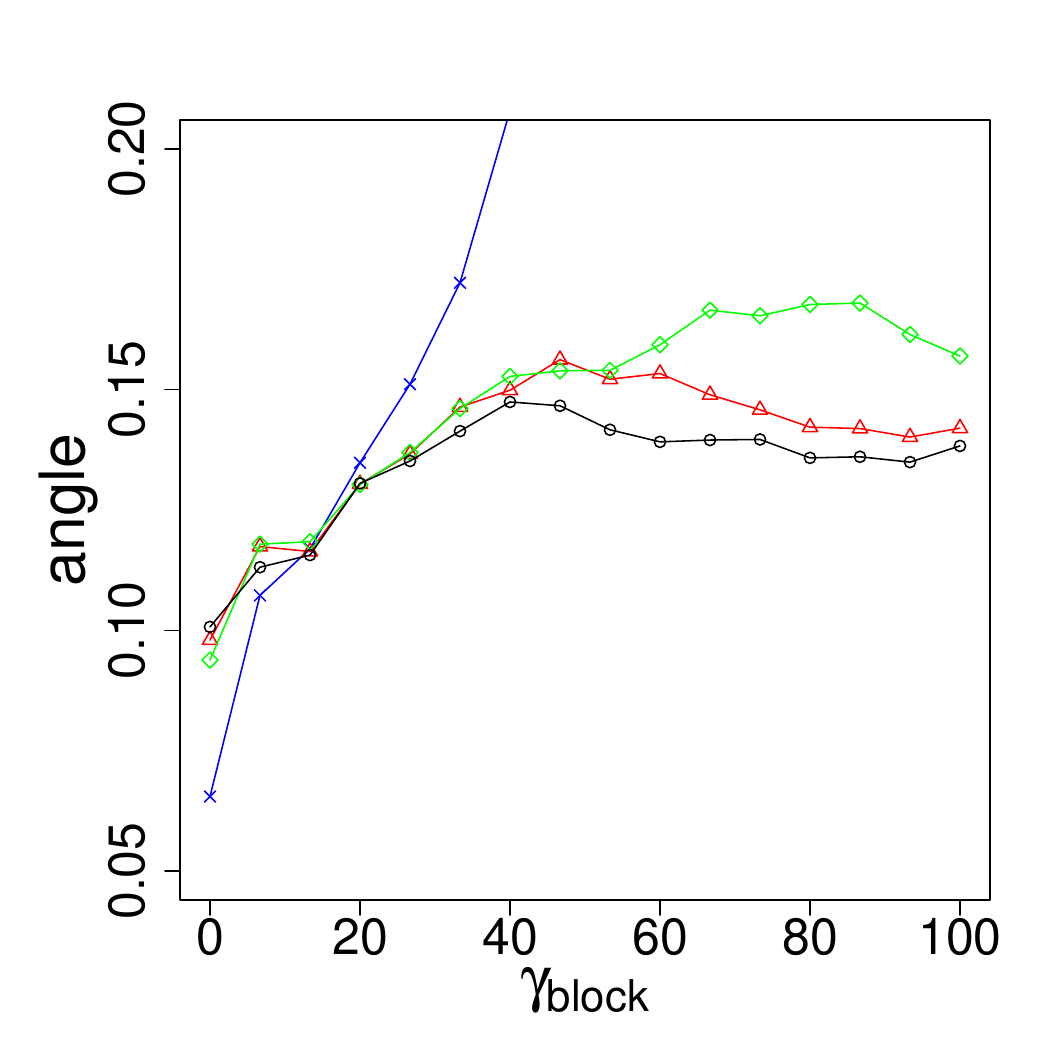} &
		\includegraphics[width=0.28\textwidth]{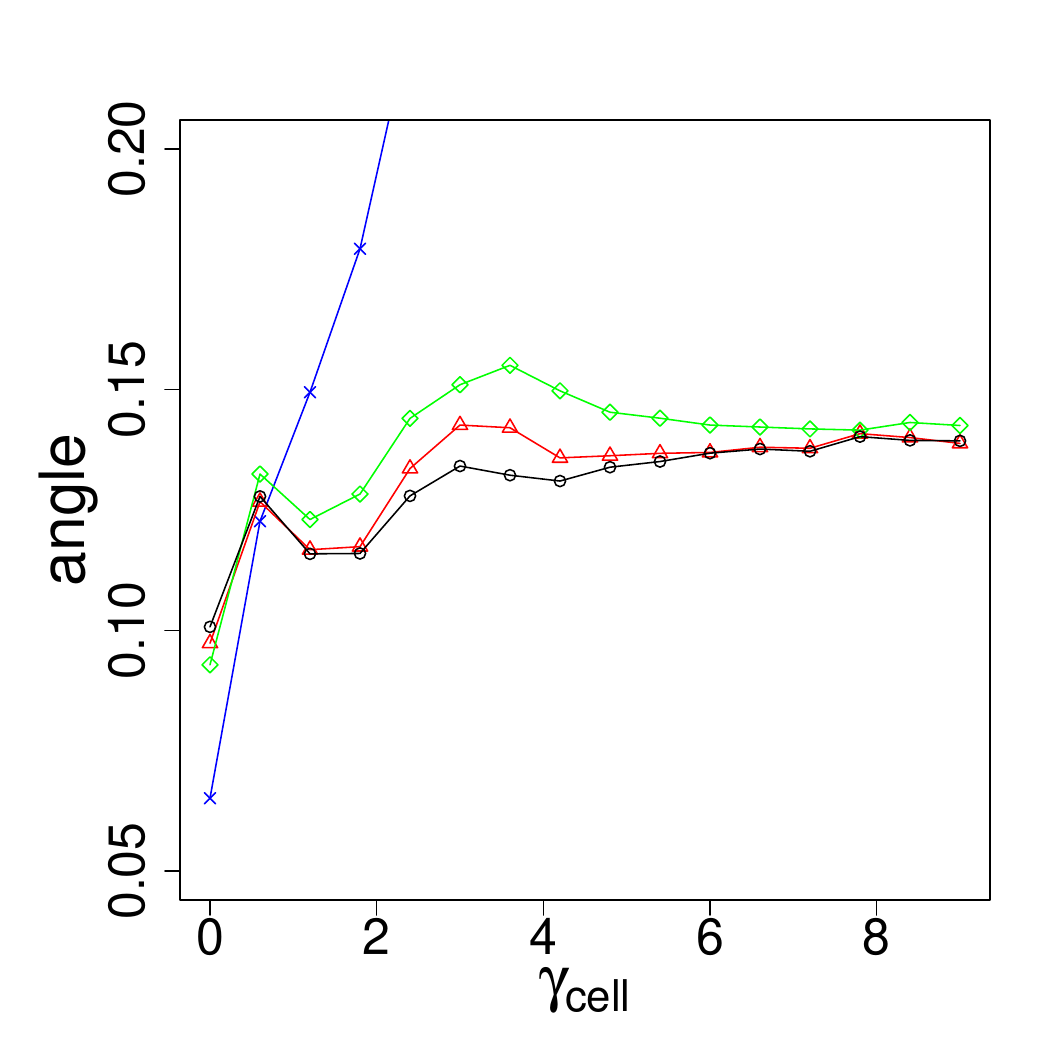}\\
	\end{tabular}
	\caption{Median angle attained by CPCA-W, MacroPCA, cellPCA, and bloccPCA under the different contamination scenarios for $n=100$, $p=200$, {$B=5$} with missing data.}
	\label{fig:sim_200_NA}
\end{figure}

Figure~\ref{fig:sim_40_NA} shows the median angle over 200 simulated datasets
for the setting with $n=100$, {$p=60$, $B=4$}, and 20\% missing values. Figures~\ref{fig:sim_200_noNA}
and~\ref{fig:sim_200_NA} report the corresponding results for the setting with
$n=100$, $p=200$, and {$B=5$}, without and with missing values, respectively.
The results are consistent with the conclusions of
Section~\ref{sec:simulation}. \\

\clearpage
\section{Additional empirical results}\label{supp:add_results}
\begin{figure}[!ht]
        \includegraphics[width=\linewidth]{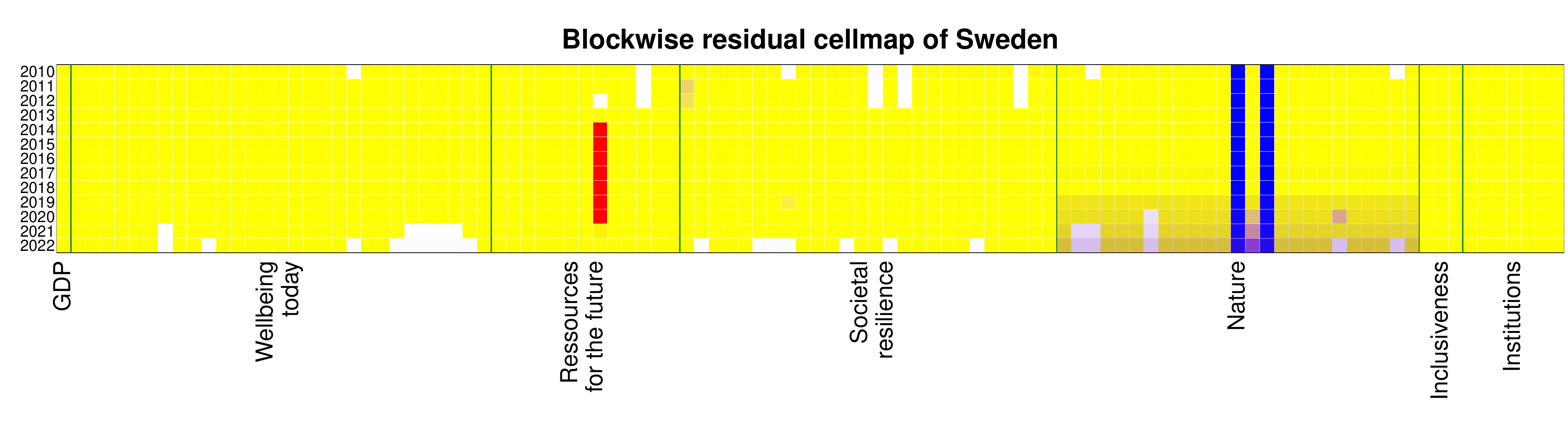}
        \caption{Blockwise residual cellmap of Sweden.}
        \label{fig:cellmap_se}
\end{figure}

This section presents additional results from the outlier
analysis of the Sustainable Wellbeing dataset, that is, 
the blockwise residual cellmap for Sweden, shown in Figure~\ref{fig:cellmap_se}.
At the casewise level, Sweden shows no outlying observations,
the profile aligns closely with the fitted multiblock structure.

At the block level, the ``Nature'' block stands out. Its blockwise outlyingness
starts in 2019 and continues onward to reach its peak in 2022.
Two variables mainly drive
the blockwise outlyingness with large negative cellwise residuals,
as displayed in dark blue in the cellmap. These two variables
are ``GHG emissions from land use, land use change and forestry''
and ``Net greenhouse gas emissions'', and are cellwise outliers
since the first year of observation. The block becomes outlying
due to the increasing outlyingness from the other variables within the block.
Similar to the cellmap of Portugal in Figure \ref{fig:cellmap_pt},
the co-occurring outlyingness of
these two variables is due to their similarity in definition.

At the cell level, in addition to the two variables in the block ``Nature'',
another variable from the block ``Resources for the future'' shows an abnormally large
positive residual between the years 2014 and 2020. Indeed, Sweden scores higher than
expected for the indicator ``Healthy life years at age 65". This suggests that the
health outcomes for older people in Sweden are considerably better
than is explained by the country's position in the
overall wellbeing structure.

\clearpage
\section{Wellbeing indicators}
\label{app:block_and_ind}
The table below reports the 104 indicators used in the analysis of
Section~\ref{sec:realdata}, together with their thematic block, a brief
description, and the sign of their relationship with wellbeing. A positive sign
indicates that larger values of the indicator are associated with higher
wellbeing, whereas a negative sign indicates that larger values are associated
with lower wellbeing.

\scriptsize
    
    \begin{longtable}{|p{4cm}|>{\centering\arraybackslash}p{1cm}|p{10cm}|}
        \hline
        \multicolumn{1}{|c|}{\rule[-1ex]{0pt}{5ex}\large\textbf{Indicator}}& \multicolumn{1}{|c|}{\rule[-1ex]{0pt}{5ex}\large\textbf{Sign}} & \multicolumn{1}{|c|}{\rule[-1ex]{0pt}{5ex}\large\textbf{Description}}\\
     \hline
\multicolumn{3}{|c|}{\rule[-1ex]{0pt}{5ex}\large\textbf{GDP}}\\
\hline
        GDP per capita & + & Gross domestic product per capita.\\
         \hline
\multicolumn{3}{|c|}{\rule[-1ex]{0pt}{5ex}\large\textbf{Wellbeing today}}\\
        \hline
        Inability to make ends meet & - & Share of the population who have difficulty or great difficulties to make ends meet.\\
        \hline
        Real gross disposable income of households per capita & + & Gross disposable income of households and Non-Profit Institutions Serving Households (NPISH) adjusted by Harmonized Index of Consumer Prices (HICP), by the total resident population.\\
        \hline
        AROPE for children, 0-17, \% & - & Share of children who are at risk of poverty, and/or severely materially and socially deprived and/or lives in a household with very low work intensity.\\
        \hline
        At risk of poverty or social exclusion rate (AROPE) & - & Share of people who are at risk of poverty, and/or severely materially and socially deprived and/or lives in a household with very low work intensity.\\
        \hline
        Inability to keep home adequately warm & - & Share of the population that cannot afford to keep the home adequately warm.\\
        \hline
        Housing cost overburden rate & - & Share of population living in households that spend 40\% or more of the household disposable income on housing (‘net' of housing allowances). Housing costs include rental or mortgage interest payments and cost of utilities (water, electricity, gas or heating).\\
        \hline
        Severe housing deprivation rate & - & Percentage of the population living in a dwelling which is considered as overcrowded, while also exhibiting at least one of the following housing deprivation measures: i) a leaking roof, ii) no bath/shower and no indoor toilet, iii) considered too dark.\\
        \hline
        In-work at-risk-of-poverty rate & - & Share of persons who are employed and have an equivalised disposable income below the risk-of-poverty threshold, which is set at 60\% of the national median equivalised disposable income (after social transfers).\\
        \hline
        Long working hours in main job & - & Persons in employment aged 15 - 64 years who usually work 49 hours or more per week  in their main job.\\
        \hline
        People killed in accidents at work & - & Number of fatal accidents that occur during the course of work and lead to the death of the victim within one year of the accident per 100 000 persons in employment.\\
        \hline
        Size of the shadow economy & - & Size of the Shadow Economy (all economic activities that are hidden from official authorities for monetary, regulatory, and institutional reasons) expressed as a percetage of GDP.\\
        \hline
        Employment rate & + & Share of the population aged 20 to 64 which is employed.\\
        \hline
        Involuntary temporary employment & - & Share of employees who could not find permanent job as a percentage of total employees.\\
        \hline
        Long-term unemployment rate & - & Share of the population in the labour force aged 15 to 74 who has been unemployed for 12 months or more.\\
        \hline
        Unemployment rate & - & Number of people unemployed (age 15-74) as a percentage of the labour force.\\
        \hline
        Inactive population due to caring responsibilities by sex & - & ‘Care responsibility’ covers care responsibilities over own children or spouse's children living inside or outside the household and over adult ill/elderly/incapacitated/disabled relatives and ‘other family reasons’.\\
        \hline
        Young people neither in employment nor in education and training (NEETS) & - & Share of people aged 15-29 meeting the following two conditions: (a) are unemployed or outside the labour force according to the ILO definition and (b)  have not received any education or training (i.e. neither formal nor non-formal) in the four weeks preceding the LFS. \\
        \hline
        Out-of-pocket expenditure on healthcare & - & Household out-of-pocket payment means a direct payment for healthcare goods and services from the household's primary income or savings, where the payment is made by the user at the time of the purchase of goods or the use of the services.\\
        \hline
        Self-reported unmet need for medical care & - & Share of the population aged 16 and over reporting unmet needs for medical care due to one of the following reasons: ‘Financial reasons’, ‘Waiting list’ and ‘Too far to travel’ (all three categories are cumulated).\\
        \hline
        Deaths from suicide & - & Death rate of the population due to suicide and intentional self-harm adjusted to a standard age distribution.\\
        \hline
        Share of people with good or very good perceived health & + & Proportion of people who assessed their health as very good or good when answering the question on self-perceived health (‘How is your health in general?’).\\
        \hline
        Average trust in EU institutions & + & Share of positive opinions (people who declare that they tend to trust) about a selection of EU institutions: the European Parliament, EC or ECB.\\
        \hline
        Average trust in national government and national parliament & + & Portion of people stating that they trust national government, national parliament\\
        \hline
        Pollution, grime or other environmental problems & - & Share of the population reporting exposure to pollution, grime or other environmental problems.\\
        \hline
        Pollution, noise, grime or other environmental problems & - & Share of the  population reporting exposure to pollution, noise, grime or other environmental problems\\
        \hline
        Crime, violence, or vandalism in the area & - & Share of the population who reported that they face the problem of crime, violence or vandalism in their local area. This perception is not necessarily based on personal experience.\\
        \hline
        Noise from neighbours or from the street & - & Refers to the percentage of the total population who face the problem of too much noise in their dwelling from neighbours or from outside.\\
        \hline
        Standardised death rate due to homicide & - & Standardised death rate of homicide and injuries inflicted by another person with the intent to injure or kill by any means, including ‘late effects’ from assault.\\
        \hline
        Traffic deaths & - & Number of fatalities caused by road accidents, including drivers and passengers of motorised vehicles and pedal cycles as well as pedestrians.\\
         \hline 
         \end{longtable}
         \clearpage
          \begin{longtable}{|p{4cm}|>{\centering\arraybackslash}p{1cm}|p{10cm}|}
           \hline
       \multicolumn{3}{|c|}{\rule[-1ex]{0pt}{5ex}\large\textbf{Resources for the future}}\\
        \hline
        Financial net worth of the total economy & + & Net foreign asset position of a country with respect to the rest of the world. The financial assets include currency, deposits, debt securities, loans, equity and investment fund shares/units, financial derivatives, employment stock options, and other accounts receivable.\\
        \hline
        Gross domestic expenditure on R\&D & + & Gross domestic expenditure on R\&D (GERD) as a percentage of the gross domestic product (GDP), also referred to as ‘R\&D intensity'.\\
        \hline
        Patent applications to the European Patent Office & + & Number of requests for patent protection of an invention filed with the European Patent Office (EPO), regardless of whether they are granted or not.\\
        \hline
        Net fixed capital formation (\%GDP) & + & Net fixed capital formation consists of gross fixed capital formation less consumption of fixed capital.\\
        \hline
        Total fixed assets & + & Fixed assets consist of a subset of produced assets (mostly machinery, equipment, buildings or other structures) that are used repeatedly or continuously in production over periods of time of more than one year.\\
        \hline
        Early leavers from education and training & - & Individuals aged 18-24 who have completed at most a lower secondary education and were not in further education or training during the four weeks preceding the labour force survey (LFS).\\
        \hline
        Spending on prevention & + & Economic resources dedicated to preventive care, excluding capital investment. Preventive care means any measure that aims to avoid or reduce the number or the severity of injuries and diseases, their sequelae and complications; it includes interventions for both individual and collective consumption.\\
        \hline
        Healthy life years at age 65 & + & Number of remaining years that a person at age 65 is expected to live without any severe or moderate health problems. HLY is a composite indicator that combines mortality data with health status data.\\
        \hline
        Life expectancy & + & Life expectancy at certain ages represents the mean number of years still to be lived by a person who has reached a certain exact age, if subjected throughout the rest of his or her life to the current mortality conditions (age-specific probabilities of dying).\\
        \hline
        Children aged less then 3 years in formal childcare & + & Percentage of children (under 3 years old) cared for by formal arrangements other than by the family.\\
        \hline
        Early childhood education & + & The indicator measures the share of the children between the age of three and the starting age of compulsory primary education who participated in early childhood education and care (ECEC) which can be classified as ISCED level 0 according to the International Standard Classification for Education (ISCED 2011). In order for ECEC programmes to be classified as ISCED level 0, they must be intentionally designed to support a child’s cognitive, physical and socio-emotional development. The starting age of compulsory primary education varies across countries. The participation of children in programmes which are not intentionally designed to support child development (such as childcare-only programmes), is not included in this indicator.\\
        \hline
        Lower-secondary completion only & - & Share of people aged 25 to 64 who have completed at most lower secondary education (levels 0-2 in ISCED 2011).\\
        \hline
        Tertiary education attainment (25-34) & + & Share of the population aged 25-34 who have successfully completed tertiary studies (e.g., at university or a higher technical institution).\\
        \hline
         \end{longtable}
         \clearpage
          \begin{longtable}{|p{4cm}|>{\centering\arraybackslash}p{1cm}|p{10cm}|}
          \hline
        \multicolumn{3}{|c|}{\rule[-1ex]{0pt}{5ex}\large\textbf{Societal resilience}}\\
        \hline
        Banking sector total capital ratio & + & The total capital ratio provides a measure of how much capital (equity + subordinated liabilities) the banking sector holds in comparison to the risks faced (credit, market and operational risks). It is calculated as: own Funds/ Total Risk Exposure Amount.\\
        \hline
        Degree of specialisation of the economy & - & Herfindahl index across sectors (NACE2), represents the sectoral concentration of domestic production. The lower the less vulnerable, as highly diversified economies are more resilient.\\
        \hline
        Government debt & - & Total consolidated gross debt at nominal value at the end of the year in the following categories of government liabilities: currency and deposits, debt securities and loans. The general government sector comprises the subsectors of central government, state government, local government and social security funds.\\
        \hline
        Household debt (\% of net disposable income) & - & Total outstanding debt of househods (including non-profit institutions serving households), which includes loans (primarily mortgage loans and consumer credit) and other accounts payable.\\
        \hline
        Adult participation in learning & + & Share of people aged 25 to 64 who stated that they received formal or non-formal education and training in the last 4 weeks preceding the survey. Adult learning covers both general and vocational formal and non-formal learning activities. Adult learning usually refers to learning activities after the end of initial education.\\
        \hline
        Antimicrobial resistance & - & Antimicrobial consumption for the community (primary care) sector expressed as the number of defined daily doses (DDD) per 1000 inhabitants per day.\\
        \hline
        Premature deaths due to exposure to PM2.5 & - & This indicator estimates the number of premature deaths attributable to long-term exposure to PM2.5. PM2.5 are particulates whose diameter is less than 2.5 micrometres and which can be carried deep into the lungs where they can cause inflammation and exacerbate the condition of people suffering heart and lung diseases.\\
        \hline
        Standardised preventable and treatable mortality (low rate) & - & Preventable mortality refers to mortality that can mainly be avoided through effective public health and primary prevention interventions (i.e. before the onset of diseases/injuries, to reduce incidence). Treatable mortality can mainly be avoided through timely and effective health care interventions, including secondary prevention and treatment (after the onset of diseases to reduce case-fatality).\\
        \hline
        Government spending on health, education, social protection & + & Total general government expenditures on education as a share of GDP.\\
        \hline
        Impact of social transfers (excluding pensions) on poverty reduction & + & Reduction in percentage of the risk of poverty rate, due to social transfers.\\
        \hline
        Income stabilisation coefficient & + & Share of a shock in market income (before taxes and transfers) which is absorbed by a country's tax and benefit system and is not transmitted into disposable income (after taxes and transfers).\\
        \hline
        Employment in energy-intensive sectors & - & The share of people employed in the following sectors, relative to total employment: B (mining and quarrying), C20 (manufacture of chemicals and chemical products), C23 (manufacture of other non-metallic mineral products), C24 (manufacture of basic metals), and C29 (manufacture of motor vehicles, trailers and semi-trailers).\\
        \hline
        Employment in manufacturing with high automation risk & - & The following types of activities have been considered: OC3: Technicians and associate professionals; OC4 Clerical support workers; OC5 Service and sales workers; OC8 Plant and machine operators and assemblers; OC9 Elementary occupations.\\
        \hline
        High-speed internet coverage & + & Share of households with fixed very high capacity network (VHCN) connection.\\
        \hline
        Digital public services for citizens & + & The share of administrative steps that can be done online for major life events (birth of a child, new residence, etc.) for citizens (unit: score from 0 to 100).\\
        \hline
        ICT graduates & + & Individuals with a degree in ICT as a share of all graduates.\\
        \hline
        Net International Investment Position & - & Difference between an economy's external financial assets and liabilities, calculated as percentage of GDP, multiplied by minus one (so that the higher the more vulnerable).\\
        \hline
        Net lending/borrowing & + & Net resources that the total economy makes available to the rest of the world (surplus or net lending) or receives from the rest of the world (deficit or net borrowing needs).\\
        \hline
        Material import dependency & - & Ratio of imports (IMP) to direct material inputs (DMI) in percentage, indicating reliance on imports for meeting material needs, ranging from 0\% to 100\%.\\
        \hline
        Supplier concentration in base metals and energy carriers & - & Supplier concentration in base metals and energy carriers.\\
        \hline
        Concentration of value chain partners & - & Average of the concentration (Herfindahl) index of each Member States' extra-EU partners for imported and re-exported content. The importing concentration is calculated as the squared sum of imports by source country as share of the importing country's GDP. The re-exporting concentration is calculated as the squared sum of re-exports by destination country as share of the re-exporting country's GDP. The resulting index has been multiplied by 1000.\\
        \hline
        Extra-EU trade partner concentration & - & Herfindahl index of each Member State's extra-EU import partners.\\
        \hline
        Trade openness & + & Trade openness is measured as the sum of a country's exports and imports as a percentage of that country's GDP, considering non-EU partners only.\\
        \hline
        Net migration rate & + & Net migration rate measured as the difference between immigration and emigration to non-EU-27 countries relative to the host country's population.\\
        \hline
        Population change & + & Change p.a. in percentage point, 2011-2020.\\
        \hline
        Total fertility rate (difference from replacement-level) & + & This indicator is calculated by subtracting the country's total fertility rate from 2.1 , which represents the replacement rate.\\
        \hline
        \multicolumn{3}{|c|}{\rule[-1ex]{0pt}{5ex}\large\textbf{Nature}}\\
        \hline
        Area under organic farming & + & This indicator measures the share of total utilised agricultural area (UAA) occupied by organic farming (existing organically-farmed areas and areas in process of conversion).\\
        \hline
        Farmland bird index & + & Average population trend of bird species in farmland habitats.\\
        \hline
        Natura 2000 & + & The Natura2000 network consists of sites designated by Member States under the Birds Directive and the proposed sites of Community importance (pSCI) that are subsequently designated under the Habitats Directive. Marine and terrestrial combined.\\
        \hline
        Soil carbon content & + & Based on the European harmonized topsoil database from LUCAS, indicating organic carbon content in relation to slope, land cover, temperature, productivity, and GPS position.\\
        \hline
        Air pollution pm10 & - & Population weighted average of annual average concentration of particle matter of size 10 micrometers (big particles) in $\mu$g/m\textsuperscript{3}, interpolated at 1 km\textsuperscript{2} grid cell level and combined with GEOSTAT 1 km\textsuperscript{2} grid population data (For NUTS2 in SPI).\\
        \hline
        Air pollution pm2.5 & - & Population weighted average of annual average concentration of particle matter of size 2.5 micrometers (small particles) in $\mu$g/m\textsuperscript{3}, interpolated at 1 km\textsuperscript{2} grid cell level and combined with GEOSTAT 1 km\textsuperscript{2} grid population data (for NUTS2 in SPI). \\
        \hline
        Water exploitation index plus (WEI+) & - & The WEI+ easures total water consumption as a percentage of renewable freshwater resources. It quantifies water abstraction, return, and consumption for a given territory, considering river basins and economic sector discharges.\\
        \hline
        Energy productivity & + & Amount of economic output that is produced per unit of gross available energy (defined as the quantity of energy products necessary to satisfy all demand of entities in the geographical area under consideration). \\
        \hline
        Primary energy consumption & - & This indicator reflects a country's overall energy demand, encompassing energy sector consumption, losses during transformation and distribution, and final consumption by end-users. It excludes non-energy uses of carriers like natural gas for chemicals.\\
        \hline
        Share of renewable energy in gross final energy consumption & + & Share of renewable energy in gross final energy consumption (defined as gross electricity production from all energy sources plus total imports of electricity minus total exports of electricity).\\
        \hline
        Share of buses and trains in inland passenger transport & + & Share of collective transport modes in total inland passenger transport performance, expressed in passenger-kilometres (pkm). Collective transport modes refer to buses, including coaches and trolley-buses, and trains. Total inland transport includes transport by passenger cars, buses and coaches, and trains.\\
        \hline
        Share of rail and inland waterways in total freight transport & + & Share of rail and inland waterways in total inland freight transport, expressed in tonne-kilometres (tkm).\\
        \hline
        GHG emissions from land use, land use change and forestry & - & Net carbon removals in the Land Use, Land Use Change, and Forestry (LULUCF) sector, accounting for both sector emissions and removals. Results are expressed as CO\textsubscript{2} equivalents, incorporating the global warming potential (GWP) of each gas.\\
        \hline
        GHG emissions intensity of the economy & - & Greenhouse gases (GHG) emitted by the entire national economy per person employed.\\
        \hline
        Net greenhouse gas emissions & - & Man-made greenhouse gas (GHG) emissions and removals of the ‘Kyoto basket' (CO\textsubscript{2}, C\textsubscript{4}, N\textsubscript{2}O, and F-gases), expressed in CO\textsubscript{2} equivalents with consideration for each gas's global warming potential (GWP). Net GHG emissions encompass international aviation and GHG removals from LULUCF.\\
        \hline
        Climate-related economic losses & - & Weather and climate-related events are defined as meteorological events (storms), hydrological events (floods, mass movements) and climatological events (heatwaves, cold waves, droughts, forest fires). It is based on data from CATDAT of RiskLayer.\\
        \hline
        Fatalities from climate extremes & - & Number of fatalities to weather or climate-related extreme events over the periods 1980-2019 and 1980-2020, per 1 million population.\\
        \hline
        Circular material use rate & + & Proportion of recycled material contributing to overall material use. Overall material use combines domestic material consumption (DMC) with circular material use (M = DMC + U). Circular use approximates recycled waste in domestic recovery plants (RCV\_R), subtracting imported waste for recycling (IMPw) and adding exported waste for recycling abroad (EXPw).\\
        \hline
        Recycling rate of municipal waste & + & Proportion of municipal waste recycled, calculated by dividing the recycled tonnage by the total municipal waste generated. Recycling includes material recycling, composting, re-use preparation, and anaerobic digestion. Municipal waste encompasses household waste and similar waste from small businesses and public institutions, collected by municipalities\\
        \hline
        Carbon footprint & - & Estimate of the total per capita emissions of carbon dioxide associated with domestic consumption, including both CO\textsubscript{2} emitted and consumed domestically and CO\textsubscript{2} emitted abroad and embodied in imports.\\
        \hline
        Consumption footprint per capita & - & The consumption footprint quantifies the environmental impacts resulting from the consumption, including the embodied and indirect impacts. It accounts for domestic production in the EU and trade with other world regions.\\
        \hline
        Raw material consumption (Material footprint) & - & Amount of extraction required to produce the products demanded by final users in the geographical reference area, irrespective of where in the world the extraction of material from the environment took place.\\
        \hline
        Resource productivity & + & The ratio of the circular use of materials (recycled products and recovered materials) to the overall material use.\\
        \hline
        Employment in the environmental goods and services sector & + & The environmental goods and services sector (EGSS) is defined as that part of a country’s economy that is engaged in producing goods and services that are used in environmental protection and resource management activities either domestically or abroad.\\
        \hline
        Share of environmental taxes in total tax revenues & + & Environmental taxes are defined as taxes that are based on a physical unit (or proxy of it) of something that has a proven, specific negative impact on the environment. There are four types: energy taxes, transport taxes, and pollution and resource taxes.\\
         \hline
\multicolumn{3}{|c|}{\rule[-1ex]{0pt}{5ex}\large\textbf{Inclusiveness}}\\
        \hline
        Gender employment gap & - & Difference between the employment rates of men and women aged 20 to 64.\\
        \hline
        Income quintile share ratio (S80/S20) & - & Ratio of total income received by the 20\% of the population with the highest income (the top quintile) to that received by the 20\% of the population with the lowest income (the bottom quintile).\\
        \hline
        Income share of the bottom 40\% of the population & + & Total disposable household income share received by the bottom 40\% of the population.\\
         \hline
\multicolumn{3}{|c|}{\rule[-1ex]{0pt}{5ex}\large\textbf{Institutions}}\\
        \hline
        Control of corruption & + & Control of Corruption (CC) captures perceptions of the extent to which public power is used for 
        private gain, including both petty and grand corruption, as well as capture of the state by elites and 
        private interests.\\
        \hline
        Government effectiveness & + & Government Effectiveness (GE) captures perceptions of the quality of public services, the civil 
        service, policy formulation and implementation, and the credibility of a government’s decisions.\\
        \hline
        Political stability & + & Political Stability (PV) captures perceptions of the extent to which political power and governance 
        are secure from destabilization, and of the likelihood that authority will be challenged or altered 
        through violent, coercive, or unconstitutional means.\\
        \hline
        Regulatory quality & + & Regulatory Quality (RQ) captures perceptions of the government’s ability to design and implement 
        policies and regulations that promote private sector development.\\
        \hline
        Rule of Law & + & Rule of Law (RL) captures perceptions of the extent to which agents respect and follow the rules of 
        society, including contract enforcement, property rights, the police, courts, and the likelihood of
        crime and violence.\\
        \hline
        Voice and accountability & + & Voice and Accountability (VA) captures perceptions of the extent to which citizens can participate 
        in selecting their government including electoral integrity, and of accountability mechanisms for 
        citizens—reflected in the ability to access information, governmental oversight bodies, and a robust 
        traditional/digital media landscape.\\
        \hline
        Worldwide Governance Index & + & The WGI feature six aggregate governance indicators for over 200 countries and territories over the period 1996–2022: Voice and Accountability,  Political Stability and Absence of Violence/Terrorism, Government Effectiveness, Regulatory Quality, Rule of Law and Control of Corruption.\\
        \hline
    \caption{List of the 104 indicators used in the empirical analysis, with their
thematic block, brief description, and sign of association with wellbeing.}
    \end{longtable}
    \label{tab:var_and_block}

\clearpage
\bibliographystyle{chicago}
\setlength{\bibsep}{5pt plus 0.2ex}
{\small
\spacingset{1}
\bibliography{reference}
}